\documentclass[aps,prx,superscriptaddress,twocolumn,showpacs,longbibliography]{revtex4-1}

\usepackage{amssymb}
\usepackage{amsmath}
\usepackage{amsthm}
\usepackage{graphicx}
\usepackage{amsfonts}
\usepackage{color}
\usepackage{times}
\usepackage{natbib}
\usepackage{comment}
\usepackage{soul}
\usepackage{booktabs}
\usepackage{pifont}
\usepackage{enumitem}
\usepackage[normalem]{ulem}
\usepackage{hyperref}
\hypersetup{
        colorlinks=true,linkcolor=blue,citecolor=blue,urlcolor=blue
}

\AtBeginDocument{%
    \newwrite\bibnotes
    \def\bibnotesext{Notes.bib}
    \immediate\openout\bibnotes=\jobname\bibnotesext
    \immediate\write\bibnotes{@CONTROL{REVTEX41Control}}
    \immediate\write\bibnotes{@CONTROL{%
    apsrev41Control,author="08",editor="1",pages="1",title="0",year="1"}}
     \if@filesw
     \immediate\write\@auxout{\string\citation{apsrev41Control}}%
    \fi
}%

\newcommand{\kBT}{k_{\text{B}}T}
\newcommand{\rmL}{{\text{L}}}
\newcommand{\rmR}{{\text{R}}}
\newcommand{\rmC}{{\text{C}}}
\newcommand{\rmH}{{\text{H}}}
\newcommand{\rmS}{{\text{S}}}
\newcommand{\rmd}{{\text{d}}}
\newcommand{\rme}{{\text{e}}}
\newcommand{\rmh}{{\text{h}}}
\newcommand{\rmc}{{\text{c}}}
\newcommand{\rms}{{\text{s}}}

\newcommand{\e}{{\rm e}}

\newcommand{\beq}{\begin{equation}}
\newcommand{\eeq}{\end{equation}}
\newcommand{\bea}{\begin{eqnarray}}
\newcommand{\eea}{\end{eqnarray}}
\newcommand{\ba}{\begin{align}}
\newcommand{\ea}{\end{align}}

\renewcommand{\AA}

\begin{document}
\title{
\vspace*{-1.25cm}
\textnormal{{\small  PHYSICAL REVIEW RESEARCH {\bf 1}, 033066 (2019)}}\\
\vspace*{-0.2cm}
\rule[0.1cm]{18cm}{0.02cm}\\
\vspace*{0.285cm}
Autonomous conversion of information to work in quantum dots}

\author{Rafael S\'anchez}
\affiliation{Departamento de F\'isica Te\'orica de la Materia Condensada, Condensed Matter Physics Center (IFIMAC) and Instituto Nicol\'as Cabrera, Universidad  Aut\'onoma de Madrid, 28049 Madrid, Spain\looseness=-1}
\author{Peter Samuelsson}
\author{Patrick P. Potts}
\affiliation{Physics Department and NanoLund, Lund University, Box 118,  22100 Lund, Sweden}

\begin{abstract}
We consider an autonomous implementation of Maxwell's demon in a quantum dot architecture acting on a system without changing its number of particles or its energy. As in the original thought experiment, only the second law of thermodynamics is seemingly violated when disregarding the demon. The autonomous architecture allows us to compare descriptions in terms of information to a more traditional, thermoelectric characterization. Our detailed investigation of information-to-work conversion is based on fluctuation relations and second law like inequalities in addition to the average heat and charge currents. By introducing a time-reversal on the level of individual electrons, we find a novel fluctuation relation that is not connected to any symmetry of the moment generating function of heat and particle flows. Furthermore, we show how an effective Markovian master equation with broken detailed balance for the system alone can emerge from a full description, allowing for an investigation of the entropic cost associated with breaking detailed balance. Interestingly, while the entropic cost of performing a perfect measurement diverges, the entropic cost of breaking detailed balance does not. Our results connect various approaches and idealized scenarios found in the literature and can be tested experimentally with present day technology.
\end{abstract}

\maketitle

\section{Introduction}
\begin{figure}[b]
	\includegraphics[width=\columnwidth]{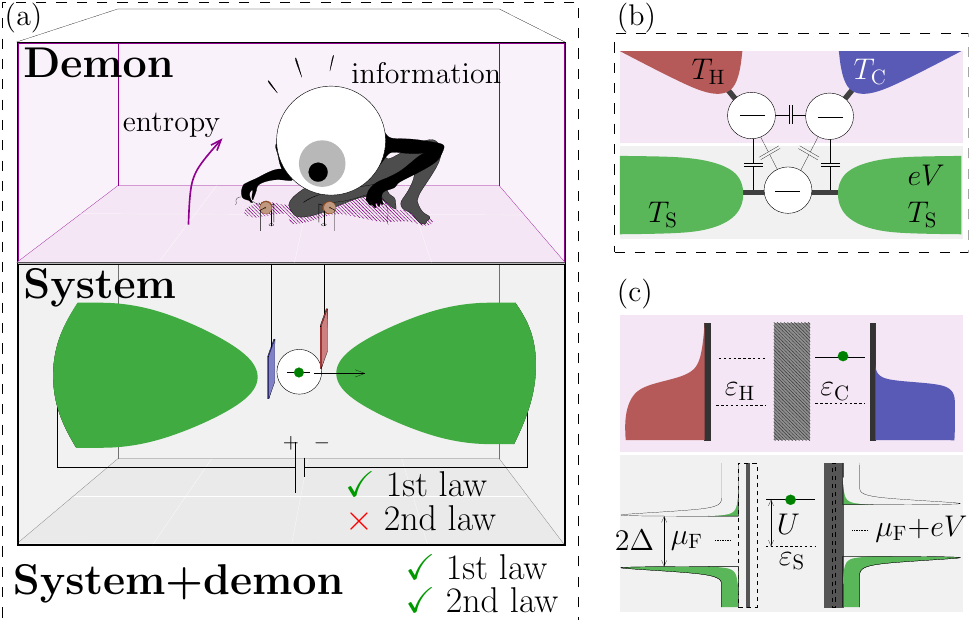}
	\caption{Composite system under investigation. (a) Illustration of the demon, measuring the occupation of a quantum dot and manipulating the transition probabilities such that electrons enter the dot from the left reservoir and leave the dot towards the right reservoir. Within the system alone, the first law of thermodynamics holds while the second law is violated. (b) Real space schematics and (c) spectral properties of an implementation with quantum dots. Both the system as well as the demon can host at most one electron. The demon includes a temperature gradient which is used to overcome the voltage bias within the system. Capacitive couplings (with equal charging energy $U$) between the system dot and the demon dots serve for \textit{measuring} the occupation of the system dot. Capacitive couplings between the demon dots and the tunnel barriers within the system result in the desired modulation of transition rates.}
	\label{fig:system}
\end{figure}

Many aspects of the theory of thermodynamics are intricately related to the concept of information \cite{lindblad:1974,parrondo:2015,goold:2016}. For instance, entropy can be understood as a lack of microscopic information about the system and its environment \cite{esposito:2010njp,sagawa:inbook}. The second law then merely states that, on average, information will be lost to a thermal environment. Similarly, heat can be understood as the change in energy of degrees of freedom that cannot be observed \cite{callen:book}. From this point of view, it is natural that measurements, which provide information, allow for decreasing entropy and for converting heat into work. Historically, the idea of converting information to work originated from Maxwell's well known thought experiment \cite{maxwell:1871}, where a \textit{``being whose faculties are so sharpened that he can follow every molecule in its course''} was introduced as an agent that obtains information about microscopic degrees of freedom. This fictional being is usually addressed as Maxwell's demon \cite{maxwell:book}. Acting on the obtained information, the demon can seemingly violate the second law of thermodynamics. The reason this does not result in any practical device that can overcome the laws of thermodynamics is best expressed in Landauer's famous quote: \textit{``Information is physical''} \cite{landauer:1992}. Any device that performs a measurement and stores the outcome must itself be a physical device and should therefore be taken into account in the thermodynamic bookkeeping \cite{landauer:1961,bennett:1982}. It then follows that any measurement-feedback device, or demon, that is itself limited by the laws of thermodynamics will generate a sufficient amount of entropy such that the second law is restored \cite{funo:2013,horowitz:2014prx}, as illustrated in Fig.~\ref{fig:system}.

In macroscopic thermodynamics, the distinction between accessible and inaccessible degrees of freedom is very clear \cite{callen:book}. The former could for instance be a position of a weight and the latter the random movements of molecules in a gas. Any device like Maxwell's demon therefore acts on a scale that is completely different from the observable degrees of freedom. This picture changes drastically in nanoscopic systems, where all processes are happening on a similar scale and fluctuations can no longer be neglected. In recent years, this regime, described by stochastic thermodynamics \cite{harris:2007,esposito:2009rmp,jarzynski:2011,seifert:2012,mansour:2017}, has seen tremendous progress both theoretically as well as experimentally. Theoretically, strong and exact results such as the Jarzynski relation \cite{jarzynski:1997,jarzynski:1997pre} and the Crooks fluctuation theorem \cite{crooks:1998,crooks:1999,crooks:2000,kurchan:2000,tasaki:2000} have deepened our understanding of fluctuating thermodynamic processes far from equilibrium. This investigation has furthermore clarified how the second law arises from microscopic equations of motion that are time-reversal symmetric \cite{esposito:2010njp,sagawa:inbook,kawai:2007,jarzynski:2011} and how logical information should be taken into account in thermodynamic bookkeeping \cite{touchette:2000,sagawa:2008,maruyama:2009,cao:2009,sagawa:2010,horowitz:2010,ponmurugan:2010,morikuni:2011,horowitz:2011,sagawa:2012,sagawa:2012prl,sagawa:2013,lahiri:2012,abreu:2012,funo:2013,deffner:2013,ito:2013,barato:2014,barato:2014pre,sagawa:2014,ashida:2014,horowitz:2014,horowitz:2014prx,um:2015,strasberg:2015,shiraishi:2015,funo:2015,shiraishi:2015,koski:2016,wachtler:2016,gong:2016,kwon:2017,potts:2018}. Experimental advances in controlling small systems have opened up the possibility of investigating stochastic thermodynamics in a variety of platforms including electronic systems \cite{giazotto:2006,joubaud:2008,ciliberto:2013,koski:2014,koski:2014prl,
thierschmann:2015,roche:2015,koski:2015,thierschmann_thermoelectrics_2016,hofmann:2016,
hofmann:2017,chida:2017,pekola:2015,pekola:2019,hartman_direct_2018,singh_extreme_2019,
kleeorin_measuring_2019,gonzalez:2019}, DNA molecules \cite{alemany:2012,dieterich:2016}, photons \cite{vidrighin:2016}, Brownian particles \cite{toyabe:2010,berut:2012}, and ultracold atoms \cite{kumar:2018}. Extensions to the quantum regime \cite{thermo:book}, where additional subtleties and challenges are encountered, have already led to many exciting insights both from theory \cite{kosloff:2013,vinjanampathy:2016,campisi:2011,goold:2016} as well as from experiment \cite{maslennikov:2019,cottet:2017,masuyama:2018,naghiloo:2018}.

Even centuries after its conception, Maxwell's (and Szilard's \cite{szilard:1929}) thought experiment provides a prime concept for illustrating novel ideas and insights in the thermodynamics of information. Studies on Maxwell's demon can broadly be grouped into two categories. Demons which rely on external control \cite{toyabe:2010,schaller:2011,esposito:2012,bergli:2013,koski:2014,koski:2014prl,sandberg:2014,vidrighin:2016,kutvonen:2016pre,chida:2017,masuyama:2018,naghiloo:2018,schaller:2018,engelhardt:2018,kumar:2018} and autonomous demons \cite{mandal:2012,mandal:2013,strasberg:2013,barato:2013,deffner:2013,deffner:2013pre,hartich:2014,cao:2015,koski:2015,shiraishi:2015njp,rana:2016,kutvonen:2016,boyd:2016,whitney:2016,cottet:2017,rossello:2017,spinney:2018,ptaszynski:2018,strasberg:2018,erdman_absorption_2018,sanchez:2018}, where no external control is needed. Connections between autonomous and non-autonomous implementations of Maxwell's demon were investigated in Refs.~\cite{horowitz:2013,strasberg:2014,barato:2014,shiraishi:2015njp,shiraishi:2016,
strasberg:2017}. Autonomous demons offer the possibility to keep track of information flows and to investigate the necessary entropy production associated with a certain level of performance. Such devices often rely either on an information reservoir \cite{mandal:2012,mandal:2013,barato:2013,deffner:2013,deffner:2013pre,cao:2015,boyd:2016,rana:2016,spinney:2018}, providing a storage medium for the measurement outcomes, or on thermal reservoirs \cite{strasberg:2013,hartich:2014,koski:2015,kutvonen:2016,whitney:2016,rossello:2017,ptaszynski:2018,strasberg:2018,sanchez:2018}. The latter are of particular interest as they only require standard thermodynamic resources and are thus within the paradigm that addresses the question \textit{what can be achieved by coupling a small system to thermal reservoirs?} These devices can give insight into how descriptions in terms of information relate to a more traditional description in terms of energy flows alone, where it is well established how to account for the required resources.

Here we consider autonomous demons based on quantum dots that only require thermal reservoirs. Quantum dots and metallic islands, where electrons can hop between well defined regions, provide promising architectures because of multiple reasons \cite{sothmann:2015,pekola:2019}: First, charging energies confine the system to few states, resulting in a tractable behavior that is well described by Markovian master equations. Furthermore, these systems are comparably stable over time \cite{pekola:2019}. Second, tunneling rates and on-site energies can be controlled by external gates which allows for tuning the relevant time- and energy-scales in situ~\cite{sanchez:2011,thierschmann:2015}. Third, all relevant ingredients for stochastic thermodynamics, such as a temperature gradient, have already been implemented experimentally~\cite{staring_coulomb-blockade_1993,dzurak_observation_1993,dzurak_thermoelectric_1997,
scheibner_sequential_2007,thierschmann_diffusion_2013,
svensson_nonlinear_2013,thierschmann_thermal_2015,josefsson_quantum-dot_2018}. Indeed, experiments based on Maxwell's thought experiment have been reported in Refs.~\cite{koski:2014,koski:2014prl,koski:2015,hofmann:2016,hofmann:2017,chida:2017}.

The system under investigation is sketched in Fig.~\ref{fig:system}. Before looking into the working principle and the details of the information-to-work conversion, we briefly summarize our main results and the main merits of the considered system:
\begin{enumerate}[nolistsep]
	\item In the spirit of Maxwell's original thought experiment, and in contrast to previous investigations \cite{strasberg:2013,koski:2015,kutvonen:2016}, the energy of the system does not change. The first law is hence respected by both the demon as well as the system alone. Only the second law is seemingly violated when disregarding the entropy production of the demon. 
	\item Our model allows for a detailed investigation of information-to-work conversion keeping track of the entropy produced by the demon. This allows for comparing a description based on information to a description in terms of a machine that uses only conventional thermal resources.
	\item In the limit of a fast demon, we find a Markovian master equation for the system alone, where detailed balance is explicitly broken. This allows for investigating the thermodynamic cost of breaking detailed balance.
	\item Considering different forms of time-reversal, we find different fluctuation relations. Notably, a novel form of time-reversal on the single-particle level results in a novel fluctuation relation which cannot be found from the full counting statistics of heat and charge currents.
\end{enumerate}

The rest of this article is structured as follows: In Sec.~\ref{sec:II}, we introduce the system and illustrate how information is used to convert heat into work. Section \ref{sec:III} discusses the system as a heat engine. A description based on information flows is given in Sec.~\ref{sec:IV}. In Sec.~\ref{sec:V} we go beyond mean values and introduce fluctuation relations. Efficiencies and second law like inequalities are discussed in Sec.~\ref{sec:VI}. Section \ref{sec:VII} is devoted to experimental considerations, and we conclude in Sec.~\ref{sec:VIII}.

\section{System and working principle}
\label{sec:II}
The system under investigation is sketched in Fig.~\ref{fig:system}. It consists of three quantum dots and four fermionic reservoirs. The system can be divided into two parts that are only coupled through the Coulomb repulsion between electrons. The lower part [in Fig.~\ref{fig:system}(b)] of the system will be referred to as the \textit{system} while the upper part will be referred to as the \textit{demon}. The combination of system and demon will henceforth be referred to as the \textit{composite system}. Related bipartite systems are interesting for Coulomb drag~\cite{sanchez:2010,bischoff:2015,kaasbjerg:2016,keller:2016}, heat engines~\cite{sanchez:2011,thierschmann:2015,dare:2017,walldorf:2017}, fluctuation theorems~\cite{sanchez:2010,schaller:2010,bulnes:2011}, thermal drag~\cite{bhandari:2018}, and refrigerators~\cite{zhang:2015,sanchez:2017,erdman_absorption_2018}. The system contains a single quantum dot which hosts a single energy level and is tunnel-coupled to two superconducting leads. Throughout the paper, these leads will be referred to as the \textit{left} and \textit{right} reservoir. They are described by the common temperature $T_{\rm S}=T$ and possibly different chemical potentials $\mu_{\rm L}$, $\mu_{\rm R}$, where we define
\begin{equation}
\label{eq:voltage}
eV=\mu_{\rm R}-\mu_{\rm L}.
\end{equation}
For positive voltages, there is thus a tendency for electrons to flow from the right to the left reservoir. Since no charge transfer in the absence of the demon is desired, the energy level of the dot, $\varepsilon_S$, is placed within the superconducting gap of both leads, see Eq.~\eqref{eq:ingapes} below. We assume the intra-dot Coulomb interaction to be so strong that the quantum dot can host at most one electron~\cite{kouwenhoven:1997}.

The demon consists of two quantum dots, each hosting a single energy level ($\varepsilon_{\rm C/H}$) and coupled to a normal conducting lead. The temperatures of these leads are different $T_{\rm H}\geq T_{\rm C}$ where the subscripts stand for \textit{hot} and \textit{cold}. In the following, we will refer to the demon quantum dots as the \textit{cold dot} and \textit{hot dot}, reflecting their respective reservoir. We assume both the intra-dot Coulomb repulsion as well as the inter-dot Coulomb repulsion within the demon to be so strong that at most one electron can occupy the demon. It also suppresses pair tunneling in the system.

Within the demon, there is a tendency of heat to flow from hot to cold.
As in conventional thermoelectrics, this tendency will be used to drive a charge flow against the voltage bias in the system. To this end, the demon and the system are coupled to each other through the Coulomb interaction. Unlike in conventional thermoelectrics, this capacitive coupling connects the system to the (demon) heat sources while keeping it electrically isolated~\cite{sanchez:2011}. We assume that the couplings between the system dot and the two demon dots are identical and captured by a charging energy $U$ (deviations from this are discussed in Sec.~\ref{sec:V}). We then demand
\begin{equation}
\label{eq:ingapes}
\mu_{\rm L/R}-\Delta<\varepsilon_S<\mu_{\rm L/R}+\Delta<\varepsilon_S+U,
\end{equation}
where $\Delta$ denotes the superconducting gap assumed to be equal for both leads. Then, tunneling in the system dot is suppressed by the gap when the demon is empty. This ensures that electrons can only enter or leave the system dot at energy $\varepsilon_S+U$, preventing any energy flow between the system reservoirs and the demon~\cite{sanchez_all-thermal_2017}. To obtain an implementation of Maxwell's demon, we choose the chemical potential of the cold reservoir such that
\begin{equation}
\label{eq:muc}
\varepsilon_{\rm C}<\mu_{\rm C}<\varepsilon_{\rm C}+U.
\end{equation}
At low temperatures, the cold dot will thus tend to be filled if the system dot is empty and vice versa. This anti-correlation provides the demon with information on the system state and is illustrated by the demon's eye in Fig.~\ref{fig:system}(a).

The final ingredient that is required is an effect of the repulsive Coulomb interaction on the tunnel barriers between the system dot and its reservoirs. In particular, we assume that if the cold dot is occupied, tunneling between the system dot and the right reservoir is suppressed. Similarly, an occupied hot dot is assumed to result in a suppressed tunneling between the system dot and the left reservoir. This effect is analogous to the current suppression resulting from single electrons which is exploited in charge counting experiments \cite{lu:2003,vandersypen:2004,fujisawa:2006,gustavsson:2006,ubbelohde:2012}. Through this effect, the demon effectively opens and closes the connections between the system dot and the corresponding reservoirs. This is illustrated by the hands of the demon in Fig.~\ref{fig:system}(a).

\begin{figure*}
	\centering
	\includegraphics[width=.8\textwidth]{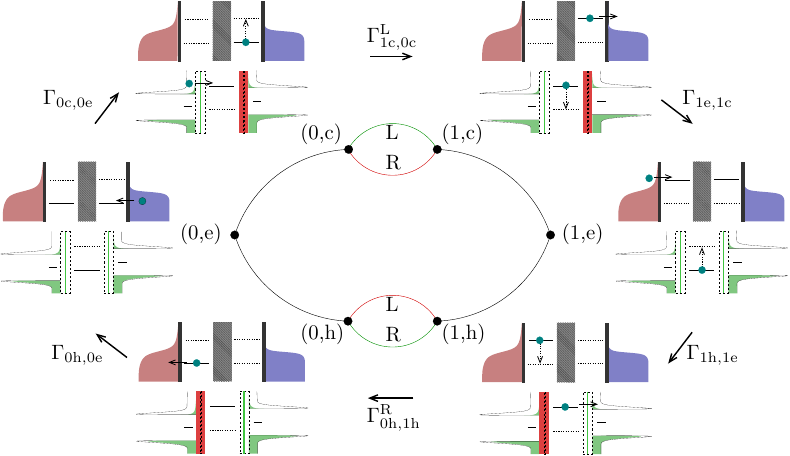}
	\caption{Working principle and graph representation. Inner part: graph representation of Eq.~\eqref{eq:rateeq}. Each node corresponds to a state $(s,d)$ and each line corresponds to possible transitions between states. The transitions illustrated in dashed-red are suppressed by the demon while the transitions illustrated in green are not. Outer part: desired behavior of the composite system. Starting with three empty dots (left-hand side), the system dot cannot be filled due to the superconducting gaps. The cold dot can be filled, lifting the energy level in the system above the superconducting gaps. When the cold dot is filled, transitions involving the right reservoir are suppressed. The system dot can be filled from the left. This raises the energy level of the cold dot above $\mu_{\rm C}$, such that the electron can leave. The electron in the system dot is now trapped due to the superconducting gaps until the hot dot is being filled. When the hot dot is full, transitions involving the left reservoir are suppressed and the system dot is emptied to the right. The cycle is closed when the hot dot is emptied. In one cycle, one electron is transported against the voltage bias and one quantum of heat of size $U$ is transported from hot to cold.}
	\label{fig:cycle}
\end{figure*}

The desired effect of these ingredients is to move charges against the voltage bias as illustrated in detail in Fig.~\ref{fig:cycle}: For an empty system dot, the cold dot is occupied, blocking charge transfer from the right reservoir. Once the system is filled from the left, the cold dot will be emptied and the energy level of the system dot will drop inside the superconducting gap, preventing any charge transport. The hot dot has a chance of becoming occupied even when the system dot is occupied. In this case, charge transfer back to the left reservoir is blocked and the system dot can only be emptied to the right. Emptying the hot dot and filling the cold dot then closes the cycle. For every cycle, one electron is moved against the voltage bias and an amount of heat $U$ flows from hot to cold. In this way, the temperature gradient within the demon can drive a charge current against the voltage bias within the system.

Before delving into a quantitative account of the dynamics, we summarize the necessary ingredients:
\begin{enumerate}[nolistsep]
	\item Coulomb repulsion between system and demon results in an empty cold dot if the system dot is occupied and vice versa. This anti-correlation constitutes the ``eye" of the demon.
	\item Coulomb repulsion between the demon dots and the barriers suppresses tunneling to the right/left reservoir if the cold/hot dot is occupied. This constitutes the ``hands" of the demon.
	\item Superconducting gaps prevent charge transport through the system when the demon is empty.
	\item Equal Coulomb interactions of strength $U$ between the system dot and the two demon dots prevent energy flow between the demon and the system.
\end{enumerate}

\subsection{Master equation}
The composite system is described by six different states that are labeled $(s,d)$. The system can be empty ($s=0$) or occupied ($s=1$); the demon can be empty $(d=\rm{e})$, the cold dot can be occupied $(d=\rm{c})$ or the hot dot can be occupied $(d=\rm{h})$. In the limit of weak coupling between the quantum dots and the reservoirs, the composite system can be described by a Markovian rate equation \cite{breuer:book,schaller:book}
\begin{equation}
\label{eq:rateeq}
\partial_t P_{sd} =\sum_{s',d',\alpha}\left[\Gamma^{\alpha}_{sd,s'd'}P_{s'd'}-\Gamma^{\alpha}_{s'd',sd}P_{sd}\right],
\end{equation}
where $P_{sd}$ denotes the probability to be in the state $(s,d)$ and $\Gamma^{\alpha}_{sd,s'd'}$ denotes the transition rate from state $(s',d')$ to state $(s,d)$ induced by reservoir $\alpha\in\{{\rm L},{\rm R},{\rm H}, {\rm C}\}$. The transition rates can be found by Fermi's golden rule and read (for all non-vanishing transitions)
\begin{equation}
\label{eq:rates}
\begin{aligned}
&\Gamma_{s{\rm c},s\rm{e}}=\Gamma_{\rm C} f_{\rm C}^s,&& \Gamma_{1\rm{c},0\rm{c}}^{\rm L}=\Gamma_{\rm L} f_{\rm L}, && \Gamma_{1\rm{c},0\rm{c}}^{\rm R}=\Gamma_{\rm R}e^{-\delta_{\rm R}} f_{\rm R},\\
&\Gamma_{s{\rm h},s\rm{e}}=\Gamma_{\rm H} f_{\rm H}^s,&&\Gamma_{1\rm{h},0\rm{h}}^{\rm R}=\Gamma_{\rm R} f_{\rm R},&&\Gamma_{1\rm{h},0\rm{h}}^{\rm L}=\Gamma_{\rm L}e^{-\delta_{\rm L}} f_{\rm L},
\end{aligned}
\end{equation}
with their reversed transitions obtained by replacing $f=f_\alpha,f_\alpha^s$ by $1-f$.
Here $\Gamma_\alpha$ denotes the tunneling rate associated with reservoir $\alpha$ and $\delta_{\rm L/R}$ characterizes the suppression of tunneling induced by the occupation of the demon dots (i.e., $\delta_{\rm L/R}\rightarrow\infty$ corresponds to complete suppression and $\delta_{\rm L/R}=0$ to no effect). For ease of notation, we omitted the superscript $\alpha$ whenever there is only one reservoir which can induce the corresponding transition. The different Fermi-Dirac distributions read
\begin{equation}
\label{eq:fermich}
f_{\rm C/H}^s = \frac{1}{e^{\beta_{{\rm C/H}}(\xi_{\rm C/H}+sU)}+1},
\end{equation}
and
\begin{equation}
\label{eq:fermilr}
f_{\rm L/R} = \frac{1}{e^{\beta(\xi_{\rm L/R}+U)}+1},
\end{equation}
with the inverse temperature $\beta_\alpha=1/T_\alpha$ and $\xi_\alpha=\varepsilon_\alpha-\mu_\alpha$ (interpreting $\varepsilon_{\rm L}=\varepsilon_{\rm R}=\varepsilon_{\rm S}$). We further included the (normalized) superconducting density of states in the transition rates, such that
\begin{equation}
\label{eq:dos}
\begin{aligned}
\Gamma_\alpha &= \Gamma_\alpha^{\rm N}\mathcal{N}_\alpha,\hspace{.75cm}\mathcal{N}_{\rm C/H}=1,\\
\mathcal{N}_{\rm L/R}&=\left|{\rm Re}\left(\frac{\xi_{\rm L/R}+U+i\gamma}{\sqrt{(\xi_{\rm L/R}+U+i\gamma	)^2-\Delta^2}}\right)\right|,
\end{aligned}
\end{equation}
where the superscript N denotes the tunnel rate for a normal conductor. We have considered a tiny but finite inverse quasiparticle lifetime, $\gamma$, to avoid numerical discontinuities~\cite{dynes:1978}.
The transition rates in Eq.~\eqref{eq:rates} fulfill detailed balance, ensuring that, in equilibrium, each transition is compensated for by its reverse \cite{breuer:book,schaller:book}. We thus find
\begin{equation}
\label{eq:detailbal}
\frac{\Gamma_{1\rm{c},0\rm{c}}^{\rm R}}{\Gamma_{0\rm{c},1\rm{c}}^{\rm R}}=e^{-\beta(\xi_{\rm R}+U)},
\end{equation}
and a similar expression for all other transitions.

From Eq.~\eqref{eq:rates}, it can be seen that many transition rates in Eq.~\eqref{eq:rateeq} vanish. This becomes particularly apparent when expressing Eq.~\eqref{eq:rateeq} as a graph \cite{schnakenberg:1976}, see the inner part of Fig.~\ref{fig:cycle}. Every state $(s,d)$ corresponds to a node in the graph and all non-vanishing transitions are denoted by a line connecting two nodes.

\subsection{Limiting cases}
As discussed above, the desired behavior of the composite system is described by the directed cycle shown in the outer part of Fig.~\ref{fig:cycle}. In general however, the behavior will deviate from this because of two reasons. First, for finite $\delta_{\rm L/R}$, undesired tunneling events, illustrated by the red lines in the inner part of Fig.~\ref{fig:cycle}, are not completely suppressed. Second, the anti-correlation between the system dot and the cold dot is in general not perfect. It is illustrative to consider the limiting cases where these deviations from the ideal behavior are suppressed one by one, cf. Ref.~\cite{strasberg:2013}. To this end, we first consider the limit of a \textit{strong demon}
\begin{equation}
\label{eq:strong}
\delta_{\rm L},\, \delta_{\rm R}\rightarrow \infty.
\end{equation} 
In this case, no tunneling to the right/left dot is allowed whenever the cold/hot dot is occupied. In the graph representation, this limit removes the two lines shown in red in Fig.~\ref{fig:cycle}, resulting in a single loop. In the long time limit, the trajectories of the composite system are then characterized by a single stochastic variable: the number of cycles along the loop. Note that in contrast to the ideal behavior, cycles in the wrong direction may be completed. The strong demon limit thus significantly facilitates the analysis and, as shown in more detail below, the heat and charge currents are tightly coupled to each other as each electron that traverses the system corresponds to a single quantum of heat that traverses the demon.

The anti-correlation between the cold dot and the system dot may be imperfect for two reasons. First, the demon can only react to changes in the system state on the time scale of $1/\Gamma_{\rm C/H}$. There is thus a delay in the {measurement} performed by the demon. Second, a finite $T_{\rm C}$ induces thermal fluctuations in the cold dot which result in a {noisy measurement}. A particularly illuminating limit is the limit of a \textit{fast demon}
\begin{equation}
\label{eq:fast}
\Gamma_{\rm C},\,\Gamma_{\rm H}\gg\Gamma_{\rm L},\,\Gamma_{\rm R}.
\end{equation}
In this limit, we can assume that the demon is at all times described by a steady state which depends on the occupation of the system dot. This steady state, denoted by $\boldsymbol{\tau}^s$, can be calculated by setting $\Gamma_{\rm L/R}=0$. In this case, the master equation in Eq.~\eqref{eq:rateeq} decouples into two blocks corresponding to a filled and an empty system dot. The steady states of those blocks read
\begin{equation}
\label{eq:steadyfast}
\boldsymbol{\tau}^s=\begin{pmatrix}
\tau_{\rm e}^s\\\tau_{\rm c}^s\\\tau_{\rm h}^s
\end{pmatrix}
 = \frac{1}{Z_s}\begin{pmatrix}
 1\\e^{-\beta_{\rm C}(\xi_{\rm C}+sU)}\\e^{-\beta_{\rm H}(\xi_{\rm H}+sU)}
 \end{pmatrix},
\end{equation}
where $Z_s$ looks like a partition function for the demon given the system state and ensures the normalization of $\boldsymbol{\tau}^s$. Under the assumption of a fast demon [cf.~Eq.~\eqref{eq:fast}], the separation of time-scales between system and demon ensures
\begin{equation}
\label{eq:psdfast}
P_{sd}=P_s^{\rm S}\tau_d^s,
\end{equation}
at all times. Here, $P_s^{\rm S}=\sum_dP_{sd}$ denotes the probability of the system being in state $s$. From Eqs.~\eqref{eq:rateeq} and \eqref{eq:psdfast}, we derive a rate equation for the transitions between $\boldsymbol{\tau}^0\leftrightarrow\boldsymbol{\tau}^1$
\begin{equation}
\label{eq:ratefast}
\partial_tP_0^{\rm S}=-(\Gamma_{10}^{\rm L}+\Gamma_{10}^{\rm R})P_0^{\rm S}+(\Gamma_{01}^{\rm L}+\Gamma_{01}^{\rm R})P_1^{\rm S},
\end{equation}
and $P_0^{\rm S}+P_1^{\rm S}=1$. Here we introduced the rates
\begin{equation}
\label{eq:ratesfast}
\begin{aligned}
&\Gamma_{10}^{\rm L}=\frac{\Gamma_{\rm L}}{Z_0}\left[e^{-\beta_{\rm C}\xi_{\rm C}}{+}e^{-\beta_{\rm H}\xi_{\rm H}-\delta_{\rm L}}\right]f_{\rm L},\\
&\Gamma_{01}^{\rm L}=\frac{\Gamma_{\rm L}}{Z_1}\left[e^{-\beta_{\rm C}(\xi_{\rm C}+U)}{+}e^{-\beta_{\rm H}(\xi_{\rm H}+U)-\delta_{\rm L}}\right](1-f_{\rm L}),\\
&\Gamma_{10}^{\rm R}=\frac{\Gamma_{\rm R}}{Z_0}\left[e^{-\beta_{\rm C}\xi_{\rm C}-\delta_{\rm R}}{+}e^{-\beta_{\rm H}\xi_{\rm H}}\right]f_{\rm R},\\
&\Gamma_{01}^{\rm R}=\frac{\Gamma_{\rm R}}{Z_1}\left[e^{-\beta_{\rm C}(\xi_{\rm C}+U)-\delta_{\rm R}}{+}e^{-\beta_{\rm H}(\xi_{\rm H}+U)}\right](1-f_{\rm R}),
\end{aligned}
\end{equation}
that account for charging and uncharging of the system dot.
These rates explicitly break detailed balance. Following Ref.~\cite{esposito:2012}, we write
\begin{equation}
\label{eq:rdet}
\ln\frac{\Gamma_{10}^{\rm L/R}}{\Gamma_{01}^{\rm L/R}}=-\beta(\xi_{\rm L/R}+U)+r_{\rm L/R},
\end{equation}
where $r_{\rm L/R}$ quantifies the breaking of detailed balance.

Deriving a rate equation with broken detailed balance allows for connecting to previous works which use such an equation as a starting point \cite{schaller:2011,esposito:2012}. For instance, upon a redefinition of the parameters, Eq.~\eqref{eq:ratefast} is equivalent to the system discussed in Ref.~\cite{schaller:2011}, where a single quantum dot is considered, with tunnel couplings to reservoirs that depend on the occupation of the dot. The dependence of the tunnel couplings is argued to result from an external measurement and feedback loop. Here, we find how such a model emerges from a completely autonomous implementation.

We note that in the fast demon limit, the state of the demon contains no memory of any previous state of the system. Tracing out a memory-less part of a composite system generally results in a Markovian master equation for the reduced system \cite{breuer:book,schaller:book}. However, in contrast to thermal environments, the demon is not in thermal equilibrium, cf.~Eq.~\eqref{eq:steadyfast}. Therefore, the transition rates between system states induced by the demon do not fulfill detailed balance. This is in agreement with Ref.~\cite{sanchez:2018}, where an environment out of equilibrium is shown to act as a demon. We note that the method outlined here can be applied to any Markovian master equation where a separation of time-scales can be found. 

As discussed above, the imperfect anti-correlation between the system dot and the demon dot may result from a delay of the demon as well as from thermal fluctuations of the cold bath. It is thus illustrative to consider the limit of an \textit{error-free demon}
\begin{equation}
\label{eq:errorfree}
\Gamma_{\rm C},\,\Gamma_{\rm H}\gg\Gamma_{\rm L},\,\Gamma_{\rm R},\hspace{.25cm}T_{\rm C}\rightarrow 0.
\end{equation}
In this case, we find from Eq.~\eqref{eq:steadyfast} $\tau_{\rm c}^s=\delta_{s,0}$, i.e., perfect anti-correlation between the system dot and the cold dot. The dynamics of the system is governed by Eq.~\eqref{eq:ratefast} with the rates
\begin{equation}
\label{eq:ratesef}
\begin{aligned}
&\Gamma_{10}^{\rm L,ef}=\Gamma_{\rm L}f_{\rm L},\\
&\Gamma_{01}^{\rm L,ef}=e^{-\delta_{\rm L}}\Gamma_{\rm L}(1-f_{\rm L})f_{\rm H}^1,\\
&\Gamma_{10}^{\rm R,ef}=e^{-\delta_{\rm R}}\Gamma_{\rm R}f_{\rm R},\\
&\Gamma_{01}^{\rm R,ef}=\Gamma_{\rm R}(1-f_{\rm R})f_{\rm H}^1.
\end{aligned}
\end{equation}
Note that the transition rates for emptying the system are reduced by the occupation probability of the hot dot which is below one. Only for a hot reservoir with a strong population inversion (i.e., $T_{\rm H}\rightarrow-\infty$) could this reduction be removed. In this case, the equivalence with Ref.~\cite{schaller:2011} becomes particularly transparent.

Finally, we consider the limit of a \textit{perfect demon}
\begin{equation}
\label{eq:perfectdemon}
\delta_{\rm L},\, \delta_{\rm R}\rightarrow \infty,\hspace{.25cm}\Gamma_{\rm C},\,\Gamma_{\rm H}\gg\Gamma_{\rm L},\,\Gamma_{\rm R},\hspace{.25cm}T_{\rm C}\rightarrow 0.
\end{equation}
In this case, Eq.~\eqref{eq:ratefast} reduces to
\begin{equation}
\label{eq:rateperf}
\partial_tP_0^{\rm S}=-\Gamma_{10}^{\rm L,ef}P_0^{\rm S}+\Gamma_{01}^{\rm R,ef}P_1^{\rm S},
\end{equation}
with rates given in Eq.~\eqref{eq:ratesef}. We thus obtain the desired behavior where the system dot can only be filled from the left and emptied to the right. The composite system thus moves along the cycle illustrated in Fig.~\ref{fig:cycle} (along the desired direction), where transitions involving the demon happen infinitely fast.

The limiting cases introduced in this section will serve as benchmarks and allow for analytic progress. In particular, we will find a trade-off between the performance of the demon and its entropy production. We summarize the considered limiting cases in Table~\ref{tab:limits}.

\begin{table}
\caption{\label{tab:limits}Limiting cases and corresponding stall voltages.}
	\begin{ruledtabular}
	\begin{tabular}{lll}
	Strong demon: &
	$\delta_{\rm L},\, \delta_{\rm R}\rightarrow \infty$ & $\beta eV_\rms = U(\beta_\rmC-\beta_\rmH)$ \\
	Fast demon: &
	$\Gamma_{\rm C},\,\Gamma_{\rm H}\gg\Gamma_{\rm L},\,\Gamma_{\rm R}$& $\beta eV_\rms = r_\rmL-r_\rmR$\\
	Error-free demon: &
	Fast \& $T_{\rm C}\rightarrow 0$& $\beta eV_\rms = \delta_\rmL+\delta_\rmR$ \\
	Perfect demon: &Strong \& Error-free& $\beta eV_\rms \rightarrow\infty$ \\
\end{tabular}	 
\end{ruledtabular}
\end{table}

\section{The demon as a heat engine}
\label{sec:III}

Let us now consider the general case described by Eqs.~\eqref{eq:rateeq} and \eqref{eq:rates}.
In the long time limit, the system reaches a stationary regime where the currents are given by the following expressions:
\begin{align}
I_l&=-e\sum_{d\neq\rme}\left(\Gamma_{0d,1d}^lP_{1d}-\Gamma_{1d,0d}^lP_{0d}\right),\\
J_l&=\sum_{d\neq\rme}(\xi_l+U)\left(\Gamma_{0d,1d}^lP_{1d}-\Gamma_{1d,0d}^lP_{0d}\right),
\end{align}
for the charge and heat currents through terminal $l\in\{\rmL,\rmR\}$ of the system ($I_\rmL=-I_\rmR$, due to charge conservation), and
\begin{align}
J_\rmH&=\sum_{s}(\xi_H+sU)\left(\Gamma_{s\rme,s\rmh}P_{s\rmh}-\Gamma_{s\rmh,s\rme}P_{s\rme}\right),\\
J_\rmC&=\sum_{s}(\xi_C+sU)\left(\Gamma_{s\rme,s\rmc}P_{s\rmc}-\Gamma_{s\rmc,s\rme}P_{s\rme}\right),
\end{align}
for heat flowing into the hot and cold terminals, respectively. In the demon dots, $I_\rmH=I_\rmC=0$. We use a convention with charge (heat) currents defined as positive when flowing out of (into) the reservoirs. {An interesting aspect of our system is that the transport of heat through the demon is fully determined by charge fluctuations (i.e., the time-dependent occupation of the quantum dots)~\cite{sanchez:2011}. The demon heat currents and their fluctuations can therefore be measured by time-resolved charge detection using, e.g., two quantum point contacts~\cite{sanchez_detection_2012,*sanchez_erratum_2013}. This scheme has recently been implemented to detect entropy flows in related configurations~\cite{singh_extreme_2019}.

For the parameters of interest, the gap forbids transitions (0,e)$\leftrightarrow$(1,e). In that case, electrons enter and leave the system dot at the same energy, implying that the system does not absorb energy from the demon. Energy is thus conserved both in the system and in the demon. As there are no charge currents in the demon, this implies
\beq
\label{eq:1stlawD}
J_\rmC+J_\rmH=0.
\eeq
The fact that the system does not absorb any heat from the demon is useful for heat management~\cite{sanchez_all-thermal_2017,sanchez_single-electron_2017} and it allows us to define the heat current flowing in the demon as $J_\rmd=J_\rmC=-J_\rmH$. 
We find that a charge current can be generated in the system at zero applied voltage
\beq
\label{eq:currV0}
I_\rmR(V{=}0)=-\frac{e}{U}\frac{\Gamma_\rmL\Gamma_\rmR\left[1-e^{-(\delta_\rmL+\delta_\rmR)}\right]}{(\Gamma_\rmR{+}e^ {-\delta_\rmL}\Gamma_\rmL)(\Gamma_\rmL{+}e^ {-\delta_\rmR}\Gamma_\rmR)}J_\rmd,
\eeq
which depends on the heat flow through the demon and the demon's ability to act on the system. 
From energy conservation we also find
\beq
\label{eq:1stlawS}
J_\rmL+J_\rmR=P,
\eeq
with the power $P=-I_\rmR V$. When $P>0$ power is generated in the system due to a charge current flowing against a voltage bias. Equations \eqref{eq:1stlawD} and \eqref{eq:1stlawS} express the separation of the first law in the two partitions, as sketched in Fig.~\ref{fig:system}\,(a). 

Using Clausius' expression for the steady-state entropy production in reservoirs in local equilibrium, $\dot{S}_\alpha=-J_\alpha/T_\alpha$, we define the entropy production associated with the system
\begin{equation}
\label{eq:entropyS}
\dot{S}_\rms=-(J_\rmL+J_\rmR)\frac{1}{T}=-\frac{P}{T},
\end{equation}
and the demon
\begin{equation}
\label{eq:entropyD}
\dot{S}_\rmd=-J_\rmd\left(\frac{1}{T_\rmC}-\frac{1}{T_\rmH}\right).
\end{equation}
Note that the entropy production associated with the demon diverges for $T_\rmC\rightarrow0$. This implies that an error free demon (where the cold dot is perfectly anti-correlated with the system dot) necessarily produces an infinite amount of entropy.
From Eqs.~\eqref{eq:1stlawS} and \eqref{eq:entropyS}, we find that the second law of thermodynamics allows for a positive power generated in the system provided that the entropy reduction in the system is compensated by the entropy produced in the demon~\cite{sanchez:2018}
\beq
\label{eq:secondlaw}
\dot{S}_\rms+\dot{S}_\rmd\ge0\hspace{.5cm}\Leftrightarrow\hspace{.5cm}\dot{S}_\rmd\ge\frac{P}{T}.
\eeq

This motivates us to define the efficiency of the demon similarly to what is done for a heat engine~\cite{benenti:2017}
\beq
\eta=\frac{P}{J_\rmd}.
\eeq
The interpretation is clear as $J_\rmd=-J_\rmH$ is the heat current emitted by the hot reservoir, which is used as a resource.
We stress that, differently from usual heat engines, the heat current is not absorbed by the system but flows into the cold demon reservoir which is spatially separated from the system. In the limit where the demon works reversibly, i.e., where the inequality in Eq.~\eqref{eq:secondlaw} becomes an equality, the efficiency is maximal and equal to
\beq
\eta_0=\frac{T}{T_\rmC}\left(1-\frac{T_\rmC}{T_\rmH}\right).
\eeq
This generalizes the expression for the efficiency bound of a heat engine coupled to a heat source and dissipating heat into a cold bath. When the system is thermalized with the cold bath (the typical situation for a thermocouple), i.e., when $T_\rmC=T$, $\eta_0$ reduces to the Carnot efficiency. We note that for $T>T_\rmC$, $\eta_0$ is not bounded by one. The reason for this is that the temperature bias between the system and the cold bath then acts as an additional resource, allowing for power production without consuming heat from the hot bath. Indeed, for an error-free demon, where $T_\rmC\rightarrow0$, $\eta_0$ diverges.

\subsection{Limiting cases}
\label{sec:currentlimits}
Let us first discuss the currents expected in the limiting cases listed in Table~\ref{tab:limits}.
Of particular interest is the strong demon, for which the demon heat flow and system charge current are maximally correlated (sometimes denoted by tight-coupling)
\beq
\label{eq:currstrong}
I_\rmR=-\frac{e}{U}J_\rmd.
\eeq
Note that similar cross-correlations occur in  Carnot efficient heat engines~\cite{sanchez:2013,hofer:2016prb}.
In this regime, we furthermore get that
\beq
J_\rmd\propto U\left(1-e^{-U(\beta_\rmC-\beta_\rmH)+eV\beta}\right).
\eeq
In this case, both $I_\rmR$ and $J_\rmd$ are stalled at a voltage
\beq
\label{eq:stall}
\beta eV_{\rm s}=U(\beta_{\rm C}-\beta_{\rm H}),
\eeq
where the demon stops producing power. The efficiency increases linearly in voltage, $\eta=eV/U$ until the stall voltage, where it reaches the maximal efficiency $\eta=\eta_0$. 

For the fast demon, we can write
\begin{equation}
\label{eq:fdcurr}
I_\rmR = -eC\left[f_{\rm L}(1-f_{\rm R})-e^{-\beta eV_{\rm s}}f_{\rm R}(1-f_{\rm L})\right],
\end{equation}
where $C\geq 0$ and the stall voltage is related to the breaking of detailed balance 
\begin{equation}
\label{eq:fevs}
\beta eV_{\rm s}=r_{\rm L}-r_{\rm R}.
\end{equation}
In the limiting case of an error-free demon, where $T_{\rm C}\rightarrow 0$, the stall voltage simplifies to
\begin{equation}
\label{eq:evs}
\beta eV_{\rm s}=
\delta_{\rm L}+\delta_{\rm R}.
\end{equation}
Evidently, in the perfect demon limit, the stall voltage diverges.  In this case, the current is given by the simple expression
\beq
I_\rmR=-e\frac{\Gamma_{01}^{\rm L,ef}\Gamma_{10}^{\rm R,ef}}{\Gamma_{01}^{\rm L,ef}+\Gamma_{10}^{\rm R,ef}}.
\eeq
As $I_\rmR<0$, electrons are always driven from the left to the right reservoir, as expected in the limit of a perfect demon. Note however that this expression is only valid as long as Eq.~\eqref{eq:ingapes} holds. The stall voltages in the limiting cases are summarized in Table~\ref{tab:limits}.

\subsection{Performance}
\begin{figure}
	\centering
	\includegraphics[width=\linewidth]{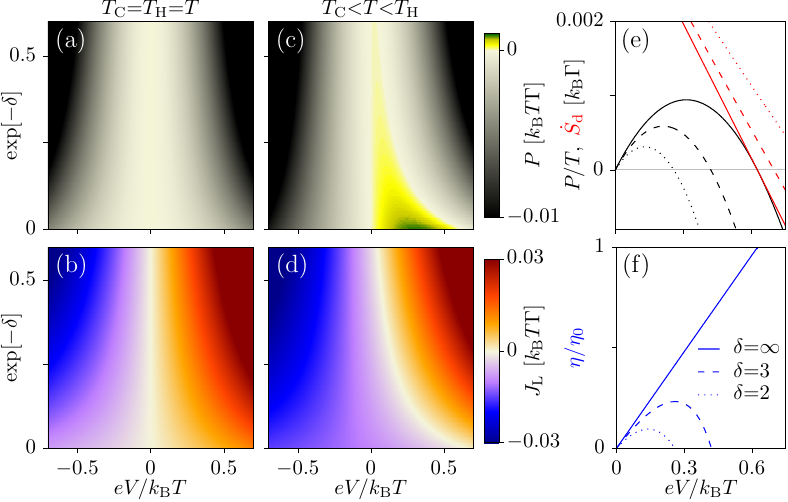}
	\caption{Performance of the demon. The generated power and the heat flow out of terminal L are compared for the cases where the demon is (a,b) in equilibrium ($T_\rmC=T_\rmH=T$) and (c,d) out of equilibrium ($T_\rmC<T<T_\rmH$), as functions of the applied voltage $V$ and the tunneling asymmetry $\delta=\delta_\rmL=\delta_\rmR$. For positive voltages, the non-equilibrium demon generates power while cooling the left reservoir at the same time. (e) The power generation reduces the entropy of the system by $P/T$, which is always smaller than the entropy production in the demon, $\dot S_\rmd$, except for the case at the stall voltage with $\delta\rightarrow\infty$, where they are equal. At this point, (f) the demon performs in the reversible limit and the maximal efficiency $\eta_0$ is achieved. Here we explicitly included the effect of a finite superconducting gap, $\Delta=0.8\kBT$, which does not influence the shown plots.  Parameters: $T_\rmH=1.2T$, $T_\rmC=0.8T$, $U=1.5\kBT$, $\varepsilon_\rmS=\xi_\rmH=0$, $\xi_\rmC=-0.4\kBT$, $\mu_\rmL=-eV/2$, $\mu_\rmR=eV/2$, $\Gamma_\alpha=0.1\kBT$, and $\gamma=10^{-8}\kBT$.}
	\label{fig:currents}
\end{figure}

The performance of the demon is illustrated in Fig.~\ref{fig:currents}, showing that the demon must be out of equilibrium in order to generate power. If $T_\rmC=T_\rmH$, even if different from $T$, the demon dots form an environment which is in local equilibrium and with which the system exchanges no energy. Therefore it is not surprising that transport in the system is only due to the applied voltage, the heat currents change sign at $V=0$,  and hence $P\le0$, as shown in Figs.~\ref{fig:currents}(a) and (b). This is the behavior expected for an isolated two terminal quantum dot~\cite{beenakker_theory_1992}: The demon sleeps.

The demon wakes up when $T_\rmH>T_\rmC$. In this case, a positive power is generated for voltages $0<V<V_{\rm s}$, where the stall voltage depends on the parameters of the composite system. In this region, the system generates power at the expense of its own heat, resulting in terminal L being cooled down ($J_\rmL>0$), as shown in Fig.~\ref{fig:currents}(d). The second law is illustrated in Fig.~\ref{fig:currents}(e), where the validity of Eq.~\eqref{eq:secondlaw} can clearly be seen. Only at the stall voltage in the strong demon limit, we have $\dot{S}_\rmd=P/T$ (the process is reversible). At that point, the efficiency attains its bound $\eta=\eta_0$, see Fig.~\ref{fig:currents}(f). Demons operating at finite $\delta$ produce an excess of entropy, less power, and operate at smaller efficiencies. 

The strong demon is dual: As shown in Fig.~\ref{fig:currents}(e), for voltages larger than $V_\rms$, the system dissipates power accompanied by the demon reducing its entropy. The roles are then exchanged, with the system acting as an electrically-driven demon that refrigerates the original demon. This is due to the high correlation of charge fluctuations in the composite system, which is maximal in the strong demon limit ($\delta_\rmL,\delta_\rmR\rightarrow\infty$). Then, reversing the charge flow through the system implies the reversal of the heat currents in the demon. This becomes clear by considering the graph in Fig.~\ref{fig:cycle}. The strong demon limit avoids the inner (red) transitions so the sign of the current determines whether the cycle runs clockwise or anticlockwise. 

\begin{figure}
	\includegraphics[width=\linewidth]{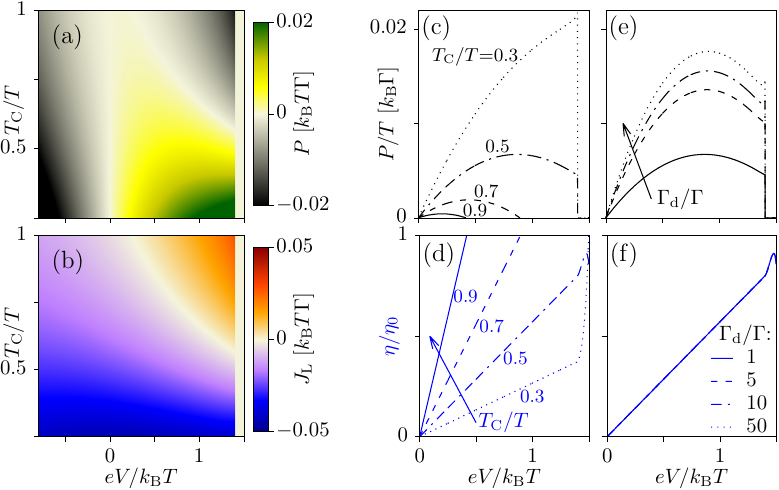}
	\caption{Performance of the strong demon. (a) Generated power and (b) heat flow out of terminal L as functions of the applied voltage $V$ and the temperature of the cold reservoir. $P\le0$ for $T_\rmC=T_\rmH$. The stall voltage increases for lower $T_\rmC$, but is cut off by the gap. (c) The generated power also increases when lowering $T_\rmC$. (d) The superconducting gap however avoids reaching the highest efficiencies. (e) By increasing the ratio $\Gamma_\rmd/\Gamma$ for the case $T_\rmC=0.5T$, the power increases toward the fast demon limit, (f) but the efficiency is not affected. Close to the gap, a small peak appears in $P$ reflecting the superconducting density of states~\eqref{eq:dos}. Same parameters as in Fig.~\ref{fig:currents} except for those explicitly indicated.}
	\label{fig:strong}
\end{figure}

Let us further explore the different limits of the strong demon configuration. These are shown in Fig.~\ref{fig:strong} by decreasing $T_\rmC$ and by increasing the ratio between the demon and system tunneling rates (setting $\Gamma_\rmH=\Gamma_\rmC=\Gamma_\rmd$ and $\Gamma_\rmL=\Gamma_\rmR=\Gamma$), while keeping $\delta_\rmL=\delta_\rmR=\infty$. As discussed in Sec.~\ref{sec:currentlimits}, the extracted power and the cooling power of terminal L, as well as the stall voltage increase when lowering $T_\rmC$, see Fig.~\ref{fig:strong}\,(a-f). However $P$ and $J_\rmL$ are limited by the gap: for large voltages such that $\mu_\rmR>\varepsilon_\rmS+U-\Delta$, the transitions through the right barrier are suppressed, and transport through the system drops to zero. At the same time, the heat current in the demon is suppressed. Up to that voltage, the efficiency grows linearly as $\eta/\eta_0=V/V_\rms$, as expected from Eq.~\eqref{eq:stall}, see Fig.~\ref{fig:strong}\,(d). 

By additionally increasing the tunneling rates in the demon, one approaches the perfect demon limit. The increase of the power is plotted in Fig.~\ref{fig:strong}\,(e), which shows the saturation of the maximal power for $\Gamma_\rmd\gg\Gamma$. Making the demon fast does not change the stall voltage. Hence, the efficiency is unaffected by changing the ratio $\Gamma_\rmd/\Gamma$, see Fig.~\ref{fig:strong}\,(f).


\section{The demon as an information engine}
\label{sec:IV}
In this section, we investigate descriptions of the composite system in terms of information. We consider two descriptions introduced in Refs.~\cite{horowitz:2014prx} and \cite{esposito:2012}. Following Ref.~\cite{horowitz:2014prx}, we find the generalized second law (in the steady state)
\begin{equation}
\label{eq:infolaw}
\dot{S}_{\rm s}+\mathcal{I}\geq 0,\hspace{1.5cm}\dot{S}_{\rm d}-\mathcal{I}\geq 0,
\end{equation} 
where the information flow $\mathcal{I}$ quantifies the average information that the demon obtains on the system. This can be understood by inspecting the time-derivative of the mutual information between the system and the demon \cite{horowitz:2014prx}. Here the information flow can be written as the product of an information current $J_I$ times an information affinity $\mathcal{F}_{\rm I}$. These quantities explicitly read
\begin{equation}
\label{eq:infoflow}
\mathcal{I}=J_{\rm I}\mathcal{F}_{\rm I},\hspace{.75cm}J_{\rm I}=\frac{J_{\rm d}}{U},\hspace{.75cm}\mathcal{F}_{\rm I}={k_{\rm B}}\ln\frac{P_{\rm 0c}P_{\rm 1h}}{P_{\rm 1c}P_{\rm 0h}}.
\end{equation}
Interestingly, the information current is determined by the heat quanta that traverse the demon, making it detectable in an experiment. For every quantum of heat, the demon obtains information about the system. The amount of information is determined by the information affinity which provides a measure for the anti-correlation between the system dot and the cold dot. 

In the limit of a fast demon, we find 
\begin{equation}
\mathcal{F}_{\rm I}=U\left(\frac{1}{T_\rmC}-\frac{1}{T_\rmH}\right).
\end{equation}
This implies that $\mathcal{I}=\dot{S}_{\rm d}$ and Eqs.~\eqref{eq:infolaw} reduce to the standard second law and a trivial equality, respectively. In this limit, the information description given in Ref.~\cite{horowitz:2014prx} does not provide any bounds that differ from the standard second law. Note that in the error-free (and in the perfect) demon limit, the information affinity diverges together with $\dot{S}_{\rm d}$~\cite{strasberg:2013}.

The second description in terms of information that we consider is based on a Markovian master equation with broken detailed balance to account for a measurement and feedback scheme \cite{esposito:2012}. In our case, this description only works in the fast demon limit, which is exactly the limit where the previous information description does not provide any additional constraint. Following Ref.~\cite{esposito:2012}, we find the generalized second law
\begin{equation}
\label{eq:infolaw2}
\dot{S}_{\rm s}+\mathcal{I}_{\rm f}\geq 0,
\end{equation}
where the subscript should remind the reader that this approach only works for the fast demon. The information flow is determined by the breaking of local detailed balance and reads
\begin{equation}
\label{eq:iflowf}
\begin{aligned}
\mathcal{I}_{\rm f} =& k_{\rm B}\sum_{l={\rm L,R}}r_l\left(P_0^{\rm S}\Gamma_{10}^l-P_1^{\rm S}\Gamma_{01}^l\right)\\&=\frac{k_{\rm B}}{-e}(r_{\rmL}-r_{\rmR})I_{\rmR}=\frac{P}{T}\frac{V_{\rms}}{V}.
\end{aligned}
\end{equation}
The quantities related to the Markovian master equation with broken detailed balance are defined in Eqs.~(\ref{eq:ratefast}-\ref{eq:rdet}). Furthermore, we made use of $eV_\rms=r_\rmL-r_\rmR$ which holds in the fast demon limit. Equation \eqref{eq:infolaw2} then reduces to the inequality
\begin{equation}
\label{eq:infolaw3}
\frac{P}{T}\frac{V_\rms-V}{V}\geq 0.
\end{equation}
Because the power is only positive for voltages that fulfill $0<V<V_\rms$, this inequality is always fulfilled, not only in the fast demon limit. 
Taking the strong demon limit in addition to the fast demon limit, we find $\mathcal{I}_{\rm f}=\dot{S}_{\rm d}$ and Eq.~\eqref{eq:infolaw2} reduces to the standard second law.
\begin{figure}
	\includegraphics[width=\linewidth]{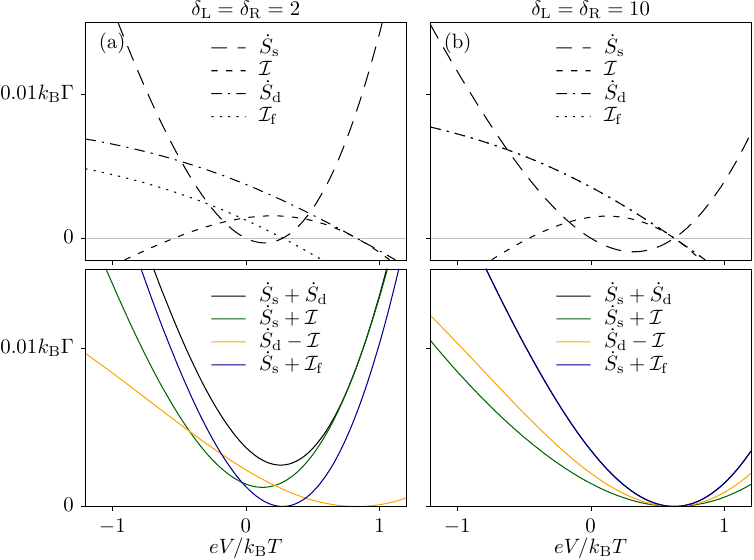}
	\caption{
	\label{fig:infolaws} Second-law like inequalities relating the rates for system and demon entropy change $\dot{S}_\rms$ and $\dot{S}_\rmd$, the information flow ${\cal I}$  and the fast information flow ${\cal I}_{\rm f}$ for different strengths of the demon: (a) $\delta_\rmL=\delta_\rmR=2$ and (b) $\delta_\rmL=\delta_\rmR=10$. In the strong demon limit, ${\dot S}_{\rm d}={\cal I}_{\rm f}$, and all inequalities are saturated at the stall voltage. Away from the fast demon limit, $\mathcal{I}_{\rm f}$ is defined as the right hand side of Eq.~\eqref{eq:iflowf}. Other parameters are as in Fig.~\ref{fig:currents}.}
\end{figure}
In the error-free demon limit, the information flow reduces to
\begin{equation}
\mathcal{I}_{\rm f} = k_{\rm B}(\delta_\rmL+\delta_\rmR)\frac{J_\rmd}{U}.
\end{equation}
As for the last description, the information flow thus diverges in the limit of a perfect demon. The inequalities in Eqs.~\eqref{eq:infolaw} and \eqref{eq:infolaw3} are illustrated in Fig.~\ref{fig:infolaws}.

We thus find that a description in terms of information flows \textit{can} complement a thermodynamic analysis and result in additional constraints. However, each information flow reduces to the entropy production of the demon, and the corresponding constraint to the second law, in a (different) limiting case.

\section{Fluctuation relations}
\label{sec:V}
In this section, we go beyond mean values and investigate fluctuations in the steady-state heat and charge currents. We first consider a fluctuation relation that is related to time-reversing the composite system and provides the standard extension to the second law \cite{andrieux:2009}. We then consider time-reversal of the system only, which results in a novel fluctuation relation that applies even when the entropy production associated with the demon diverges.

Let us denote by $X$ a trajectory of the composite system, which specifies the state $(s,d)$ at each point in time during the time interval $[0,t]$. We denote the entropy that is produced during the trajectory by $S(X)$. In the long-time limit, the entropy is fully determined by the number of charges that traversed the system ($w$) and the number of heat quanta that traversed the demon ($q$). We can thus write
\begin{equation}
S(X)=-w\frac{eV}{T}+q\left(\frac{1}{T_\rmC}-\frac{1}{T_\rmH}\right).
\end{equation}
Since the transitions in the composite system fulfill detailed balance, we find
\begin{equation}
\label{eq:ft0}
\frac{P(X^\dagger)}{P(X)}=e^{-S(X)/k_{\rm B}},
\end{equation}
where $P(X)$ denotes the probability that the composite system follows trajectory $X$ and $X^\dagger$ denotes the time-reversed of $X$. Summing over all trajectories that have the same values for $w$ and $q$, we find
\begin{equation}
\label{eq:ft1}
\frac{P(-w,-q)}{P(w,q)}=e^{w\beta eV-qU(\beta_\rmC-\beta_\rmH)}.
\end{equation}
Alternatively, this fluctuation relation can be obtained by considering the symmetry of the cumulant generating function that characterizes the charge and heat transport through the composite system \cite{sanchez_detection_2012}.

From Eq.~\eqref{eq:ft1}, one can recover the second law using Jensen's inequality which results in
\begin{equation}
\langle w\rangle \beta eV\leq \langle q\rangle U(\beta_\rmC-\beta_\rmH),
\end{equation}
where the averages are taken over the distribution $P(w,q)$. The second law is recovered by identifying $P=\partial_t\langle w\rangle eV$ and $J_\rmd = \partial_t\langle q\rangle U$.

In the strong demon limit, charge and heat transport are tightly coupled, enforcing $w=q$ on each trajectory. The fluctuation relation then reduces to
\begin{equation}
\label{eq:ft2}
\frac{P(-w)}{P(w)}=e^{w\beta e(V-V_\rms)},
\end{equation}
with $\beta eV_\rms=U(\beta_\rmC-\beta_\rmH)$. In the limit where $T_\rmC\rightarrow0$, heat can only flow into the cold bath enforcing $q\geq0$ in every trajectory. The fluctuation relation in Eq.~\eqref{eq:ft1} is then reduced to the trivial equality $0=0$ for all terms where $q\neq0$.

We now consider a novel type of fluctuation relation which is based on time-reversal of the system only. A naive time reversal of only the system state will in general result in trajectories that cannot occur in the composite system. To circumvent this problem, we consider a time-reversal on the single electron level. To this end, we note that a trajectory $X$ describes electrons that enter and leave the system dot one after the other. Time-reversing only the system is now defined by reversing the path of each electron that traverses the system: An electron that enters the system from the left bath and leaves it to the right bath is replaced by an electron that enters the system from the right bath and leaves it to the left bath. Electrons that leave the system to the same bath they originate from are not affected. Therefore, only electrons that contribute to transport are affected. The trajectory that is obtained from $X$ in this way is denoted $X^+$. The different forms of time-reversal considered here are illustrated in Fig.~\ref{fig:time-reversal}.

We now introduce the number $n$ which counts the electrons that contribute to transport, weighted by the amount of open barriers they traverse. An electron that contributes to transport by traversing two open barriers increases $n$ by one, irrespective of the sign with which it contributes to the charge current. An electron that contributes to transport by traversing two closed barriers reduces $n$ by one. All other electrons do not contribute to $n$. As $n$ counts electrons that behave according to the feedback effected by the demon, we call the corresponding current $F=\partial_t\langle n\rangle$ the \textit{feedback-assisted current}. As illustrated in Fig.~\ref{fig:time-reversal}, the transformation $X\rightarrow X^+$ inverses the feedback-assisted current as well as the charge current but leaves the demon heat current invariant. This is in contrast to complete time-reversal, $X\rightarrow X^\dagger$, which inverses charge and heat currents but leaves the feedback-assisted current invariant.

With these definitions, we find the fluctuation relation
\begin{equation}
\label{eq:ft3}
\frac{P(X^+)}{P(X)}=\frac{P(-w,-n)}{P(w,n)}=e^{w\beta eV-n(\delta_\rmL+\delta_\rmR)}.
\end{equation}
Using Jensen's inequality, we find the constraint on the produced work
\begin{equation}
\label{eq:ineqf}
\langle w\rangle \beta eV\leq \langle n\rangle(\delta_\rmL+\delta_\rmR).
\end{equation}
In contrast to Eq.~\eqref{eq:ft1}, this fluctuation relation provides a constraint even when the entropy production associated with the demon diverges. Indeed, in the limit of an error-free demon, all electrons that flow against the voltage bias traverse open barriers and all electrons that flow with the bias traverse closed barriers. This enforces $n=w$ on all trajectories and Eq.~\eqref{eq:ft3} reduces to Eq.~\eqref{eq:ft2} with $\beta eV_\rms =\delta_\rmL+\delta_\rmR$, in agreement with Ref.~\cite{schaller:2011}, where the same fluctuation relation was found for a non-autonomous demon which does not make any measurement errors. We further remark that Eq.~\eqref{eq:ft3} breaks down in the strong demon limit, where it reduces to the trivial equality $0=0$ for all terms with $n\neq0$.

\begin{figure}
	\includegraphics[width=\linewidth]{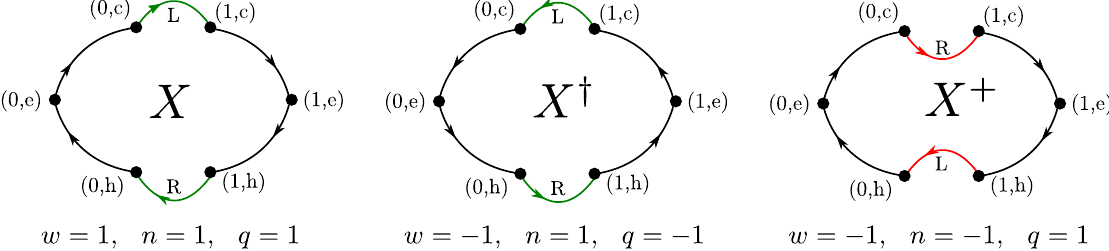}
	\caption{Illustration of time-reversed trajectories. Left panel: desired trajectory, where an electron enters the system dot from the left through an open (green) barrier and leaves the system dot to the right through an open barrier. Completing this loop once results in a single charge transported against the voltage bias $w=1$, a single quantum of heat transported from hot to cold $q=1$, and a single electron contributing positively to the feedback-assisted current $n$. Middle panel: Time-reversing the composite system results in changing the direction of all arrows. While $w$ and $q$ change sign, the contribution to the feedback-assisted current remains positive because the electron passes through two open barriers. Right panel: Time-reversing only the system exchanges origin and destination of the electron that passes through the system dot. This inverts $w$ and $n$ while $q$ remains invariant since the demon is not affected by the transformation.}
	\label{fig:time-reversal}
\end{figure}

We note that Eq.~\eqref{eq:ft3}, as well as the feedback-assisted current, cannot generally be obtained by considering the symmetries of the cumulant generating function of heat and charge currents. The reason for this is that in contrast to the heat and charge currents, the feedback-assisted current of a trajectory depends on the \textit{order} of the transition rates. The order is necessary to determine both the origin as well as the destination of each electron. One can however derive Eq.~\eqref{eq:ft3} from the cumulant generating function obtained from an extended master equation, where one explicitly keeps track of the origin of the electrons that occupy the system dot, see Appendix~\ref{sec:demonchannel}.

We close this section by considering the fast demon limit which is described by a Markovian master equation for the system alone. In this case, a standard time-reversal of the system trajectories is possible. This results in the fluctuation relation given in Eq.~\eqref{eq:ft2} with the stall voltage $\beta eV_\rms = r_\rmL-r_\rmR$. Therefore, in all limiting cases, except in the perfect demon limit where the stall voltage diverges, the fluctuation relation in Eq.~\eqref{eq:ft2} holds. In addition, the fluctuations are constrained by the relations given in Eq.~\eqref{eq:ft1} and Eq.~\eqref{eq:ft3}. Equation \eqref{eq:ft1} differs from Eq.~\eqref{eq:ft2} for finite $\delta_l$ and breaks down in the limit $T_\rmC\rightarrow0$. Equation \eqref{eq:ft3} differs from Eq.~\eqref{eq:ft2} as long as the demon is not error-free and breaks down in the strong demon limit. 

\section{Efficiencies: comparing heat and information engines}
\label{sec:VI}
In the sections above, we found a number of second law like inequalities, each of which motivates the introduction of an efficiency. Here we discuss and compare these efficiencies, focusing on the regime where $P\geq0$. We first consider the (normalized) thermal efficiency introduced above which follows from the standard second law
\begin{equation}
\label{eq:efftherm}
{\eta}_{\rm T}=\frac{-\dot{S}_s}{\dot{S}_\rmd}=\frac{\eta}{\eta_0}=\frac{\beta P}{J_\rmd(\beta_\rmC-\beta_\rmH)}\leq 1.
\end{equation}
This efficiency quantifies how well heat is converted into work.
In the strong demon limit, it reduces to ${\eta}_{\rm T}=V/V_\rms$, reaching its maximum value at the stall voltage. The thermal efficiency vanishes in the limit $T_\rmC\rightarrow0$ because $\dot{S}_\rmd$ diverges in this case.
The second efficiency we consider is based on Eq.~\eqref{eq:infolaw} and reads
\begin{equation}
\label{eq:effinfo1}
\eta_{\rm I} = \frac{-\dot{S}_s}{\mathcal{I}}=\frac{\beta P}{J_{\rm I}\mathcal{F}_{\rm I}}\leq1,
\end{equation}
where the information quantities are defined in Eq.~\eqref{eq:infoflow}. This efficiency can be understood as quantifying how the information flow is converted into power. It reduces to ${\eta}_{\rm T}$ in the fast demon limit. In this limit, we can use Eq.~\eqref{eq:infolaw2} to introduce a second information efficiency
\begin{equation}
\label{eq:effinfo2}
\eta_{\rm f} = \frac{-\dot{S}_s}{\mathcal{I}_{\rm f}}=\frac{V}{V_\rms}\leq 1.
\end{equation}
This efficiency reaches its maximum value at the stall voltage and reduces to the thermal efficiency in the strong demon limit. Finally, we introduce an efficiency based on Eq.~\eqref{eq:ineqf}
\begin{equation}
\label{eq:efffeedback}
\eta_{\rm F} = \frac{\beta P}{F(\delta_\rmL+\delta_\rmR)}\leq 1.
\end{equation}
We recall that $F=\partial_t\langle n\rangle$ denotes the feedback assisted current. This efficiency can be understood as how well the demon uses feedback, determined by the asymmetry in the tunnel barriers, to produce work. For an error-free demon, the efficiency reduces to $V/V_\rms$ and thus to $\eta_{\rm f}$.

\begin{figure}
	\includegraphics[width=\linewidth]{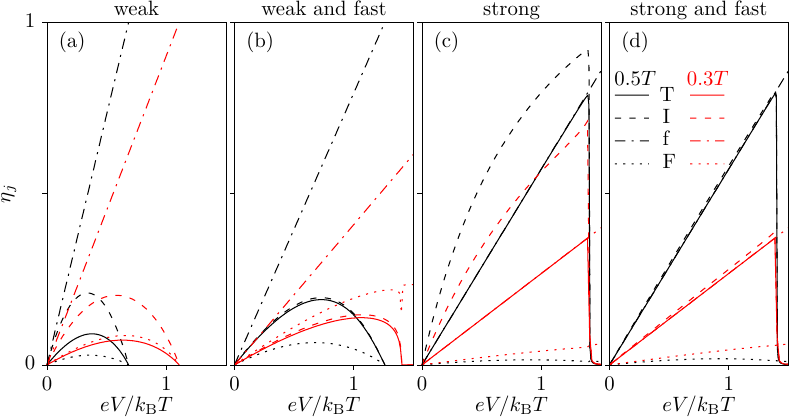}
	\caption{Information related efficiencies, $\eta_j$, with $j$=\{T,I,f,F\}, calculated from the right hand side of Eqs.~\eqref{eq:efftherm}--\eqref{eq:efffeedback}, for (a) a weak, (b) a weak and fast, (c) a strong and (d) a strong and fast demon. The cases with $T_\rmC=0.5T$ (black) and $T_\rmC=0.3T$ (red lines) are compared. We set $\delta_\rmL=\delta_\rmR=2$ for weak, $\delta_\rmL=\delta_\rmR=10$ for strong, $\Gamma_\rmd=50\Gamma$ for fast, and $\Gamma_\rmd=\Gamma$ otherwise. For the fast demon (b) and (d), $\eta_{\rm T}$ and $\eta_{\rm I}$ overlap. For a strong demon, $\eta_{\rm T}$ coincides with $\eta_{\rm f}$, panels (c) and (d), except for being limited by the gap.}
	\label{fig:efficiencies}
\end{figure}

The conversion of thermal resources and information can thus be characterized with different efficiencies. In the limiting cases, where these efficiencies can reach their maximum value, they reduce to the simple expression $V/V_\rms$. A perfect demon thus works very inefficiently with respect to all those efficiencies because the stall voltage diverges in this limit. This reflects the fact that the perfect demon cannot work in a reversible manner. Furthermore, the maximal efficiency  will in this case be limited by the gap. Hence counter-intuitively, demons that make errors are in general more efficient. 
The different efficiencies are compared in Fig.~\ref{fig:efficiencies}. For strong and fast demons $\eta_{\rm T}$, $\eta_{\rm I}$ and $\eta_{\rm f}$ coincide for voltages below the gap, see Figs.~\ref{fig:efficiencies}\,(c) and (d).

\section{Experimental considerations}
\label{sec:VII}
In this section, we briefly consider the consequences of unequal charging energies and we discuss an alternative implementation of the device using metallic islands instead of quantum dots.

\subsection{Unequal charging energies}
We have so far considered the case $U_\rmC=U_\rmH$ which ensures the energy conservation in the demon and in the system.
In an experimental realization, the charging energies $U_\rmC$, $U_\rmH$ depend on the  geometrical capacitance of the two demon dots as well as on their respective capacitive coupling to the system dot. They are in principle different, which affects the demon heat currents: in the desired cycle illustrated in Fig.~\ref{fig:cycle}, the energy  $U_\rmH$ is extracted from the hot terminal and $U_\rmC$ is absorbed by the cold one. The difference $U_\rmH-U_\rmC$ flows into the system: The electron tunnels from the left reservoir with an energy $\varepsilon_\rmS+U_\rmC$ and tunnels out to the right one with $\varepsilon_\rmS+U_\rmH$. Hence, the system alone does no longer satisfy the first law.

In this case, we have a more general relation for the heat currents in the demon
\beq
\frac{J_\rmC}{U_\rmC}+\frac{J_\rmH}{U_\rmH}=0,
\eeq
showing a tight coupling which is mediated by the charge fluctuations in the system. From conservation of energy in the composite system: $J_\rmL+J_\rmR=(U_\rmH{-}U_\rmC)J_\rmC/U_\rmC-P$, we find that heat leaks into the system, even when $P=0$.
However this does not have a fundamental impact on the demon operation. At $V=0$ (no Joule heating), the current has the same expression as in Eq.~\eqref{eq:currV0} replacing $J_\rmd/U$ by $J_\rmC/U_\rmC$.

\begin{figure}
	\includegraphics[width=\linewidth]{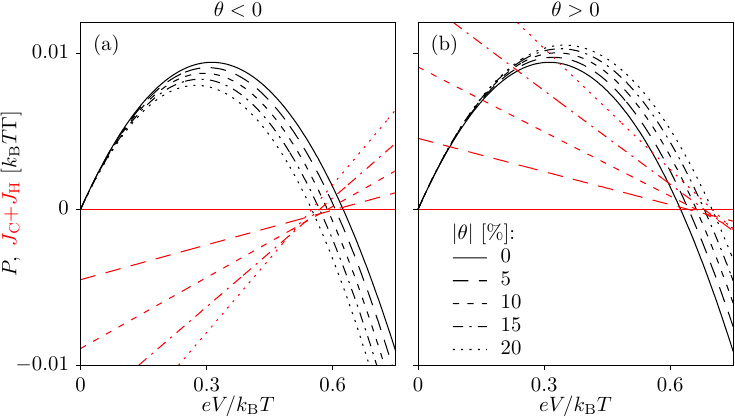}
	\caption{Effect of unequal charging energies. Generated power (black lines) and heat current absorbed by the demon (red lines) for different factors of asymmetry, $\theta$, having (a) $U_{\rmC}>U_{\rmH}$ and (b) $U_{\rmC}<U_{\rmH}$ (in both cases fixing $U_\rmH=1.5\kBT$). Same parameters as in Fig.~\ref{fig:currents} for the strong demon limit with $\delta_\rmL=\delta_\rmR=\infty$, except for those explicitly indicated.}
	\label{fig:unequal}
\end{figure}

This is clearly demonstrated in Fig.~\ref{fig:unequal}, parametrizing the difference of charging energies in terms of the parameter
\beq
\theta=\frac{U_{\rmC}-U_\rmH}{U_\rmH}\times 100.
\eeq
It shows that the generated power is not affected by a finite $J_\rmC+J_\rmH$, even when it is of the same order as $P$. Indeed, we see in Fig.~\ref{fig:unequal}(a) that having $U_\rmC>U_\rmH$ is beneficial for increasing the power and the stall voltage. This is understood as it helps the demon mechanisms: larger $U_C$ reduces the errors in the same way as lowering $T_\rmC$ would do. At the same time, smaller $U_\rmH$ helps the hot dot to react better, requiring lower $T_\rmH$ to be effective. Note furthermore that in this case $J_\rmC+J_\rmH>0$ i.e., heat is flowing into the demon and out of the system. In the opposite case, $U_\rmC<U_\rmH$, the demon injects heat into the system and power is reduced, see Fig.~\ref{fig:unequal}(b). This is exactly the opposite behavior that one expects from a conventional heat engine, marking the unique operation of our system being driven by information. Fluctuation theorems for $U_\rmC\neq U_\rmH$ are discussed in App.~\ref{sec:ftuneq}.

\subsection{Hard gap}
We have considered that the superconducting gap is hard, except for a small inverse quasiparticle lifetime, $\gamma$. For a long time, achieving a hard gap in superconductor-semiconductor interfaces has been an experimental challenge. However, fast improvements have been achieved recently in junctions forming nanowires~\cite{krogstrup:2015}, 2D structures~\cite{shabani:2016}, and even quantum point contacts~\cite{kjaergaard:2016}. 

Imperfections in the gap would result in leakage heat currents into the system. This effect is captured in our model by increasing $\gamma$. It is nevertheless expected to be tiny in realistic configurations as soon as $\varepsilon_\rmS$ is well within the gap. In the same sense as in the previous subsection, these leakage currents can be easily dientangled from the relevant information flows.

An alternative way to filter transitions without involving superconducting leads involves the use of triple quantum dot arrays~\cite{sanchez:2011,sanchez_single-electron_2017}: the central one is coupled to the demon dots, while the outermost ones serve as energy filters at $\varepsilon_\rmS+U$. This scheme has the advantage of reducing the effect of unequal charging energies by appropriately tuning the quantum dot levels.

\subsection{Metallic islands}

We have so far discussed an implementation based on single-level quantum dots. Recent experiments~\cite{koski:2015} motivate configurations with the quantum dots replaced by metallic islands, which are also affected by Coulomb blockade effects~\cite{averin:1986}. We briefly discuss a plausible configuration here, a more detailed analysis being beyond the aim of this manuscript.
The main differences with respect to semiconductor quantum dots are that (i) the tunneling barriers (formed by insulating layers) are not affected by gate potentials and that (ii) they have a dense spectrum. These two effects compromise the exact cancellation of the energy currents in the system as well as the action of the demon on the system dynamics.

\begin{figure}[t]
	\includegraphics[width=\linewidth]{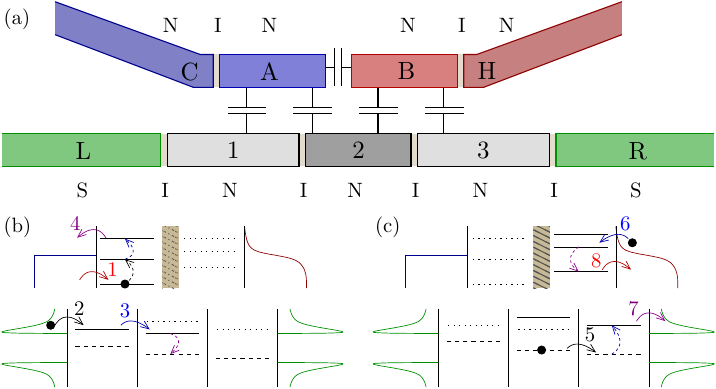}
	\caption{Sketch of an analogue device formed by metallic islands. (a) The pale grey islands are used to avoid/permit transport into the system island (dark grey) depending on the occupation of the demon islands. Labels indicate superconducting leads (S), insulating tunneling barriers (I), and normal metal islands (N). The transitions in (b) and (c) picture the sequence of the demon action. Dashed arrows mark the change in the electrochemical potentials due to a tunneling transition in a different island (numbered full arrows).}
	\label{fig:metaldemon}
\end{figure}

These issues can be partially overcome by adding two islands to the left and right of the system island, each of them capacitively coupled to one of the demon islands, see Fig.~\ref{fig:metaldemon}\,(a). Note that the spatial arrangement of the hot and cold terminals is opposite to the quantum dot case in Fig.~\ref{fig:system}\,(b). The electrostatic energy of the composite system reads
\beq
\label{eq:Hmetal}
{\cal U}=\sum_i E_i(n_i-n_{{\rm g}i})^2+\sum_{i,k\neq i}J_{ik}(n_i-n_{{\rm g}i})(n_k-n_{{\rm g}k}),
\eeq
where $i$=1,2,3 label the system islands, and $i$=A,B the cold and hot island, $E_i$ is the on-site charging energy and $J_{ik}=J_{ki}$ describes the Coulomb interaction between islands. Let us define $\mu_i$ as the energy given by Eq.~\eqref{eq:Hmetal} when every island is empty except for $i$ having one electron. It depends on the control parameters $n_{{\rm g}j}$ which can be tuned by gate voltages. 
Assuming $J_{\rm AB}\gg J_{i{\rm A}},J_{i{\rm B}}$, with $i{\neq}$A,B, only one of the demon islands can be occupied at a time. We further consider $J_{1{\rm B}}$ and $J_{3{\rm A}}$ to be negligible. 
By making $\mu_{\rm A}+J_{\rm 1A}<\mu_\rmC<\mu_{\rm A}+J_{\rm 2A}$, the occupation of the cold island is sensitive only to the state of island 2.
The chemical potential of islands 1 and 3 depend on the occupation of the demon through $J_{1{\rm A}}$ and $J_{3{\rm B}}$. 
Having $\mu_{1}>\mu_2>\mu_3$ prevents an electron from tunneling from 3 to 1 when the demon islands are empty if the respective differences are not small compared to $\kBT$. In order to suppress the contribution of undesired transitions, one requires that the cold demon rate is much faster than the system ones such that, e.g., the cold island is immediately occupied when the system is empty.
Once A is occupied, an electron tunneling into 1 will be transferred to 2 if $\mu_1+J_{\rm 1A}>\mu_2+J_{\rm 2A}$. Then, the cold island is emptied, see Fig.~\ref{fig:metaldemon}\,(b), the system electron can tunnel to 3 and, upon the charging of B, tunnel out to the right terminal. This is favored if $\mu_2+J_{\rm 2B}>\mu_3+J_{\rm 3B}$, see Fig.~\ref{fig:metaldemon}\,(c). 

Other processes are possible that contribute to the wrong direction. Their contribution can be reduced if, e.g., the tunneling rate from L to 1 is larger than that from R to 3, but they cannot be totally suppressed. Hence, the range of parameters where the action of the demon is effective is reduced compared to the quantum dot setup, making it hard, e.g., to achieve a strong demon. 
 


The system energy can still be conserved on average by gate tuning the island chemical potentials. This is for example the case for $V=0$ if $\mu_1+J_{\rm 1A}=\mu_3+J_{\rm 3B}$. In any case, as discussed above, a finite energy flow is easily disentangled from the relevant information currents.

\section{Outlook and conclusions}
\label{sec:VIII}
We presented a detailed investigation of information-to-work conversion in an autonomous implementation of Maxwell's demon based on three quantum dots coupled to separate thermal reservoirs. Since only reservoirs in local thermal equilibrium are used as resources, the entropy production associated with each reservoir can be accounted for. This allows for describing the information-to-work conversion process as a conventional thermoelectric heat engine. In addition, we investigated the process in terms of information flows. Such a description can result in additional constraints, depending on the regime of operation. From all descriptions, we find that in order for the demon to perform an error-free measurement, its associated entropy production must diverge. This is in agreement with previous results which found a diverging mutual information for error-free continuous measurements \cite{horowitz:2014,potts:2018}. Nevertheless, fluctuation relations and second law like inequalities can be found in this regime, cf.~Eqs.~\eqref{eq:ft3} and \eqref{eq:ineqf}.

Most of our results are illustrated using limiting cases, cf.~Table~\ref{tab:limits}, which result in analytically tractable and intuitive results. A particularly useful limit is given by the fast demon limit, where we derive a master equation for the system alone with explicitly broken detailed balance. Interestingly, this is not necessarily accompanied by a diverging entropy production associated with the demon. From the graph representation of the original master equation (cf.~Fig.~\ref{fig:cycle}) we find that any heat flow from hot to cold requires transitions in the system dot. Increasing the rates of the demon will thus not result in a diverging heat flow. While a perfect measurement is associated with a diverging entropic cost, breaking of detailed balance is not.

Fluctuation relations provide particularly useful constraints on the dynamics when operated far from equilibrium. To describe the composite system, we derived two different fluctuation relations. While the first is related to ordinary time-reversal and takes on the standard form [cf.~Eq.~\eqref{eq:ft0}], the second is obtained by a novel kind of time-reversal that reverses the paths of all the electrons which traverse the system. Interestingly, this results in a fluctuation theorem which includes a current variable that has no simple relation to either heat or charge currents, see Eq.~\eqref{eq:ft3}. This implies that this fluctuation relation cannot be obtained from the heat and charge statistics alone, except in the limit of an error-free demon where it reduces to the fluctuation relation of Ref.~\cite{schaller:2011}. It can however be obtained from an extended master equation, where the origin of the electrons is explicitly accounted for. We thus provide an example where extending the master equation can uncover hidden symmetries that result in useful fluctuation relations. Using this approach to find novel fluctuation theorems in different systems provides a promising avenue for future research. We have restricted ourselves to the weak coupling regime. Achieveing the strong coupling limit relaxes the conditions required for the fast demon limit, and opens interesting questions on the energy that the demon invests on the operation of the system barriers~\cite{strasberg:2018}.

In addition to providing fundamental insights, the proposed device can be implemented with current day technology. While such an implementation might result in deviations from the ideal behavior, where the first law is respected within the system alone, the underlying information-to-work conversion process should not be affected.

\acknowledgments
We acknowledge fruitful discussions with J. Splettstoesser, R. S. Whitney, V. Maisi, P. Strasberg, and D. S\'anchez. R.S. acknowledges support from the Ram\'on y Cajal program RYC-2016-20778 and the ``Mar\'ia de Maeztu\textquotedblright{} Programme for Units of Excellence in R\&D (MDM-2014-0377). P.S. acknowledges support by the Swedish VR. P.P.P. acknowledges funding from the European Union's Horizon 2020 research and innovation programme under the Marie Sk{\l}odowska-Curie Grant Agreement No. 796700.

\appendix

\section{Extended space master equation}
\label{sec:demonchannel}

In this appendix we introduce a generalized master equation that accounts for feedback-assisted transitions by counting the number of electrons that contribute to transport by traversing the system only through {\it open} barriers, $n$.
To this end, the states with one electron  in the system need to be distinguished depending on through which barrier ($l$=L,R) the electron tunneled in, and whether this barrier was open (o) or closed ($\nu$=o,c). This way, the configuration space includes 15 states: (0,\rme), (1,\rme,$\nu l$), (0,\rmh), (0,\rmC), (1,\rmh,$\nu l$), (1,\rmc,$\nu l$). For example, the sequence
\beq
\label{ndlr}
\text{(0,\rmc)\textcolor{black}{$\xrightarrow{\rmL}$}(1,\rmc,oL)$\rightarrow$(1,\rme,oL)
$\rightarrow$(1,\rmh,oL)\textcolor{black}{$\xrightarrow{\rmR}$}(0,\rmh)},
\eeq
(where the label over the arrows mark the barrier of the system involved in the transition), corresponding to the $X$ cycle in Fig.~\ref{fig:time-reversal}, must count $n$=1, while 
\beq
\label{ndrl}
\text{(0,c)\textcolor{black}{$\xrightarrow{\rmR}$}(1,c,cR)$\rightarrow$(1,e,cR)
$\rightarrow$(1,h,cR)\textcolor{black}{$\xrightarrow{\rmL}$}(0,h)},
\eeq
counts $n=-1$, see cycle $X^+$ in Fig.~\ref{fig:time-reversal}. 

This is done by introducing a counting variable $z_{\rm F}=e^{i\chi_{\rm F}}$ in the appropriate transitions: It counts $n=1$ at the occurrence of transitions (1,c,oR)$\xrightarrow{\rmL}$(0,c) and (1,h,oL)$\xrightarrow{\rmR}$(0,h). On the other hand, it counts $n=-1$ when (1,c,cL)$\xrightarrow{\rmR}$(0,c) and (1,h,cR)$\xrightarrow{\rmL}$(0,h). A second variable, $z=e^{i\chi}$, counts the number of transported particles, $w$. The resulting modified master equation reads
\begin{align}
\dot\rho_{0,\rme}&=-\Gamma_{0\rmd,0\rme}\rho_{0,\rme}
+\Gamma_{0\rme,0\rmc}\rho_{0,\rmc}+\Gamma_{0\rme,0\rmh}\rho_{0,\rmh},\nonumber\\
\dot\rho_{0,\rmh}&=\Gamma_{0\rmh,0\rme}\rho_{0,\rme}-(\Gamma_{0\rme,0\rmh}{+}\Gamma_{1\rmh,0\rmh}^{\rm s})\rho_{0,\rmh}\nonumber\\
&+\left(\frac{1}{z}\Gamma_{0\rmh,1\rmh}^{\rmL}{+}\Gamma_{0\rmh,1\rmh}^{\rmR}\right)(\rho_{1,\rmh,{\rm cL}}{+}\rho_{1,\rmh,{\rm oR}})\nonumber\\
&{+}\left(z_{\rm F}\Gamma_{0\rmh,1\rmh}^{\rm R}{+}\frac{1}{z}\Gamma_{0\rmh,1\rmh}^{\rm L}\right)\rho_{1,\rmh,{\rm oL}}\nonumber\\
&{+}\left(\Gamma_{0\rmh,1\rmh}^{\rm R}{+}\frac{1}{zz_{\rm F}}\Gamma_{0\rmh,1\rmh}^{\rm L}\right)\rho_{1,\rmh,{\rm cR}},\\
\dot\rho_{0,\rmc}&=\Gamma_{0\rmc,0\rme}\rho_{0,\rme}-(\Gamma_{0\rme,0\rmc}+\Gamma_{1\rmc,0\rmc}^{\rm s})\rho_{0,\rmc}\nonumber\\
&+\left(\frac{1}{z}\Gamma_{0\rmc,1\rmc}^{\rmL}{+}\Gamma_{0\rmc,1\rmc}^{\rmR}\right)(\rho_{1,\rmc,{\rm oL}}{+}\rho_{1,\rmc,{\rm cR}})\nonumber\\
&{+}\left(\frac{1}{z}\Gamma_{0\rmc,1\rmc}^{\rm L}{+}\frac{1}{z_{\rm F}}\Gamma_{0\rmc,1\rmc}^{\rm R}\right)\rho_{1,\rmc,{\rm cL}}\nonumber\\
&{+}\left(\frac{z_{\rm F}}{z}\Gamma_{0\rmc,1\rmc}^{\rm L}{+}\Gamma_{0\rmc,1\rmc}^{\rm R}\right)\rho_{1,\rmc,{\rm oR}},\nonumber\\
\dot\rho_{1,\rme,{\nu l}}&=-\Gamma_{1\rmd,1\rme}\rho_{1,\rme,{\alpha l}}
+\Gamma_{1\rme,1\rmc}\rho_{1,\rmc,{\alpha l}}+\Gamma_{1\rme,1\rmh}\rho_{1,\rmh,{\nu l}},\nonumber
\end{align}
for the probabilities where the composite system contains a single electron and
\begin{align}
\dot\rho_{1,\rmh,{\rm oL}}&=\Gamma_{1\rmh,1\rme}\rho_{1,\rme,{\rm oL}}-(\Gamma_{0\rmh,1\rmh}^{\rm s}{+}\Gamma_{1\rme,1\rmh})\rho_{1,\rmh,{\rm oL}},\nonumber\\
\dot\rho_{1,\rmh,{\rm cL}}&=z\Gamma_{1\rmh,0\rmh}^{\rm L}\rho_{0,\rmh}+\Gamma_{1\rmh,1\rme}\rho_{1,\rme,{\rm cL}}-(\Gamma_{0\rmh,1\rmh}^{\rm s}{+}\Gamma_{1\rme,1\rmh})\rho_{1,\rmh,{\rm cL}},\nonumber\\
\dot\rho_{1,\rmh,{\rm oR}}&=\Gamma_{1\rmh,0\rmh}^{\rm R}\rho_{0,\rmh}+\Gamma_{1\rmh,1\rme}\rho_{1,\rme,{\rm oR}}-(\Gamma_{0\rmh,1\rmh}^{\rm s}{+}\Gamma_{1\rme,1\rmh})\rho_{1,\rmh,{\rm oR}},\nonumber\\
\dot\rho_{1,\rmh,{\rm cR}}&=\Gamma_{1\rmh,1\rme}\rho_{1,\rme,{\rm cR}}-(\Gamma_{0\rmh,1\rmh}^{\rm s}{+}\Gamma_{1\rme,1\rmh})\rho_{1,\rmh,{\rm cR}},\\
\dot\rho_{1,\rmc,{\rm oL}}&=z\Gamma_{1\rmc,0\rmc}^{\rm L}\rho_{0,\rmc}+\Gamma_{1\rmc,1\rme}\rho_{1,\rme,{\rm oL}}-(\Gamma_{0\rmc,1\rmc}^{\rm s}{+}\Gamma_{1\rme,1\rmc})\rho_{1,\rmc,{\rm oL}}\nonumber,\\
\dot\rho_{1,\rmc,{\rm cL}}&=\Gamma_{1\rmc,1\rme}\rho_{1,\rme,{\rm cL}}-(\Gamma_{0\rmc,1\rmc}^{\rm s}{+}\Gamma_{1\rme,1\rmc})\rho_{1,\rmc,{\rm cL}},\nonumber\\
\dot\rho_{1,\rmc,{\rm oR}}&=\Gamma_{1\rmc,1\rme}\rho_{1,\rme,{\rm oR}}-(\Gamma_{0\rmc,1\rmc}^{\rm s}{+}\Gamma_{1\rme,1\rmc})\rho_{1,\rmc,{\rm oR}},\nonumber\\
\dot\rho_{1,\rmc,{\rm cR}}&=\Gamma_{1\rmc,0\rmc}^{\rm R}\rho_{0,\rmc}+\Gamma_{1\rmc,1\rme}\rho_{1,\rme,{\rm cR}}-(\Gamma_{0\rmc,1\rmc}^{\rm s}{+}\Gamma_{1\rme,1\rmc})\rho_{1,\rmc,{\rm cR}},\nonumber
\end{align}
for the probabilities where the composite system contains two electrons. Here we have defined $\Gamma_{m,p}^{\rm s}=\Gamma_{m,p}^\rmL+\Gamma_{m,p}^\rmR$, $\Gamma_{s\rmd,s\rme}=\Gamma_{s\rmh,s\rme}+\Gamma_{s\rmc,s\rme}$, and $\Gamma_{s\rme,s\rmd}=\Gamma_{s\rme,s\rmh}+\Gamma_{s\rme,s\rmc}$. The full counting statistics for $n$ and $w$ is obtained from the lowest eigenvalue of the matrix ${\cal M}$ associated with the previous set of equations $\pmb{\dot\rho}={\cal M}\pmb{\rho}$. For more details, see e.g, Ref.~\cite{sanchez_detection_2012}. With this, one can verify the fluctuation theorem expressed in Eq.~\eqref{eq:ft3}.

Setting $z=z_{\rm F}=1$ one can also compute the feedback-assisted current:
\beq
\begin{aligned}
F=&\Gamma_{0\rmh,1\rmh}^{\rmR}\rho_{1,\rmh,{\rm oL}}+\Gamma_{0\rmc,1\rmc}^{\rmL}\rho_{1,\rmc,{\rm oR}}\\&-\Gamma_{0\rmc,1\rmc}^{\rmR}\rho_{1,\rmc,{\rm cL}}-\Gamma_{0\rmh,1\rmh}^{\rmL}\rho_{1,\rmh,{\rm cR}},
\end{aligned}
\eeq
which cannot be obtained from knowing only the occupation of the charge states.
\newpage
\section{Fluctuation theorem for unequal charging energies}
\label{sec:ftuneq}
For unequal charging energies, $U_\rmC\neq U_\rmH$, the fluctuation theorem in Eq.~\eqref{eq:ft1} is generalized to
\begin{equation}
\label{eq:ft1uneq}
\frac{P(-w,-q)}{P(w,q)}=e^{w\beta eV-q(U_\rmC\beta_\rmC-U_\rmH\beta_\rmH)}.
\end{equation}
In the strong demon limit (where $w=q$), the stall voltage is 
\beq
V_{\rm s}=T\left(\frac{U_\rmC}{T_\rmC}-\frac{U_\rmH}{T_\rmH}\right).
\eeq

The fluctuation relation in Eq.~\eqref{eq:ft3} is modified as
\begin{eqnarray}
\label{eq:ftnon}
\frac{P(-w,-n)}{P(w,n)}=e^{w\beta eV-n(\delta_\rmL+\delta_\rmR+g)},
\end{eqnarray}
where
\begin{equation}
\label{eq:g}
e^g=\frac{\cosh\left[\beta\left(U_\rmC{+}U_\rmH{+}\xi_\rmL{+}\xi_\rmR\right)\right]+\cosh\left[\frac{\beta}{2}\left(eV{-}U_\rmC{+}U_\rmH\right)\right]}{\cosh\left[\beta\left(U_\rmC{+}U_\rmH{+}\xi_\rmL{+}\xi_\rmR\right)\right]+\cosh\left[\frac{\beta}{2}\left(eV{+}U_\rmC{-}U_\rmH\right)\right]}.
\end{equation}
The fluctuation relation in Eq.~\eqref{eq:ftnon} thus reduces to Eq.~\eqref{eq:ft3} both for equal charging energies, as well as for vanishing voltages. We note that in the error-free demon limit, where $w=n$ on each trajectory, Eq.~\eqref{eq:ftnon} implies that the stall voltage is given by the solution of the transcendental equation $\beta eV=\delta_\rmL+\delta_\rmR+g$.

\bibliography{biblio}

\begin{thebibliography}{162}%
\makeatletter
\providecommand \@ifxundefined [1]{%
 \@ifx{#1\undefined}
}%
\providecommand \@ifnum [1]{%
 \ifnum #1\expandafter \@firstoftwo
 \else \expandafter \@secondoftwo
 \fi
}%
\providecommand \@ifx [1]{%
 \ifx #1\expandafter \@firstoftwo
 \else \expandafter \@secondoftwo
 \fi
}%
\providecommand \natexlab [1]{#1}%
\providecommand \enquote  [1]{``#1''}%
\providecommand \bibnamefont  [1]{#1}%
\providecommand \bibfnamefont [1]{#1}%
\providecommand \citenamefont [1]{#1}%
\providecommand \href@noop [0]{\@secondoftwo}%
\providecommand \href [0]{\begingroup \@sanitize@url \@href}%
\providecommand \@href[1]{\@@startlink{#1}\@@href}%
\providecommand \@@href[1]{\endgroup#1\@@endlink}%
\providecommand \@sanitize@url [0]{\catcode `\\12\catcode `\$12\catcode
  `\&12\catcode `\#12\catcode `\^12\catcode `\_12\catcode `\%12\relax}%
\providecommand \@@startlink[1]{}%
\providecommand \@@endlink[0]{}%
\providecommand \url  [0]{\begingroup\@sanitize@url \@url }%
\providecommand \@url [1]{\endgroup\@href {#1}{\urlprefix }}%
\providecommand \urlprefix  [0]{URL }%
\providecommand \Eprint [0]{\href }%
\providecommand \doibase [0]{http://dx.doi.org/}%
\providecommand \selectlanguage [0]{\@gobble}%
\providecommand \bibinfo  [0]{\@secondoftwo}%
\providecommand \bibfield  [0]{\@secondoftwo}%
\providecommand \translation [1]{[#1]}%
\providecommand \BibitemOpen [0]{}%
\providecommand \bibitemStop [0]{}%
\providecommand \bibitemNoStop [0]{.\EOS\space}%
\providecommand \EOS [0]{\spacefactor3000\relax}%
\providecommand \BibitemShut  [1]{\csname bibitem#1\endcsname}%
\let\auto@bib@innerbib\@empty
\bibitem [{\citenamefont {Lindblad}(1974)}]{lindblad:1974}%
  \BibitemOpen
  \bibfield  {author} {\bibinfo {author} {\bibfnamefont {G.}~\bibnamefont
  {Lindblad}},\ }\bibfield  {title} {\enquote {\bibinfo {title} {Measurements
  and information for thermodynamic quantities},}\ }\href {\doibase
  10.1007/BF01010219} {\bibfield  {journal} {\bibinfo  {journal} {J. Stat.
  Phys.}\ }\textbf {\bibinfo {volume} {11}},\ \bibinfo {pages} {231} (\bibinfo
  {year} {1974})}\BibitemShut {NoStop}%
\bibitem [{\citenamefont {Parrondo}\ \emph {et~al.}(2015)\citenamefont
  {Parrondo}, \citenamefont {Horowitz},\ and\ \citenamefont
  {Sagawa}}]{parrondo:2015}%
  \BibitemOpen
  \bibfield  {author} {\bibinfo {author} {\bibfnamefont {J.~M.~R.}\
  \bibnamefont {Parrondo}}, \bibinfo {author} {\bibfnamefont {J.~M.}\
  \bibnamefont {Horowitz}}, \ and\ \bibinfo {author} {\bibfnamefont
  {T.}~\bibnamefont {Sagawa}},\ }\bibfield  {title} {\enquote {\bibinfo {title}
  {Thermodynamics of information},}\ }\href {\doibase 10.1038/nphys3230}
  {\bibfield  {journal} {\bibinfo  {journal} {Nat. Phys.}\ }\textbf {\bibinfo
  {volume} {11}},\ \bibinfo {pages} {131} (\bibinfo {year} {2015})}\BibitemShut
  {NoStop}%
\bibitem [{\citenamefont {Goold}\ \emph {et~al.}(2016)\citenamefont {Goold},
  \citenamefont {Huber}, \citenamefont {Riera}, \citenamefont {del Rio},\ and\
  \citenamefont {Skrzypczyk}}]{goold:2016}%
  \BibitemOpen
  \bibfield  {author} {\bibinfo {author} {\bibfnamefont {J.}~\bibnamefont
  {Goold}}, \bibinfo {author} {\bibfnamefont {M.}~\bibnamefont {Huber}},
  \bibinfo {author} {\bibfnamefont {A.}~\bibnamefont {Riera}}, \bibinfo
  {author} {\bibfnamefont {L.}~\bibnamefont {del Rio}}, \ and\ \bibinfo
  {author} {\bibfnamefont {P.}~\bibnamefont {Skrzypczyk}},\ }\bibfield  {title}
  {\enquote {\bibinfo {title} {The role of quantum information in
  thermodynamics -- a topical review},}\ }\href
  {http://stacks.iop.org/1751-8121/49/i=14/a=143001} {\bibfield  {journal}
  {\bibinfo  {journal} {J. Phys. A: Math. Theor.}\ }\textbf {\bibinfo {volume}
  {49}},\ \bibinfo {pages} {143001} (\bibinfo {year} {2016})}\BibitemShut
  {NoStop}%
\bibitem [{\citenamefont {Esposito}\ \emph {et~al.}(2010)\citenamefont
  {Esposito}, \citenamefont {Lindenberg},\ and\ \citenamefont {den
  Broeck}}]{esposito:2010njp}%
  \BibitemOpen
  \bibfield  {author} {\bibinfo {author} {\bibfnamefont {M.}~\bibnamefont
  {Esposito}}, \bibinfo {author} {\bibfnamefont {K.}~\bibnamefont
  {Lindenberg}}, \ and\ \bibinfo {author} {\bibfnamefont {C.~V.}\ \bibnamefont
  {den Broeck}},\ }\bibfield  {title} {\enquote {\bibinfo {title} {Entropy
  production as correlation between system and reservoir},}\ }\href {\doibase
  10.1088/1367-2630/12/1/013013} {\bibfield  {journal} {\bibinfo  {journal}
  {New J. Phys.}\ }\textbf {\bibinfo {volume} {12}},\ \bibinfo {pages} {013013}
  (\bibinfo {year} {2010})}\BibitemShut {NoStop}%
\bibitem [{\citenamefont {Sagawa}(2012)}]{sagawa:inbook}%
  \BibitemOpen
  \bibfield  {author} {\bibinfo {author} {\bibfnamefont {T.}~\bibnamefont
  {Sagawa}},\ }\enquote {\bibinfo {title} {Second law-like inequalities with
  quantum relative entropy: An introduction},}\ in\ \href {\doibase
  10.1142/9789814425193_0003} {\emph {\bibinfo {booktitle} {Lectures on Quantum
  Computing, Thermodynamics and Statistical Physics}}},\ \bibinfo {editor}
  {edited by\ \bibinfo {editor} {\bibfnamefont {M.}~\bibnamefont {Nakahara}}\
  and\ \bibinfo {editor} {\bibfnamefont {S.}~\bibnamefont {Tanaka}}}\ (\bibinfo
   {publisher} {World Scientific},\ \bibinfo {year} {2012})\BibitemShut
  {NoStop}%
\bibitem [{\citenamefont {Callen}(1985)}]{callen:book}%
  \BibitemOpen
  \bibfield  {author} {\bibinfo {author} {\bibfnamefont {H.~B.}\ \bibnamefont
  {Callen}},\ }\href@noop {} {\emph {\bibinfo {title} {Thermodynamics and an
  Introduction to Thermostatistics}}}\ (\bibinfo  {publisher} {Wiley, New
  York},\ \bibinfo {year} {1985})\BibitemShut {NoStop}%
\bibitem [{\citenamefont {Maxwell}(1871)}]{maxwell:1871}%
  \BibitemOpen
  \bibfield  {author} {\bibinfo {author} {\bibfnamefont {J.~C.}\ \bibnamefont
  {Maxwell}},\ }\href@noop {} {\emph {\bibinfo {title} {Theory of Heat}}}\
  (\bibinfo  {publisher} {Longmans, Green, and Co.},\ \bibinfo {year}
  {1871})\BibitemShut {NoStop}%
\bibitem [{\citenamefont {Leff}\ and\ \citenamefont
  {Rex}(2002)}]{maxwell:book}%
  \BibitemOpen
  \bibinfo {editor} {\bibfnamefont {H.}~\bibnamefont {Leff}}\ and\ \bibinfo
  {editor} {\bibfnamefont {A.~F.}\ \bibnamefont {Rex}},\ eds.,\ \href@noop {}
  {\emph {\bibinfo {title} {Maxwell's Demon 2 Entropy, Classical and Quantum
  Information, Computing}}}\ (\bibinfo  {publisher} {CRC Press},\ \bibinfo
  {year} {2002})\BibitemShut {NoStop}%
\bibitem [{\citenamefont {Landauer}(1992)}]{landauer:1992}%
  \BibitemOpen
  \bibfield  {author} {\bibinfo {author} {\bibfnamefont {R.}~\bibnamefont
  {Landauer}},\ }\bibfield  {title} {\enquote {\bibinfo {title} {Information is
  physical},}\ }in\ \href {\doibase 10.1109/PHYCMP.1992.615478} {\emph
  {\bibinfo {booktitle} {Workshop on Physics and Computation}}}\ (\bibinfo
  {year} {1992})\BibitemShut {NoStop}%
\bibitem [{\citenamefont {Landauer}(1961)}]{landauer:1961}%
  \BibitemOpen
  \bibfield  {author} {\bibinfo {author} {\bibfnamefont {R.}~\bibnamefont
  {Landauer}},\ }\bibfield  {title} {\enquote {\bibinfo {title}
  {Irreversibility and heat generation in the computing process},}\ }\href
  {\doibase 10.1147/rd.53.0183} {\bibfield  {journal} {\bibinfo  {journal} {IBM
  J. Res. Dev.}\ }\textbf {\bibinfo {volume} {5}},\ \bibinfo {pages} {183}
  (\bibinfo {year} {1961})}\BibitemShut {NoStop}%
\bibitem [{\citenamefont {Bennett}(1982)}]{bennett:1982}%
  \BibitemOpen
  \bibfield  {author} {\bibinfo {author} {\bibfnamefont {C.~H.}\ \bibnamefont
  {Bennett}},\ }\bibfield  {title} {\enquote {\bibinfo {title} {The
  thermodynamics of computation---a review},}\ }\href {\doibase
  10.1007/BF02084158} {\bibfield  {journal} {\bibinfo  {journal} {Int. J.
  Theor. Phys.}\ }\textbf {\bibinfo {volume} {21}},\ \bibinfo {pages} {905}
  (\bibinfo {year} {1982})}\BibitemShut {NoStop}%
\bibitem [{\citenamefont {Funo}\ \emph {et~al.}(2013)\citenamefont {Funo},
  \citenamefont {Watanabe},\ and\ \citenamefont {Ueda}}]{funo:2013}%
  \BibitemOpen
  \bibfield  {author} {\bibinfo {author} {\bibfnamefont {K.}~\bibnamefont
  {Funo}}, \bibinfo {author} {\bibfnamefont {Y.}~\bibnamefont {Watanabe}}, \
  and\ \bibinfo {author} {\bibfnamefont {M.}~\bibnamefont {Ueda}},\ }\bibfield
  {title} {\enquote {\bibinfo {title} {Integral quantum fluctuation theorems
  under measurement and feedback control},}\ }\href {\doibase
  10.1103/PhysRevE.88.052121} {\bibfield  {journal} {\bibinfo  {journal} {Phys.
  Rev. E}\ }\textbf {\bibinfo {volume} {88}},\ \bibinfo {pages} {052121}
  (\bibinfo {year} {2013})}\BibitemShut {NoStop}%
\bibitem [{\citenamefont {Horowitz}\ and\ \citenamefont
  {Esposito}(2014)}]{horowitz:2014prx}%
  \BibitemOpen
  \bibfield  {author} {\bibinfo {author} {\bibfnamefont {J.~M.}\ \bibnamefont
  {Horowitz}}\ and\ \bibinfo {author} {\bibfnamefont {M.}~\bibnamefont
  {Esposito}},\ }\bibfield  {title} {\enquote {\bibinfo {title} {Thermodynamics
  with continuous information flow},}\ }\href {\doibase
  10.1103/PhysRevX.4.031015} {\bibfield  {journal} {\bibinfo  {journal} {Phys.
  Rev. X}\ }\textbf {\bibinfo {volume} {4}},\ \bibinfo {pages} {031015}
  (\bibinfo {year} {2014})}\BibitemShut {NoStop}%
\bibitem [{\citenamefont {Harris}\ and\ \citenamefont
  {Sch\"utz}(2007)}]{harris:2007}%
  \BibitemOpen
  \bibfield  {author} {\bibinfo {author} {\bibfnamefont {R.~J.}\ \bibnamefont
  {Harris}}\ and\ \bibinfo {author} {\bibfnamefont {G.~M.}\ \bibnamefont
  {Sch\"utz}},\ }\bibfield  {title} {\enquote {\bibinfo {title} {Fluctuation
  theorems for stochastic dynamics},}\ }\href
  {http://stacks.iop.org/1742-5468/2007/i=07/a=P07020} {\bibfield  {journal}
  {\bibinfo  {journal} {J. Stat. Mech. Theor. Exp.}\ }\textbf {\bibinfo
  {volume} {2007}},\ \bibinfo {pages} {P07020} (\bibinfo {year}
  {2007})}\BibitemShut {NoStop}%
\bibitem [{\citenamefont {Esposito}\ \emph {et~al.}(2009)\citenamefont
  {Esposito}, \citenamefont {Harbola},\ and\ \citenamefont
  {Mukamel}}]{esposito:2009rmp}%
  \BibitemOpen
  \bibfield  {author} {\bibinfo {author} {\bibfnamefont {M.}~\bibnamefont
  {Esposito}}, \bibinfo {author} {\bibfnamefont {U.}~\bibnamefont {Harbola}}, \
  and\ \bibinfo {author} {\bibfnamefont {S.}~\bibnamefont {Mukamel}},\
  }\bibfield  {title} {\enquote {\bibinfo {title} {Nonequilibrium fluctuations,
  fluctuation theorems, and counting statistics in quantum systems},}\ }\href
  {\doibase 10.1103/RevModPhys.81.1665} {\bibfield  {journal} {\bibinfo
  {journal} {Rev. Mod. Phys.}\ }\textbf {\bibinfo {volume} {81}},\ \bibinfo
  {pages} {1665} (\bibinfo {year} {2009})}\BibitemShut {NoStop}%
\bibitem [{\citenamefont {Jarzynski}(2011)}]{jarzynski:2011}%
  \BibitemOpen
  \bibfield  {author} {\bibinfo {author} {\bibfnamefont {C.}~\bibnamefont
  {Jarzynski}},\ }\bibfield  {title} {\enquote {\bibinfo {title} {{Equalities
  and Inequalities: Irreversibility and the Second Law of Thermodynamics at the
  Nanoscale}},}\ }\href {\doibase 10.1146/annurev-conmatphys-062910-140506}
  {\bibfield  {journal} {\bibinfo  {journal} {Annu. Rev. Condens. Matter
  Phys.}\ }\textbf {\bibinfo {volume} {2}},\ \bibinfo {pages} {329--351}
  (\bibinfo {year} {2011})}\BibitemShut {NoStop}%
\bibitem [{\citenamefont {Seifert}(2012)}]{seifert:2012}%
  \BibitemOpen
  \bibfield  {author} {\bibinfo {author} {\bibfnamefont {U.}~\bibnamefont
  {Seifert}},\ }\bibfield  {title} {\enquote {\bibinfo {title} {Stochastic
  thermodynamics, fluctuation theorems and molecular machines},}\ }\href
  {\doibase 10.1088/0034-4885/75/12/126001} {\bibfield  {journal} {\bibinfo
  {journal} {Rep. Prog. Phys.}\ }\textbf {\bibinfo {volume} {75}},\ \bibinfo
  {pages} {126001} (\bibinfo {year} {2012})}\BibitemShut {NoStop}%
\bibitem [{\citenamefont {Mansour}\ and\ \citenamefont
  {Baras}(2017)}]{mansour:2017}%
  \BibitemOpen
  \bibfield  {author} {\bibinfo {author} {\bibfnamefont {M.~M.}\ \bibnamefont
  {Mansour}}\ and\ \bibinfo {author} {\bibfnamefont {F.}~\bibnamefont
  {Baras}},\ }\bibfield  {title} {\enquote {\bibinfo {title} {Fluctuation
  theorem: A critical review},}\ }\href {\doibase 10.1063/1.4986600} {\bibfield
   {journal} {\bibinfo  {journal} {Chaos}\ }\textbf {\bibinfo {volume} {27}},\
  \bibinfo {pages} {104609} (\bibinfo {year} {2017})}\BibitemShut {NoStop}%
\bibitem [{\citenamefont {Jarzynski}(1997{\natexlab{a}})}]{jarzynski:1997}%
  \BibitemOpen
  \bibfield  {author} {\bibinfo {author} {\bibfnamefont {C.}~\bibnamefont
  {Jarzynski}},\ }\bibfield  {title} {\enquote {\bibinfo {title}
  {Nonequilibrium equality for free energy differences},}\ }\href {\doibase
  10.1103/PhysRevLett.78.2690} {\bibfield  {journal} {\bibinfo  {journal}
  {Phys. Rev. Lett.}\ }\textbf {\bibinfo {volume} {78}},\ \bibinfo {pages}
  {2690} (\bibinfo {year} {1997}{\natexlab{a}})}\BibitemShut {NoStop}%
\bibitem [{\citenamefont {Jarzynski}(1997{\natexlab{b}})}]{jarzynski:1997pre}%
  \BibitemOpen
  \bibfield  {author} {\bibinfo {author} {\bibfnamefont {C.}~\bibnamefont
  {Jarzynski}},\ }\bibfield  {title} {\enquote {\bibinfo {title} {Equilibrium
  free-energy differences from nonequilibrium measurements: A master-equation
  approach},}\ }\href {\doibase 10.1103/PhysRevE.56.5018} {\bibfield  {journal}
  {\bibinfo  {journal} {Phys. Rev. E}\ }\textbf {\bibinfo {volume} {56}},\
  \bibinfo {pages} {5018} (\bibinfo {year} {1997}{\natexlab{b}})}\BibitemShut
  {NoStop}%
\bibitem [{\citenamefont {Crooks}(1998)}]{crooks:1998}%
  \BibitemOpen
  \bibfield  {author} {\bibinfo {author} {\bibfnamefont {G.~E.}\ \bibnamefont
  {Crooks}},\ }\bibfield  {title} {\enquote {\bibinfo {title} {Nonequilibrium
  measurements of free energy differences for microscopically reversible
  markovian systems},}\ }\href {\doibase 10.1023/A:1023208217925} {\bibfield
  {journal} {\bibinfo  {journal} {J. Stat. Phys.}\ }\textbf {\bibinfo {volume}
  {90}},\ \bibinfo {pages} {1481} (\bibinfo {year} {1998})}\BibitemShut
  {NoStop}%
\bibitem [{\citenamefont {Crooks}(1999)}]{crooks:1999}%
  \BibitemOpen
  \bibfield  {author} {\bibinfo {author} {\bibfnamefont {G.~E.}\ \bibnamefont
  {Crooks}},\ }\bibfield  {title} {\enquote {\bibinfo {title} {Entropy
  production fluctuation theorem and the nonequilibrium work relation for free
  energy differences},}\ }\href {\doibase 10.1103/PhysRevE.60.2721} {\bibfield
  {journal} {\bibinfo  {journal} {Phys. Rev. E}\ }\textbf {\bibinfo {volume}
  {60}},\ \bibinfo {pages} {2721} (\bibinfo {year} {1999})}\BibitemShut
  {NoStop}%
\bibitem [{\citenamefont {Crooks}(2000)}]{crooks:2000}%
  \BibitemOpen
  \bibfield  {author} {\bibinfo {author} {\bibfnamefont {G.~E.}\ \bibnamefont
  {Crooks}},\ }\bibfield  {title} {\enquote {\bibinfo {title} {Path-ensemble
  averages in systems driven far from equilibrium},}\ }\href {\doibase
  10.1103/PhysRevE.61.2361} {\bibfield  {journal} {\bibinfo  {journal} {Phys.
  Rev. E}\ }\textbf {\bibinfo {volume} {61}},\ \bibinfo {pages} {2361}
  (\bibinfo {year} {2000})}\BibitemShut {NoStop}%
\bibitem [{\citenamefont {Kurchan}()}]{kurchan:2000}%
  \BibitemOpen
  \bibfield  {author} {\bibinfo {author} {\bibfnamefont {J.}~\bibnamefont
  {Kurchan}},\ }\href@noop {} {\enquote {\bibinfo {title} {A quantum
  fluctuation theorem},}\ }\Eprint {http://arxiv.org/abs/cond-mat/0007360}
  {arXiv:cond-mat/0007360} \BibitemShut {NoStop}%
\bibitem [{\citenamefont {Tasaki}()}]{tasaki:2000}%
  \BibitemOpen
  \bibfield  {author} {\bibinfo {author} {\bibfnamefont {H.}~\bibnamefont
  {Tasaki}},\ }\href@noop {} {\enquote {\bibinfo {title} {Jarzynski relations
  for quantum systems and some applications},}\ }\Eprint
  {http://arxiv.org/abs/cond-mat/0009244} {arXiv:cond-mat/0009244} \BibitemShut
  {NoStop}%
\bibitem [{\citenamefont {Kawai}\ \emph {et~al.}(2007)\citenamefont {Kawai},
  \citenamefont {Parrondo},\ and\ \citenamefont {den Broeck}}]{kawai:2007}%
  \BibitemOpen
  \bibfield  {author} {\bibinfo {author} {\bibfnamefont {R.}~\bibnamefont
  {Kawai}}, \bibinfo {author} {\bibfnamefont {J.~M.~R.}\ \bibnamefont
  {Parrondo}}, \ and\ \bibinfo {author} {\bibfnamefont {C.~V.}\ \bibnamefont
  {den Broeck}},\ }\bibfield  {title} {\enquote {\bibinfo {title} {Dissipation:
  The phase-space perspective},}\ }\href {\doibase
  10.1103/PhysRevLett.98.080602} {\bibfield  {journal} {\bibinfo  {journal}
  {Phys. Rev. Lett.}\ }\textbf {\bibinfo {volume} {98}},\ \bibinfo {pages}
  {080602} (\bibinfo {year} {2007})}\BibitemShut {NoStop}%
\bibitem [{\citenamefont {Touchette}\ and\ \citenamefont
  {Lloyd}(2000)}]{touchette:2000}%
  \BibitemOpen
  \bibfield  {author} {\bibinfo {author} {\bibfnamefont {H.}~\bibnamefont
  {Touchette}}\ and\ \bibinfo {author} {\bibfnamefont {S.}~\bibnamefont
  {Lloyd}},\ }\bibfield  {title} {\enquote {\bibinfo {title}
  {Information-theoretic limits of control},}\ }\href {\doibase
  10.1103/PhysRevLett.84.1156} {\bibfield  {journal} {\bibinfo  {journal}
  {Phys. Rev. Lett.}\ }\textbf {\bibinfo {volume} {84}},\ \bibinfo {pages}
  {1156} (\bibinfo {year} {2000})}\BibitemShut {NoStop}%
\bibitem [{\citenamefont {Sagawa}\ and\ \citenamefont
  {Ueda}(2008)}]{sagawa:2008}%
  \BibitemOpen
  \bibfield  {author} {\bibinfo {author} {\bibfnamefont {T.}~\bibnamefont
  {Sagawa}}\ and\ \bibinfo {author} {\bibfnamefont {M.}~\bibnamefont {Ueda}},\
  }\bibfield  {title} {\enquote {\bibinfo {title} {Second law of thermodynamics
  with discrete quantum feedback control},}\ }\href {\doibase
  10.1103/PhysRevLett.100.080403} {\bibfield  {journal} {\bibinfo  {journal}
  {Phys. Rev. Lett.}\ }\textbf {\bibinfo {volume} {100}},\ \bibinfo {pages}
  {080403} (\bibinfo {year} {2008})}\BibitemShut {NoStop}%
\bibitem [{\citenamefont {Maruyama}\ \emph {et~al.}(2009)\citenamefont
  {Maruyama}, \citenamefont {Nori},\ and\ \citenamefont
  {Vedral}}]{maruyama:2009}%
  \BibitemOpen
  \bibfield  {author} {\bibinfo {author} {\bibfnamefont {K.}~\bibnamefont
  {Maruyama}}, \bibinfo {author} {\bibfnamefont {F.}~\bibnamefont {Nori}}, \
  and\ \bibinfo {author} {\bibfnamefont {V.}~\bibnamefont {Vedral}},\
  }\bibfield  {title} {\enquote {\bibinfo {title} {Colloquium: The physics of
  {M}axwell's demon and information},}\ }\href {\doibase
  10.1103/RevModPhys.81.1} {\bibfield  {journal} {\bibinfo  {journal} {Rev.
  Mod. Phys.}\ }\textbf {\bibinfo {volume} {81}},\ \bibinfo {pages} {1}
  (\bibinfo {year} {2009})}\BibitemShut {NoStop}%
\bibitem [{\citenamefont {Cao}\ and\ \citenamefont {Feito}(2009)}]{cao:2009}%
  \BibitemOpen
  \bibfield  {author} {\bibinfo {author} {\bibfnamefont {F.~J.}\ \bibnamefont
  {Cao}}\ and\ \bibinfo {author} {\bibfnamefont {M.}~\bibnamefont {Feito}},\
  }\bibfield  {title} {\enquote {\bibinfo {title} {Thermodynamics of feedback
  controlled systems},}\ }\href {\doibase 10.1103/PhysRevE.79.041118}
  {\bibfield  {journal} {\bibinfo  {journal} {Phys. Rev. E}\ }\textbf {\bibinfo
  {volume} {79}},\ \bibinfo {pages} {041118} (\bibinfo {year}
  {2009})}\BibitemShut {NoStop}%
\bibitem [{\citenamefont {Sagawa}\ and\ \citenamefont
  {Ueda}(2010)}]{sagawa:2010}%
  \BibitemOpen
  \bibfield  {author} {\bibinfo {author} {\bibfnamefont {T.}~\bibnamefont
  {Sagawa}}\ and\ \bibinfo {author} {\bibfnamefont {M.}~\bibnamefont {Ueda}},\
  }\bibfield  {title} {\enquote {\bibinfo {title} {Generalized {J}arzynski
  equality under nonequilibrium feedback control},}\ }\href {\doibase
  10.1103/PhysRevLett.104.090602} {\bibfield  {journal} {\bibinfo  {journal}
  {Phys. Rev. Lett.}\ }\textbf {\bibinfo {volume} {104}},\ \bibinfo {pages}
  {090602} (\bibinfo {year} {2010})}\BibitemShut {NoStop}%
\bibitem [{\citenamefont {Horowitz}\ and\ \citenamefont
  {Vaikuntanathan}(2010)}]{horowitz:2010}%
  \BibitemOpen
  \bibfield  {author} {\bibinfo {author} {\bibfnamefont {J.~M.}\ \bibnamefont
  {Horowitz}}\ and\ \bibinfo {author} {\bibfnamefont {S.}~\bibnamefont
  {Vaikuntanathan}},\ }\bibfield  {title} {\enquote {\bibinfo {title}
  {Nonequilibrium detailed fluctuation theorem for repeated discrete
  feedback},}\ }\href {\doibase 10.1103/PhysRevE.82.061120} {\bibfield
  {journal} {\bibinfo  {journal} {Phys. Rev. E}\ }\textbf {\bibinfo {volume}
  {82}},\ \bibinfo {pages} {061120} (\bibinfo {year} {2010})}\BibitemShut
  {NoStop}%
\bibitem [{\citenamefont {Ponmurugan}(2010)}]{ponmurugan:2010}%
  \BibitemOpen
  \bibfield  {author} {\bibinfo {author} {\bibfnamefont {M.}~\bibnamefont
  {Ponmurugan}},\ }\bibfield  {title} {\enquote {\bibinfo {title} {Generalized
  detailed fluctuation theorem under nonequilibrium feedback control},}\ }\href
  {\doibase 10.1103/PhysRevE.82.031129} {\bibfield  {journal} {\bibinfo
  {journal} {Phys. Rev. E}\ }\textbf {\bibinfo {volume} {82}},\ \bibinfo
  {pages} {031129} (\bibinfo {year} {2010})}\BibitemShut {NoStop}%
\bibitem [{\citenamefont {Morikuni}\ and\ \citenamefont
  {Tasaki}(2011)}]{morikuni:2011}%
  \BibitemOpen
  \bibfield  {author} {\bibinfo {author} {\bibfnamefont {Y.}~\bibnamefont
  {Morikuni}}\ and\ \bibinfo {author} {\bibfnamefont {H.}~\bibnamefont
  {Tasaki}},\ }\bibfield  {title} {\enquote {\bibinfo {title} {Quantum
  {J}arzynski-{S}agawa-{U}eda relations},}\ }\href {\doibase
  10.1007/s10955-011-0153-7} {\bibfield  {journal} {\bibinfo  {journal} {J.
  Stat. Phys.}\ }\textbf {\bibinfo {volume} {143}},\ \bibinfo {pages} {1}
  (\bibinfo {year} {2011})}\BibitemShut {NoStop}%
\bibitem [{\citenamefont {Horowitz}\ and\ \citenamefont
  {Parrondo}(2011)}]{horowitz:2011}%
  \BibitemOpen
  \bibfield  {author} {\bibinfo {author} {\bibfnamefont {J.~M.}\ \bibnamefont
  {Horowitz}}\ and\ \bibinfo {author} {\bibfnamefont {J.~M.~R.}\ \bibnamefont
  {Parrondo}},\ }\bibfield  {title} {\enquote {\bibinfo {title} {Designing
  optimal discrete-feedback thermodynamic engines},}\ }\href {\doibase
  10.1088/1367-2630/13/12/123019} {\bibfield  {journal} {\bibinfo  {journal}
  {New J. Phys.}\ }\textbf {\bibinfo {volume} {13}},\ \bibinfo {pages} {123019}
  (\bibinfo {year} {2011})}\BibitemShut {NoStop}%
\bibitem [{\citenamefont {Sagawa}\ and\ \citenamefont
  {Ueda}(2012{\natexlab{a}})}]{sagawa:2012}%
  \BibitemOpen
  \bibfield  {author} {\bibinfo {author} {\bibfnamefont {T.}~\bibnamefont
  {Sagawa}}\ and\ \bibinfo {author} {\bibfnamefont {M.}~\bibnamefont {Ueda}},\
  }\bibfield  {title} {\enquote {\bibinfo {title} {Nonequilibrium
  thermodynamics of feedback control},}\ }\href {\doibase
  10.1103/PhysRevE.85.021104} {\bibfield  {journal} {\bibinfo  {journal} {Phys.
  Rev. E}\ }\textbf {\bibinfo {volume} {85}},\ \bibinfo {pages} {021104}
  (\bibinfo {year} {2012}{\natexlab{a}})}\BibitemShut {NoStop}%
\bibitem [{\citenamefont {Sagawa}\ and\ \citenamefont
  {Ueda}(2012{\natexlab{b}})}]{sagawa:2012prl}%
  \BibitemOpen
  \bibfield  {author} {\bibinfo {author} {\bibfnamefont {T.}~\bibnamefont
  {Sagawa}}\ and\ \bibinfo {author} {\bibfnamefont {M.}~\bibnamefont {Ueda}},\
  }\bibfield  {title} {\enquote {\bibinfo {title} {Fluctuation theorem with
  information exchange: Role of correlations in stochastic thermodynamics},}\
  }\href {\doibase 10.1103/PhysRevLett.109.180602} {\bibfield  {journal}
  {\bibinfo  {journal} {Phys. Rev. Lett.}\ }\textbf {\bibinfo {volume} {109}},\
  \bibinfo {pages} {180602} (\bibinfo {year} {2012}{\natexlab{b}})}\BibitemShut
  {NoStop}%
\bibitem [{\citenamefont {Sagawa}\ and\ \citenamefont
  {Ueda}(2013)}]{sagawa:2013}%
  \BibitemOpen
  \bibfield  {author} {\bibinfo {author} {\bibfnamefont {T.}~\bibnamefont
  {Sagawa}}\ and\ \bibinfo {author} {\bibfnamefont {M.}~\bibnamefont {Ueda}},\
  }\bibfield  {title} {\enquote {\bibinfo {title} {Role of mutual information
  in entropy production under information exchanges},}\ }\href
  {http://stacks.iop.org/1367-2630/15/i=12/a=125012} {\bibfield  {journal}
  {\bibinfo  {journal} {New J. Phys.}\ }\textbf {\bibinfo {volume} {15}},\
  \bibinfo {pages} {125012} (\bibinfo {year} {2013})}\BibitemShut {NoStop}%
\bibitem [{\citenamefont {Lahiri}\ \emph {et~al.}(2012)\citenamefont {Lahiri},
  \citenamefont {Rana},\ and\ \citenamefont {Jayannavar}}]{lahiri:2012}%
  \BibitemOpen
  \bibfield  {author} {\bibinfo {author} {\bibfnamefont {S.}~\bibnamefont
  {Lahiri}}, \bibinfo {author} {\bibfnamefont {S.}~\bibnamefont {Rana}}, \ and\
  \bibinfo {author} {\bibfnamefont {A.~M.}\ \bibnamefont {Jayannavar}},\
  }\bibfield  {title} {\enquote {\bibinfo {title} {Fluctuation theorems in the
  presence of information gain and feedback},}\ }\href
  {http://stacks.iop.org/1751-8121/45/i=6/a=065002} {\bibfield  {journal}
  {\bibinfo  {journal} {J. Phys. A: Math. Theor.}\ }\textbf {\bibinfo {volume}
  {45}},\ \bibinfo {pages} {065002} (\bibinfo {year} {2012})}\BibitemShut
  {NoStop}%
\bibitem [{\citenamefont {Abreu}\ and\ \citenamefont
  {Seifert}(2012)}]{abreu:2012}%
  \BibitemOpen
  \bibfield  {author} {\bibinfo {author} {\bibfnamefont {D.}~\bibnamefont
  {Abreu}}\ and\ \bibinfo {author} {\bibfnamefont {U.}~\bibnamefont
  {Seifert}},\ }\bibfield  {title} {\enquote {\bibinfo {title} {Thermodynamics
  of genuine nonequilibrium states under feedback control},}\ }\href {\doibase
  10.1103/PhysRevLett.108.030601} {\bibfield  {journal} {\bibinfo  {journal}
  {Phys. Rev. Lett.}\ }\textbf {\bibinfo {volume} {108}},\ \bibinfo {pages}
  {030601} (\bibinfo {year} {2012})}\BibitemShut {NoStop}%
\bibitem [{\citenamefont {Deffner}\ and\ \citenamefont
  {Jarzynski}(2013)}]{deffner:2013}%
  \BibitemOpen
  \bibfield  {author} {\bibinfo {author} {\bibfnamefont {S.}~\bibnamefont
  {Deffner}}\ and\ \bibinfo {author} {\bibfnamefont {C.}~\bibnamefont
  {Jarzynski}},\ }\bibfield  {title} {\enquote {\bibinfo {title} {Information
  processing and the second law of thermodynamics: {A}n inclusive,
  {H}amiltonian approach},}\ }\href {\doibase 10.1103/PhysRevX.3.041003}
  {\bibfield  {journal} {\bibinfo  {journal} {Phys. Rev. X}\ }\textbf {\bibinfo
  {volume} {3}},\ \bibinfo {pages} {041003} (\bibinfo {year}
  {2013})}\BibitemShut {NoStop}%
\bibitem [{\citenamefont {Ito}\ and\ \citenamefont {Sagawa}(2013)}]{ito:2013}%
  \BibitemOpen
  \bibfield  {author} {\bibinfo {author} {\bibfnamefont {S.}~\bibnamefont
  {Ito}}\ and\ \bibinfo {author} {\bibfnamefont {T.}~\bibnamefont {Sagawa}},\
  }\bibfield  {title} {\enquote {\bibinfo {title} {Information thermodynamics
  on causal networks},}\ }\href {\doibase 10.1103/PhysRevLett.111.180603}
  {\bibfield  {journal} {\bibinfo  {journal} {Phys. Rev. Lett.}\ }\textbf
  {\bibinfo {volume} {111}},\ \bibinfo {pages} {180603} (\bibinfo {year}
  {2013})}\BibitemShut {NoStop}%
\bibitem [{\citenamefont {Barato}\ and\ \citenamefont
  {Seifert}(2014{\natexlab{a}})}]{barato:2014}%
  \BibitemOpen
  \bibfield  {author} {\bibinfo {author} {\bibfnamefont {A.~C.}\ \bibnamefont
  {Barato}}\ and\ \bibinfo {author} {\bibfnamefont {U.}~\bibnamefont
  {Seifert}},\ }\bibfield  {title} {\enquote {\bibinfo {title} {Unifying three
  perspectives on information processing in stochastic thermodynamics},}\
  }\href {\doibase 10.1103/PhysRevLett.112.090601} {\bibfield  {journal}
  {\bibinfo  {journal} {Phys. Rev. Lett.}\ }\textbf {\bibinfo {volume} {112}},\
  \bibinfo {pages} {090601} (\bibinfo {year} {2014}{\natexlab{a}})}\BibitemShut
  {NoStop}%
\bibitem [{\citenamefont {Barato}\ and\ \citenamefont
  {Seifert}(2014{\natexlab{b}})}]{barato:2014pre}%
  \BibitemOpen
  \bibfield  {author} {\bibinfo {author} {\bibfnamefont {A.~C.}\ \bibnamefont
  {Barato}}\ and\ \bibinfo {author} {\bibfnamefont {U.}~\bibnamefont
  {Seifert}},\ }\bibfield  {title} {\enquote {\bibinfo {title} {Stochastic
  thermodynamics with information reservoirs},}\ }\href {\doibase
  10.1103/PhysRevE.90.042150} {\bibfield  {journal} {\bibinfo  {journal} {Phys.
  Rev. E}\ }\textbf {\bibinfo {volume} {90}},\ \bibinfo {pages} {042150}
  (\bibinfo {year} {2014}{\natexlab{b}})}\BibitemShut {NoStop}%
\bibitem [{\citenamefont {Sagawa}(2014)}]{sagawa:2014}%
  \BibitemOpen
  \bibfield  {author} {\bibinfo {author} {\bibfnamefont {T.}~\bibnamefont
  {Sagawa}},\ }\bibfield  {title} {\enquote {\bibinfo {title} {Thermodynamic
  and logical reversibilities revisited},}\ }\href {\doibase
  10.1088/1742-5468/2014/03/P03025} {\bibfield  {journal} {\bibinfo  {journal}
  {J. Stat. Mech. Theor. Exp.}\ }\textbf {\bibinfo {volume} {2014}},\ \bibinfo
  {pages} {P03025} (\bibinfo {year} {2014})}\BibitemShut {NoStop}%
\bibitem [{\citenamefont {Ashida}\ \emph {et~al.}(2014)\citenamefont {Ashida},
  \citenamefont {Funo}, \citenamefont {Murashita},\ and\ \citenamefont
  {Ueda}}]{ashida:2014}%
  \BibitemOpen
  \bibfield  {author} {\bibinfo {author} {\bibfnamefont {Y.}~\bibnamefont
  {Ashida}}, \bibinfo {author} {\bibfnamefont {K.}~\bibnamefont {Funo}},
  \bibinfo {author} {\bibfnamefont {Y.}~\bibnamefont {Murashita}}, \ and\
  \bibinfo {author} {\bibfnamefont {M.}~\bibnamefont {Ueda}},\ }\bibfield
  {title} {\enquote {\bibinfo {title} {General achievable bound of extractable
  work under feedback control},}\ }\href {\doibase 10.1103/PhysRevE.90.052125}
  {\bibfield  {journal} {\bibinfo  {journal} {Phys. Rev. E}\ }\textbf {\bibinfo
  {volume} {90}},\ \bibinfo {pages} {052125} (\bibinfo {year}
  {2014})}\BibitemShut {NoStop}%
\bibitem [{\citenamefont {Horowitz}\ and\ \citenamefont
  {Sandberg}(2014)}]{horowitz:2014}%
  \BibitemOpen
  \bibfield  {author} {\bibinfo {author} {\bibfnamefont {J.~M.}\ \bibnamefont
  {Horowitz}}\ and\ \bibinfo {author} {\bibfnamefont {H.}~\bibnamefont
  {Sandberg}},\ }\bibfield  {title} {\enquote {\bibinfo {title}
  {Second-law-like inequalities with information and their interpretations},}\
  }\href {http://stacks.iop.org/1367-2630/16/i=12/a=125007} {\bibfield
  {journal} {\bibinfo  {journal} {New J. Phys.}\ }\textbf {\bibinfo {volume}
  {16}},\ \bibinfo {pages} {125007} (\bibinfo {year} {2014})}\BibitemShut
  {NoStop}%
\bibitem [{\citenamefont {Um}\ \emph {et~al.}(2015)\citenamefont {Um},
  \citenamefont {Hinrichsen}, \citenamefont {Kwon},\ and\ \citenamefont
  {Park}}]{um:2015}%
  \BibitemOpen
  \bibfield  {author} {\bibinfo {author} {\bibfnamefont {J.}~\bibnamefont
  {Um}}, \bibinfo {author} {\bibfnamefont {H.}~\bibnamefont {Hinrichsen}},
  \bibinfo {author} {\bibfnamefont {C.}~\bibnamefont {Kwon}}, \ and\ \bibinfo
  {author} {\bibfnamefont {H.}~\bibnamefont {Park}},\ }\bibfield  {title}
  {\enquote {\bibinfo {title} {Total cost of operating an information
  engine},}\ }\href {\doibase 10.1088/1367-2630/17/8/085001} {\bibfield
  {journal} {\bibinfo  {journal} {New J. Phys.}\ }\textbf {\bibinfo {volume}
  {17}},\ \bibinfo {pages} {085001} (\bibinfo {year} {2015})}\BibitemShut
  {NoStop}%
\bibitem [{\citenamefont {Strasberg}\ \emph {et~al.}(2015)\citenamefont
  {Strasberg}, \citenamefont {Cerrillo}, \citenamefont {Schaller},\ and\
  \citenamefont {Brandes}}]{strasberg:2015}%
  \BibitemOpen
  \bibfield  {author} {\bibinfo {author} {\bibfnamefont {P.}~\bibnamefont
  {Strasberg}}, \bibinfo {author} {\bibfnamefont {J.}~\bibnamefont {Cerrillo}},
  \bibinfo {author} {\bibfnamefont {G.}~\bibnamefont {Schaller}}, \ and\
  \bibinfo {author} {\bibfnamefont {T.}~\bibnamefont {Brandes}},\ }\bibfield
  {title} {\enquote {\bibinfo {title} {Thermodynamics of stochastic {T}uring
  machines},}\ }\href {\doibase 10.1103/PhysRevE.92.042104} {\bibfield
  {journal} {\bibinfo  {journal} {Phys. Rev. E}\ }\textbf {\bibinfo {volume}
  {92}},\ \bibinfo {pages} {042104} (\bibinfo {year} {2015})}\BibitemShut
  {NoStop}%
\bibitem [{\citenamefont {Shiraishi}\ and\ \citenamefont
  {Sagawa}(2015)}]{shiraishi:2015}%
  \BibitemOpen
  \bibfield  {author} {\bibinfo {author} {\bibfnamefont {N.}~\bibnamefont
  {Shiraishi}}\ and\ \bibinfo {author} {\bibfnamefont {T.}~\bibnamefont
  {Sagawa}},\ }\bibfield  {title} {\enquote {\bibinfo {title} {Fluctuation
  theorem for partially masked nonequilibrium dynamics},}\ }\href {\doibase
  10.1103/PhysRevE.91.012130} {\bibfield  {journal} {\bibinfo  {journal} {Phys.
  Rev. E}\ }\textbf {\bibinfo {volume} {91}},\ \bibinfo {pages} {012130}
  (\bibinfo {year} {2015})}\BibitemShut {NoStop}%
\bibitem [{\citenamefont {Funo}\ \emph {et~al.}(2015)\citenamefont {Funo},
  \citenamefont {Murashita},\ and\ \citenamefont {Ueda}}]{funo:2015}%
  \BibitemOpen
  \bibfield  {author} {\bibinfo {author} {\bibfnamefont {K.}~\bibnamefont
  {Funo}}, \bibinfo {author} {\bibfnamefont {Y.}~\bibnamefont {Murashita}}, \
  and\ \bibinfo {author} {\bibfnamefont {M.}~\bibnamefont {Ueda}},\ }\bibfield
  {title} {\enquote {\bibinfo {title} {Quantum nonequilibrium equalities with
  absolute irreversibility},}\ }\href
  {http://stacks.iop.org/1367-2630/17/i=7/a=075005} {\bibfield  {journal}
  {\bibinfo  {journal} {New J. Phys.}\ }\textbf {\bibinfo {volume} {17}},\
  \bibinfo {pages} {075005} (\bibinfo {year} {2015})}\BibitemShut {NoStop}%
\bibitem [{\citenamefont {Koski}\ and\ \citenamefont
  {Pekola}(2016)}]{koski:2016}%
  \BibitemOpen
  \bibfield  {author} {\bibinfo {author} {\bibfnamefont {J.~V.}\ \bibnamefont
  {Koski}}\ and\ \bibinfo {author} {\bibfnamefont {J.~P.}\ \bibnamefont
  {Pekola}},\ }\bibfield  {title} {\enquote {\bibinfo {title} {Maxwell's demons
  realized in electronic circuits},}\ }\href {\doibase
  10.1016/j.crhy.2016.08.011} {\bibfield  {journal} {\bibinfo  {journal} {C. R.
  Phys.}\ }\textbf {\bibinfo {volume} {17}},\ \bibinfo {pages} {1130} (\bibinfo
  {year} {2016})}\BibitemShut {NoStop}%
\bibitem [{\citenamefont {W\"achtler}\ \emph {et~al.}(2016)\citenamefont
  {W\"achtler}, \citenamefont {Strasberg},\ and\ \citenamefont
  {Brandes}}]{wachtler:2016}%
  \BibitemOpen
  \bibfield  {author} {\bibinfo {author} {\bibfnamefont {C.~W.}\ \bibnamefont
  {W\"achtler}}, \bibinfo {author} {\bibfnamefont {P.}~\bibnamefont
  {Strasberg}}, \ and\ \bibinfo {author} {\bibfnamefont {T.}~\bibnamefont
  {Brandes}},\ }\bibfield  {title} {\enquote {\bibinfo {title} {Stochastic
  thermodynamics based on incomplete information: generalized {J}arzynski
  equality with measurement errors with or without feedback},}\ }\href
  {http://stacks.iop.org/1367-2630/18/i=11/a=113042} {\bibfield  {journal}
  {\bibinfo  {journal} {New J. Phys.}\ }\textbf {\bibinfo {volume} {18}},\
  \bibinfo {pages} {113042} (\bibinfo {year} {2016})}\BibitemShut {NoStop}%
\bibitem [{\citenamefont {Gong}\ \emph {et~al.}(2016)\citenamefont {Gong},
  \citenamefont {Ashida},\ and\ \citenamefont {Ueda}}]{gong:2016}%
  \BibitemOpen
  \bibfield  {author} {\bibinfo {author} {\bibfnamefont {Z.}~\bibnamefont
  {Gong}}, \bibinfo {author} {\bibfnamefont {Y.}~\bibnamefont {Ashida}}, \ and\
  \bibinfo {author} {\bibfnamefont {M.}~\bibnamefont {Ueda}},\ }\bibfield
  {title} {\enquote {\bibinfo {title} {Quantum-trajectory thermodynamics with
  discrete feedback control},}\ }\href {\doibase 10.1103/PhysRevA.94.012107}
  {\bibfield  {journal} {\bibinfo  {journal} {Phys. Rev. A}\ }\textbf {\bibinfo
  {volume} {94}},\ \bibinfo {pages} {012107} (\bibinfo {year}
  {2016})}\BibitemShut {NoStop}%
\bibitem [{\citenamefont {Kwon}\ \emph {et~al.}(2017)\citenamefont {Kwon},
  \citenamefont {Um},\ and\ \citenamefont {Park}}]{kwon:2017}%
  \BibitemOpen
  \bibfield  {author} {\bibinfo {author} {\bibfnamefont {C.}~\bibnamefont
  {Kwon}}, \bibinfo {author} {\bibfnamefont {J.}~\bibnamefont {Um}}, \ and\
  \bibinfo {author} {\bibfnamefont {H.}~\bibnamefont {Park}},\ }\bibfield
  {title} {\enquote {\bibinfo {title} {Information thermodynamics for a
  multi-feedback process with time delay},}\ }\href
  {http://stacks.iop.org/0295-5075/117/i=1/a=10011} {\bibfield  {journal}
  {\bibinfo  {journal} {EPL}\ }\textbf {\bibinfo {volume} {117}},\ \bibinfo
  {pages} {10011} (\bibinfo {year} {2017})}\BibitemShut {NoStop}%
\bibitem [{\citenamefont {Potts}\ and\ \citenamefont
  {Samuelsson}(2018)}]{potts:2018}%
  \BibitemOpen
  \bibfield  {author} {\bibinfo {author} {\bibfnamefont {P.~P.}\ \bibnamefont
  {Potts}}\ and\ \bibinfo {author} {\bibfnamefont {P.}~\bibnamefont
  {Samuelsson}},\ }\bibfield  {title} {\enquote {\bibinfo {title} {Detailed
  fluctuation relation for arbitrary measurement and feedback schemes},}\
  }\href {\doibase 10.1103/PhysRevLett.121.210603} {\bibfield  {journal}
  {\bibinfo  {journal} {Phys. Rev. Lett.}\ }\textbf {\bibinfo {volume} {121}},\
  \bibinfo {pages} {210603} (\bibinfo {year} {2018})}\BibitemShut {NoStop}%
\bibitem [{\citenamefont {Giazotto}\ \emph {et~al.}(2006)\citenamefont
  {Giazotto}, \citenamefont {Heikkil\"a}, \citenamefont {Luukanen},
  \citenamefont {Savin},\ and\ \citenamefont {Pekola}}]{giazotto:2006}%
  \BibitemOpen
  \bibfield  {author} {\bibinfo {author} {\bibfnamefont {F.}~\bibnamefont
  {Giazotto}}, \bibinfo {author} {\bibfnamefont {T.~T.}\ \bibnamefont
  {Heikkil\"a}}, \bibinfo {author} {\bibfnamefont {A.}~\bibnamefont
  {Luukanen}}, \bibinfo {author} {\bibfnamefont {A.~M.}\ \bibnamefont {Savin}},
  \ and\ \bibinfo {author} {\bibfnamefont {J.~P.}\ \bibnamefont {Pekola}},\
  }\bibfield  {title} {\enquote {\bibinfo {title} {Opportunities for
  mesoscopics in thermometry and refrigeration: Physics and applications},}\
  }\href {\doibase 10.1103/RevModPhys.78.217} {\bibfield  {journal} {\bibinfo
  {journal} {Rev. Mod. Phys.}\ }\textbf {\bibinfo {volume} {78}},\ \bibinfo
  {pages} {217} (\bibinfo {year} {2006})}\BibitemShut {NoStop}%
\bibitem [{\citenamefont {Joubaud}\ \emph {et~al.}(2008)\citenamefont
  {Joubaud}, \citenamefont {Garnier},\ and\ \citenamefont
  {Ciliberto}}]{joubaud:2008}%
  \BibitemOpen
  \bibfield  {author} {\bibinfo {author} {\bibfnamefont {S.}~\bibnamefont
  {Joubaud}}, \bibinfo {author} {\bibfnamefont {N.~B.}\ \bibnamefont
  {Garnier}}, \ and\ \bibinfo {author} {\bibfnamefont {S.}~\bibnamefont
  {Ciliberto}},\ }\bibfield  {title} {\enquote {\bibinfo {title} {{Fluctuations
  of the total entropy production in stochastic systems}},}\ }\href {\doibase
  10.1209/0295-5075/82/30007} {\bibfield  {journal} {\bibinfo  {journal} {EPL}\
  }\textbf {\bibinfo {volume} {82}},\ \bibinfo {pages} {30007} (\bibinfo {year}
  {2008})}\BibitemShut {NoStop}%
\bibitem [{\citenamefont {Ciliberto}\ \emph {et~al.}(2013)\citenamefont
  {Ciliberto}, \citenamefont {Imparato}, \citenamefont {Naert},\ and\
  \citenamefont {Tanase}}]{ciliberto:2013}%
  \BibitemOpen
  \bibfield  {author} {\bibinfo {author} {\bibfnamefont {S.}~\bibnamefont
  {Ciliberto}}, \bibinfo {author} {\bibfnamefont {A.}~\bibnamefont {Imparato}},
  \bibinfo {author} {\bibfnamefont {A.}~\bibnamefont {Naert}}, \ and\ \bibinfo
  {author} {\bibfnamefont {M.}~\bibnamefont {Tanase}},\ }\bibfield  {title}
  {\enquote {\bibinfo {title} {Heat flux and entropy produced by thermal
  fluctuations},}\ }\href {\doibase 10.1103/PhysRevLett.110.180601} {\bibfield
  {journal} {\bibinfo  {journal} {Phys. Rev. Lett.}\ }\textbf {\bibinfo
  {volume} {110}},\ \bibinfo {pages} {180601} (\bibinfo {year}
  {2013})}\BibitemShut {NoStop}%
\bibitem [{\citenamefont {Koski}\ \emph
  {et~al.}(2014{\natexlab{a}})\citenamefont {Koski}, \citenamefont {Maisi},
  \citenamefont {Pekola},\ and\ \citenamefont {Averin}}]{koski:2014}%
  \BibitemOpen
  \bibfield  {author} {\bibinfo {author} {\bibfnamefont {J.~V.}\ \bibnamefont
  {Koski}}, \bibinfo {author} {\bibfnamefont {V.~F.}\ \bibnamefont {Maisi}},
  \bibinfo {author} {\bibfnamefont {J.~P.}\ \bibnamefont {Pekola}}, \ and\
  \bibinfo {author} {\bibfnamefont {D.~V.}\ \bibnamefont {Averin}},\ }\bibfield
   {title} {\enquote {\bibinfo {title} {Experimental realization of a {S}zilard
  engine with a single electron},}\ }\href {\doibase 10.1073/pnas.1406966111}
  {\bibfield  {journal} {\bibinfo  {journal} {Proc. Natl. Acad. Sci. USA}\
  }\textbf {\bibinfo {volume} {111}},\ \bibinfo {pages} {13786} (\bibinfo
  {year} {2014}{\natexlab{a}})}\BibitemShut {NoStop}%
\bibitem [{\citenamefont {Koski}\ \emph
  {et~al.}(2014{\natexlab{b}})\citenamefont {Koski}, \citenamefont {Maisi},
  \citenamefont {Sagawa},\ and\ \citenamefont {Pekola}}]{koski:2014prl}%
  \BibitemOpen
  \bibfield  {author} {\bibinfo {author} {\bibfnamefont {J.~V.}\ \bibnamefont
  {Koski}}, \bibinfo {author} {\bibfnamefont {V.~F.}\ \bibnamefont {Maisi}},
  \bibinfo {author} {\bibfnamefont {T.}~\bibnamefont {Sagawa}}, \ and\ \bibinfo
  {author} {\bibfnamefont {J.~P.}\ \bibnamefont {Pekola}},\ }\bibfield  {title}
  {\enquote {\bibinfo {title} {Experimental observation of the role of mutual
  information in the nonequilibrium dynamics of a {M}axwell demon},}\ }\href
  {\doibase 10.1103/PhysRevLett.113.030601} {\bibfield  {journal} {\bibinfo
  {journal} {Phys. Rev. Lett.}\ }\textbf {\bibinfo {volume} {113}},\ \bibinfo
  {pages} {030601} (\bibinfo {year} {2014}{\natexlab{b}})}\BibitemShut
  {NoStop}%
\bibitem [{\citenamefont {Thierschmann}\ \emph
  {et~al.}(2015{\natexlab{a}})\citenamefont {Thierschmann}, \citenamefont
  {S\'anchez}, \citenamefont {Sothmann}, \citenamefont {Arnold}, \citenamefont
  {Heyn}, \citenamefont {Hansen}, \citenamefont {Buhmann},\ and\ \citenamefont
  {Molenkamp}}]{thierschmann:2015}%
  \BibitemOpen
  \bibfield  {author} {\bibinfo {author} {\bibfnamefont {H.}~\bibnamefont
  {Thierschmann}}, \bibinfo {author} {\bibfnamefont {R.}~\bibnamefont
  {S\'anchez}}, \bibinfo {author} {\bibfnamefont {B.}~\bibnamefont {Sothmann}},
  \bibinfo {author} {\bibfnamefont {F.}~\bibnamefont {Arnold}}, \bibinfo
  {author} {\bibfnamefont {C.}~\bibnamefont {Heyn}}, \bibinfo {author}
  {\bibfnamefont {W.}~\bibnamefont {Hansen}}, \bibinfo {author} {\bibfnamefont
  {H.}~\bibnamefont {Buhmann}}, \ and\ \bibinfo {author} {\bibfnamefont
  {L.~W.}\ \bibnamefont {Molenkamp}},\ }\bibfield  {title} {\enquote {\bibinfo
  {title} {Three-terminal energy harvester with coupled quantum dots},}\ }\href
  {http://dx.doi.org/10.1038/nnano.2015.176} {\bibfield  {journal} {\bibinfo
  {journal} {Nat. Nano.}\ }\textbf {\bibinfo {volume} {10}},\ \bibinfo {pages}
  {854} (\bibinfo {year} {2015}{\natexlab{a}})}\BibitemShut {NoStop}%
\bibitem [{\citenamefont {Roche}\ \emph {et~al.}(2015)\citenamefont {Roche},
  \citenamefont {Roulleau}, \citenamefont {Jullien}, \citenamefont {Jompol},
  \citenamefont {Farrer}, \citenamefont {Ritchie},\ and\ \citenamefont
  {Glattli}}]{roche:2015}%
  \BibitemOpen
  \bibfield  {author} {\bibinfo {author} {\bibfnamefont {B.}~\bibnamefont
  {Roche}}, \bibinfo {author} {\bibfnamefont {P.}~\bibnamefont {Roulleau}},
  \bibinfo {author} {\bibfnamefont {T.}~\bibnamefont {Jullien}}, \bibinfo
  {author} {\bibfnamefont {Y.}~\bibnamefont {Jompol}}, \bibinfo {author}
  {\bibfnamefont {I.}~\bibnamefont {Farrer}}, \bibinfo {author} {\bibfnamefont
  {D.}~\bibnamefont {Ritchie}}, \ and\ \bibinfo {author} {\bibfnamefont
  {D.}~\bibnamefont {Glattli}},\ }\bibfield  {title} {\enquote {\bibinfo
  {title} {Harvesting dissipated energy with a mesoscopic ratchet},}\ }\href
  {http://dx.doi.org/10.1038/ncomms7738} {\bibfield  {journal} {\bibinfo
  {journal} {Nat. Commun.}\ }\textbf {\bibinfo {volume} {6}},\ \bibinfo {pages}
  {--} (\bibinfo {year} {2015})}\BibitemShut {NoStop}%
\bibitem [{\citenamefont {Koski}\ \emph {et~al.}(2015)\citenamefont {Koski},
  \citenamefont {Kutvonen}, \citenamefont {Khaymovich}, \citenamefont
  {Ala-Nissila},\ and\ \citenamefont {Pekola}}]{koski:2015}%
  \BibitemOpen
  \bibfield  {author} {\bibinfo {author} {\bibfnamefont {J.~V.}\ \bibnamefont
  {Koski}}, \bibinfo {author} {\bibfnamefont {A.}~\bibnamefont {Kutvonen}},
  \bibinfo {author} {\bibfnamefont {I.~M.}\ \bibnamefont {Khaymovich}},
  \bibinfo {author} {\bibfnamefont {T.}~\bibnamefont {Ala-Nissila}}, \ and\
  \bibinfo {author} {\bibfnamefont {J.~P.}\ \bibnamefont {Pekola}},\ }\bibfield
   {title} {\enquote {\bibinfo {title} {On-chip {M}axwell's demon as an
  information-powered refrigerator},}\ }\href {\doibase
  10.1103/PhysRevLett.115.260602} {\bibfield  {journal} {\bibinfo  {journal}
  {Phys. Rev. Lett.}\ }\textbf {\bibinfo {volume} {115}},\ \bibinfo {pages}
  {260602} (\bibinfo {year} {2015})}\BibitemShut {NoStop}%
\bibitem [{\citenamefont {Thierschmann}\ \emph {et~al.}(2016)\citenamefont
  {Thierschmann}, \citenamefont {S\'anchez}, \citenamefont {Sothmann},
  \citenamefont {Buhmann},\ and\ \citenamefont
  {Molenkamp}}]{thierschmann_thermoelectrics_2016}%
  \BibitemOpen
  \bibfield  {author} {\bibinfo {author} {\bibfnamefont {H.}~\bibnamefont
  {Thierschmann}}, \bibinfo {author} {\bibfnamefont {R.}~\bibnamefont
  {S\'anchez}}, \bibinfo {author} {\bibfnamefont {B.}~\bibnamefont {Sothmann}},
  \bibinfo {author} {\bibfnamefont {H.}~\bibnamefont {Buhmann}}, \ and\
  \bibinfo {author} {\bibfnamefont {L.~W.}\ \bibnamefont {Molenkamp}},\
  }\bibfield  {title} {\enquote {\bibinfo {title} {Thermoelectrics with
  {C}oulomb-coupled quantum dots},}\ }\href {\doibase
  https://doi.org/10.1016/j.crhy.2016.08.001} {\bibfield  {journal} {\bibinfo
  {journal} {C. R. Phys.}\ }\textbf {\bibinfo {volume} {17}},\ \bibinfo {pages}
  {1109} (\bibinfo {year} {2016})}\BibitemShut {NoStop}%
\bibitem [{\citenamefont {Hofmann}\ \emph {et~al.}(2016)\citenamefont
  {Hofmann}, \citenamefont {Maisi}, \citenamefont {R\"ossler}, \citenamefont
  {Basset}, \citenamefont {Kr\"ahenmann}, \citenamefont {M\"arki},
  \citenamefont {Ihn}, \citenamefont {Ensslin}, \citenamefont {Reichl},\ and\
  \citenamefont {Wegscheider}}]{hofmann:2016}%
  \BibitemOpen
  \bibfield  {author} {\bibinfo {author} {\bibfnamefont {A.}~\bibnamefont
  {Hofmann}}, \bibinfo {author} {\bibfnamefont {V.~F.}\ \bibnamefont {Maisi}},
  \bibinfo {author} {\bibfnamefont {C.}~\bibnamefont {R\"ossler}}, \bibinfo
  {author} {\bibfnamefont {J.}~\bibnamefont {Basset}}, \bibinfo {author}
  {\bibfnamefont {T.}~\bibnamefont {Kr\"ahenmann}}, \bibinfo {author}
  {\bibfnamefont {P.}~\bibnamefont {M\"arki}}, \bibinfo {author} {\bibfnamefont
  {T.}~\bibnamefont {Ihn}}, \bibinfo {author} {\bibfnamefont {K.}~\bibnamefont
  {Ensslin}}, \bibinfo {author} {\bibfnamefont {C.}~\bibnamefont {Reichl}}, \
  and\ \bibinfo {author} {\bibfnamefont {W.}~\bibnamefont {Wegscheider}},\
  }\bibfield  {title} {\enquote {\bibinfo {title} {Equilibrium free energy
  measurement of a confined electron driven out of equilibrium},}\ }\href
  {\doibase 10.1103/PhysRevB.93.035425} {\bibfield  {journal} {\bibinfo
  {journal} {Phys. Rev. B}\ }\textbf {\bibinfo {volume} {93}},\ \bibinfo
  {pages} {035425} (\bibinfo {year} {2016})}\BibitemShut {NoStop}%
\bibitem [{\citenamefont {Hofmann}\ \emph {et~al.}(2017)\citenamefont
  {Hofmann}, \citenamefont {Maisi}, \citenamefont {Basset}, \citenamefont
  {Reichl}, \citenamefont {Wegscheider}, \citenamefont {Ihn}, \citenamefont
  {Ensslin},\ and\ \citenamefont {Jarzynski}}]{hofmann:2017}%
  \BibitemOpen
  \bibfield  {author} {\bibinfo {author} {\bibfnamefont {A.}~\bibnamefont
  {Hofmann}}, \bibinfo {author} {\bibfnamefont {V.~F.}\ \bibnamefont {Maisi}},
  \bibinfo {author} {\bibfnamefont {J.}~\bibnamefont {Basset}}, \bibinfo
  {author} {\bibfnamefont {C.}~\bibnamefont {Reichl}}, \bibinfo {author}
  {\bibfnamefont {W.}~\bibnamefont {Wegscheider}}, \bibinfo {author}
  {\bibfnamefont {T.}~\bibnamefont {Ihn}}, \bibinfo {author} {\bibfnamefont
  {K.}~\bibnamefont {Ensslin}}, \ and\ \bibinfo {author} {\bibfnamefont
  {C.}~\bibnamefont {Jarzynski}},\ }\bibfield  {title} {\enquote {\bibinfo
  {title} {Heat dissipation and fluctuations in a driven quantum dot},}\ }\href
  {\doibase 10.1002/pssb.201600546} {\bibfield  {journal} {\bibinfo  {journal}
  {Phys. Status Solidi B}\ }\textbf {\bibinfo {volume} {254}},\ \bibinfo
  {pages} {1600546} (\bibinfo {year} {2017})}\BibitemShut {NoStop}%
\bibitem [{\citenamefont {Chida}\ \emph {et~al.}(2017)\citenamefont {Chida},
  \citenamefont {Desai}, \citenamefont {Nishiguchi},\ and\ \citenamefont
  {Fujiwara}}]{chida:2017}%
  \BibitemOpen
  \bibfield  {author} {\bibinfo {author} {\bibfnamefont {K.}~\bibnamefont
  {Chida}}, \bibinfo {author} {\bibfnamefont {S.}~\bibnamefont {Desai}},
  \bibinfo {author} {\bibfnamefont {K.}~\bibnamefont {Nishiguchi}}, \ and\
  \bibinfo {author} {\bibfnamefont {A.}~\bibnamefont {Fujiwara}},\ }\bibfield
  {title} {\enquote {\bibinfo {title} {Power generator driven by {M}axwell’s
  demon},}\ }\href {http://dx.doi.org/10.1038/ncomms15301} {\bibfield
  {journal} {\bibinfo  {journal} {Nat. Commun.}\ }\textbf {\bibinfo {volume}
  {8}},\ \bibinfo {pages} {15310} (\bibinfo {year} {2017})}\BibitemShut
  {NoStop}%
\bibitem [{\citenamefont {Pekola}(2015)}]{pekola:2015}%
  \BibitemOpen
  \bibfield  {author} {\bibinfo {author} {\bibfnamefont {J.~P.}\ \bibnamefont
  {Pekola}},\ }\bibfield  {title} {\enquote {\bibinfo {title} {Towards quantum
  thermodynamics in electronic circuits},}\ }\href
  {http://dx.doi.org/10.1038/nphys3169} {\bibfield  {journal} {\bibinfo
  {journal} {Nat. Phys.}\ }\textbf {\bibinfo {volume} {11}},\ \bibinfo {pages}
  {118} (\bibinfo {year} {2015})}\BibitemShut {NoStop}%
\bibitem [{\citenamefont {Pekola}\ and\ \citenamefont
  {Khaymovich}(2019)}]{pekola:2019}%
  \BibitemOpen
  \bibfield  {author} {\bibinfo {author} {\bibfnamefont {J.}~\bibnamefont
  {Pekola}}\ and\ \bibinfo {author} {\bibfnamefont {I.}~\bibnamefont
  {Khaymovich}},\ }\bibfield  {title} {\enquote {\bibinfo {title}
  {Thermodynamics in single-electron circuits and superconducting qubits},}\
  }\href {\doibase 10.1146/annurev-conmatphys-033117-054120} {\bibfield
  {journal} {\bibinfo  {journal} {Annu. Rev. Condens. Matter Phys.}\ }\textbf
  {\bibinfo {volume} {10}},\ \bibinfo {pages} {193} (\bibinfo {year}
  {2019})}\BibitemShut {NoStop}%
\bibitem [{\citenamefont {Hartman}\ \emph {et~al.}(2018)\citenamefont
  {Hartman}, \citenamefont {Olsen}, \citenamefont
  {L{\ifmmode\ddot{u}\else\"{u}\fi}scher}, \citenamefont {Samani},
  \citenamefont {Fallahi}, \citenamefont {Gardner}, \citenamefont {Manfra},\
  and\ \citenamefont {Folk}}]{hartman_direct_2018}%
  \BibitemOpen
  \bibfield  {author} {\bibinfo {author} {\bibfnamefont {N.}~\bibnamefont
  {Hartman}}, \bibinfo {author} {\bibfnamefont {C.}~\bibnamefont {Olsen}},
  \bibinfo {author} {\bibfnamefont {S.}~\bibnamefont
  {L{\ifmmode\ddot{u}\else\"{u}\fi}scher}}, \bibinfo {author} {\bibfnamefont
  {M.}~\bibnamefont {Samani}}, \bibinfo {author} {\bibfnamefont
  {S.}~\bibnamefont {Fallahi}}, \bibinfo {author} {\bibfnamefont {G.~C.}\
  \bibnamefont {Gardner}}, \bibinfo {author} {\bibfnamefont {M.}~\bibnamefont
  {Manfra}}, \ and\ \bibinfo {author} {\bibfnamefont {J.}~\bibnamefont
  {Folk}},\ }\bibfield  {title} {\enquote {\bibinfo {title} {{Direct entropy
  measurement in a mesoscopic quantum system}},}\ }\href {\doibase
  10.1038/s41567-018-0250-5} {\bibfield  {journal} {\bibinfo  {journal} {Nat.
  Phys.}\ }\textbf {\bibinfo {volume} {14}},\ \bibinfo {pages} {1083} (\bibinfo
  {year} {2018})}\BibitemShut {NoStop}%
\bibitem [{\citenamefont {Singh}\ \emph {et~al.}(2019)\citenamefont {Singh},
  \citenamefont {Rold\'an}, \citenamefont {Neri}, \citenamefont {Khaymovich},
  \citenamefont {Golubev}, \citenamefont {Maisi}, \citenamefont {Peltonen},
  \citenamefont {J\"ulicher},\ and\ \citenamefont
  {Pekola}}]{singh_extreme_2019}%
  \BibitemOpen
  \bibfield  {author} {\bibinfo {author} {\bibfnamefont {S.}~\bibnamefont
  {Singh}}, \bibinfo {author} {\bibfnamefont {E.}~\bibnamefont {Rold\'an}},
  \bibinfo {author} {\bibfnamefont {I.}~\bibnamefont {Neri}}, \bibinfo {author}
  {\bibfnamefont {I.~M.}\ \bibnamefont {Khaymovich}}, \bibinfo {author}
  {\bibfnamefont {D.~S.}\ \bibnamefont {Golubev}}, \bibinfo {author}
  {\bibfnamefont {V.~F.}\ \bibnamefont {Maisi}}, \bibinfo {author}
  {\bibfnamefont {J.~T.}\ \bibnamefont {Peltonen}}, \bibinfo {author}
  {\bibfnamefont {F.}~\bibnamefont {J\"ulicher}}, \ and\ \bibinfo {author}
  {\bibfnamefont {J.~P.}\ \bibnamefont {Pekola}},\ }\bibfield  {title}
  {\enquote {\bibinfo {title} {Extreme reductions of entropy in an electronic
  double dot},}\ }\href {\doibase 10.1103/PhysRevB.99.115422} {\bibfield
  {journal} {\bibinfo  {journal} {Phys. Rev. B}\ }\textbf {\bibinfo {volume}
  {99}},\ \bibinfo {pages} {115422} (\bibinfo {year} {2019})}\BibitemShut
  {NoStop}%
\bibitem [{\citenamefont {Kleeorin}\ \emph {et~al.}()\citenamefont {Kleeorin},
  \citenamefont {Thierschmann}, \citenamefont {Buhmann}, \citenamefont
  {Georges}, \citenamefont {Molenkamp},\ and\ \citenamefont
  {Meir}}]{kleeorin_measuring_2019}%
  \BibitemOpen
  \bibfield  {author} {\bibinfo {author} {\bibfnamefont {Y.}~\bibnamefont
  {Kleeorin}}, \bibinfo {author} {\bibfnamefont {H.}~\bibnamefont
  {Thierschmann}}, \bibinfo {author} {\bibfnamefont {H.}~\bibnamefont
  {Buhmann}}, \bibinfo {author} {\bibfnamefont {A.}~\bibnamefont {Georges}},
  \bibinfo {author} {\bibfnamefont {L.~W.}\ \bibnamefont {Molenkamp}}, \ and\
  \bibinfo {author} {\bibfnamefont {Y.}~\bibnamefont {Meir}},\ }\href@noop {}
  {\enquote {\bibinfo {title} {{Measuring the Entropy of a Mesoscopic System
  via Thermoelectric Transport}},}\ }\Eprint {http://arxiv.org/abs/1904.08948}
  {arXiv:1904.08948 [cond-mat]} \BibitemShut {NoStop}%
\bibitem [{\citenamefont {Gonzalez}\ \emph {et~al.}(2019)\citenamefont
  {Gonzalez}, \citenamefont {Neu},\ and\ \citenamefont
  {Teitsworth}}]{gonzalez:2019}%
  \BibitemOpen
  \bibfield  {author} {\bibinfo {author} {\bibfnamefont {J.~P.}\ \bibnamefont
  {Gonzalez}}, \bibinfo {author} {\bibfnamefont {J.~C.}\ \bibnamefont {Neu}}, \
  and\ \bibinfo {author} {\bibfnamefont {S.~W.}\ \bibnamefont {Teitsworth}},\
  }\bibfield  {title} {\enquote {\bibinfo {title} {Experimental metrics for
  detection of detailed balance violation},}\ }\href {\doibase
  10.1103/PhysRevE.99.022143} {\bibfield  {journal} {\bibinfo  {journal} {Phys.
  Rev. E}\ }\textbf {\bibinfo {volume} {99}},\ \bibinfo {pages} {022143}
  (\bibinfo {year} {2019})}\BibitemShut {NoStop}%
\bibitem [{\citenamefont {Alemany}\ \emph {et~al.}(2012)\citenamefont
  {Alemany}, \citenamefont {Mossa}, \citenamefont {Junier},\ and\ \citenamefont
  {Ritort}}]{alemany:2012}%
  \BibitemOpen
  \bibfield  {author} {\bibinfo {author} {\bibfnamefont {A.}~\bibnamefont
  {Alemany}}, \bibinfo {author} {\bibfnamefont {A.}~\bibnamefont {Mossa}},
  \bibinfo {author} {\bibfnamefont {I.}~\bibnamefont {Junier}}, \ and\ \bibinfo
  {author} {\bibfnamefont {F.}~\bibnamefont {Ritort}},\ }\bibfield  {title}
  {\enquote {\bibinfo {title} {Experimental free-energy measurements of kinetic
  molecular states using fluctuation theorems},}\ }\href
  {http://dx.doi.org/10.1038/nphys2375} {\bibfield  {journal} {\bibinfo
  {journal} {Nat. Phys.}\ }\textbf {\bibinfo {volume} {8}},\ \bibinfo {pages}
  {688} (\bibinfo {year} {2012})}\BibitemShut {NoStop}%
\bibitem [{\citenamefont {Dieterich}\ \emph {et~al.}(2016)\citenamefont
  {Dieterich}, \citenamefont {Camunas-Soler}, \citenamefont
  {Ribezzi-Crivellari}, \citenamefont {Seifert},\ and\ \citenamefont
  {Ritort}}]{dieterich:2016}%
  \BibitemOpen
  \bibfield  {author} {\bibinfo {author} {\bibfnamefont {E.}~\bibnamefont
  {Dieterich}}, \bibinfo {author} {\bibfnamefont {J.}~\bibnamefont
  {Camunas-Soler}}, \bibinfo {author} {\bibfnamefont {M.}~\bibnamefont
  {Ribezzi-Crivellari}}, \bibinfo {author} {\bibfnamefont {U.}~\bibnamefont
  {Seifert}}, \ and\ \bibinfo {author} {\bibfnamefont {F.}~\bibnamefont
  {Ritort}},\ }\bibfield  {title} {\enquote {\bibinfo {title} {Control of force
  through feedback in small driven systems},}\ }\href {\doibase
  10.1103/PhysRevE.94.012107} {\bibfield  {journal} {\bibinfo  {journal} {Phys.
  Rev. E}\ }\textbf {\bibinfo {volume} {94}},\ \bibinfo {pages} {012107}
  (\bibinfo {year} {2016})}\BibitemShut {NoStop}%
\bibitem [{\citenamefont {Vidrighin}\ \emph {et~al.}(2016)\citenamefont
  {Vidrighin}, \citenamefont {Dahlsten}, \citenamefont {Barbieri},
  \citenamefont {Kim}, \citenamefont {Vedral},\ and\ \citenamefont
  {Walmsley}}]{vidrighin:2016}%
  \BibitemOpen
  \bibfield  {author} {\bibinfo {author} {\bibfnamefont {M.~D.}\ \bibnamefont
  {Vidrighin}}, \bibinfo {author} {\bibfnamefont {O.}~\bibnamefont {Dahlsten}},
  \bibinfo {author} {\bibfnamefont {M.}~\bibnamefont {Barbieri}}, \bibinfo
  {author} {\bibfnamefont {M.~S.}\ \bibnamefont {Kim}}, \bibinfo {author}
  {\bibfnamefont {V.}~\bibnamefont {Vedral}}, \ and\ \bibinfo {author}
  {\bibfnamefont {I.~A.}\ \bibnamefont {Walmsley}},\ }\bibfield  {title}
  {\enquote {\bibinfo {title} {Photonic {M}axwell's demon},}\ }\href {\doibase
  10.1103/PhysRevLett.116.050401} {\bibfield  {journal} {\bibinfo  {journal}
  {Phys. Rev. Lett.}\ }\textbf {\bibinfo {volume} {116}},\ \bibinfo {pages}
  {050401} (\bibinfo {year} {2016})}\BibitemShut {NoStop}%
\bibitem [{\citenamefont {Toyabe}\ \emph {et~al.}(2010)\citenamefont {Toyabe},
  \citenamefont {Sagawa}, \citenamefont {Ueda}, \citenamefont {Muneyuki},\ and\
  \citenamefont {Sano}}]{toyabe:2010}%
  \BibitemOpen
  \bibfield  {author} {\bibinfo {author} {\bibfnamefont {S.}~\bibnamefont
  {Toyabe}}, \bibinfo {author} {\bibfnamefont {T.}~\bibnamefont {Sagawa}},
  \bibinfo {author} {\bibfnamefont {M.}~\bibnamefont {Ueda}}, \bibinfo {author}
  {\bibfnamefont {E.}~\bibnamefont {Muneyuki}}, \ and\ \bibinfo {author}
  {\bibfnamefont {M.}~\bibnamefont {Sano}},\ }\bibfield  {title} {\enquote
  {\bibinfo {title} {Experimental demonstration of information-to-energy
  conversion and validation of the generalized {J}arzynski equality},}\ }\href
  {http://dx.doi.org/10.1038/nphys1821} {\bibfield  {journal} {\bibinfo
  {journal} {Nat. Phys.}\ }\textbf {\bibinfo {volume} {6}},\ \bibinfo {pages}
  {988} (\bibinfo {year} {2010})}\BibitemShut {NoStop}%
\bibitem [{\citenamefont {B\'erut}\ \emph {et~al.}(2012)\citenamefont
  {B\'erut}, \citenamefont {Arakelyan}, \citenamefont {Petrosyan},
  \citenamefont {Ciliberto}, \citenamefont {Dillenschneider},\ and\
  \citenamefont {Lutz}}]{berut:2012}%
  \BibitemOpen
  \bibfield  {author} {\bibinfo {author} {\bibfnamefont {A.}~\bibnamefont
  {B\'erut}}, \bibinfo {author} {\bibfnamefont {A.}~\bibnamefont {Arakelyan}},
  \bibinfo {author} {\bibfnamefont {A.}~\bibnamefont {Petrosyan}}, \bibinfo
  {author} {\bibfnamefont {S.}~\bibnamefont {Ciliberto}}, \bibinfo {author}
  {\bibfnamefont {R.}~\bibnamefont {Dillenschneider}}, \ and\ \bibinfo {author}
  {\bibfnamefont {E.}~\bibnamefont {Lutz}},\ }\bibfield  {title} {\enquote
  {\bibinfo {title} {Experimental verification of {L}andauer’s principle
  linking information and thermodynamics},}\ }\href {\doibase
  10.1038/nature10872} {\bibfield  {journal} {\bibinfo  {journal} {Nature}\
  }\textbf {\bibinfo {volume} {483}},\ \bibinfo {pages} {187} (\bibinfo {year}
  {2012})}\BibitemShut {NoStop}%
\bibitem [{\citenamefont {Kumar}\ \emph {et~al.}(2018)\citenamefont {Kumar},
  \citenamefont {Wu}, \citenamefont {Giraldo},\ and\ \citenamefont
  {Weiss}}]{kumar:2018}%
  \BibitemOpen
  \bibfield  {author} {\bibinfo {author} {\bibfnamefont {A.}~\bibnamefont
  {Kumar}}, \bibinfo {author} {\bibfnamefont {T.-Y.}\ \bibnamefont {Wu}},
  \bibinfo {author} {\bibfnamefont {F.}~\bibnamefont {Giraldo}}, \ and\
  \bibinfo {author} {\bibfnamefont {D.~S.}\ \bibnamefont {Weiss}},\ }\bibfield
  {title} {\enquote {\bibinfo {title} {Sorting ultracold atoms in a
  three-dimensional optical lattice in a realization of {M}axwell{'}s demon},}\
  }\href {\doibase 10.1038/s41586-018-0458-7} {\bibfield  {journal} {\bibinfo
  {journal} {Nature}\ }\textbf {\bibinfo {volume} {561}},\ \bibinfo {pages}
  {83} (\bibinfo {year} {2018})}\BibitemShut {NoStop}%
\bibitem [{\citenamefont {Binder}\ \emph {et~al.}(2019)\citenamefont {Binder},
  \citenamefont {Correa}, \citenamefont {Gogolin}, \citenamefont {Anders},\
  and\ \citenamefont {Adesso}}]{thermo:book}%
  \BibitemOpen
  \bibinfo {editor} {\bibfnamefont {F.}~\bibnamefont {Binder}}, \bibinfo
  {editor} {\bibfnamefont {L.~A.}\ \bibnamefont {Correa}}, \bibinfo {editor}
  {\bibfnamefont {C.}~\bibnamefont {Gogolin}}, \bibinfo {editor} {\bibfnamefont
  {J.}~\bibnamefont {Anders}}, \ and\ \bibinfo {editor} {\bibfnamefont
  {G.}~\bibnamefont {Adesso}},\ eds.,\ \href {\doibase
  10.1007/978-3-319-99046-0} {\emph {\bibinfo {title} {Thermodynamics in the
  Quantum Regime}}}\ (\bibinfo  {publisher} {Springer},\ \bibinfo {year}
  {2019})\BibitemShut {NoStop}%
\bibitem [{\citenamefont {Kosloff}(2013)}]{kosloff:2013}%
  \BibitemOpen
  \bibfield  {author} {\bibinfo {author} {\bibfnamefont {R.}~\bibnamefont
  {Kosloff}},\ }\bibfield  {title} {\enquote {\bibinfo {title} {Quantum
  thermodynamics: A dynamical viewpoint},}\ }\href {\doibase 10.3390/e15062100}
  {\bibfield  {journal} {\bibinfo  {journal} {Entropy}\ }\textbf {\bibinfo
  {volume} {15}},\ \bibinfo {pages} {2100} (\bibinfo {year}
  {2013})}\BibitemShut {NoStop}%
\bibitem [{\citenamefont {Vinjanampathy}\ and\ \citenamefont
  {Anders}(2016)}]{vinjanampathy:2016}%
  \BibitemOpen
  \bibfield  {author} {\bibinfo {author} {\bibfnamefont {S.}~\bibnamefont
  {Vinjanampathy}}\ and\ \bibinfo {author} {\bibfnamefont {J.}~\bibnamefont
  {Anders}},\ }\bibfield  {title} {\enquote {\bibinfo {title} {Quantum
  thermodynamics},}\ }\href {\doibase 10.1080/00107514.2016.1201896} {\bibfield
   {journal} {\bibinfo  {journal} {Contemp. Phys.}\ }\textbf {\bibinfo {volume}
  {57}},\ \bibinfo {pages} {545} (\bibinfo {year} {2016})}\BibitemShut
  {NoStop}%
\bibitem [{\citenamefont {Campisi}\ \emph {et~al.}(2011)\citenamefont
  {Campisi}, \citenamefont {H\"anggi},\ and\ \citenamefont
  {Talkner}}]{campisi:2011}%
  \BibitemOpen
  \bibfield  {author} {\bibinfo {author} {\bibfnamefont {M.}~\bibnamefont
  {Campisi}}, \bibinfo {author} {\bibfnamefont {P.}~\bibnamefont {H\"anggi}}, \
  and\ \bibinfo {author} {\bibfnamefont {P.}~\bibnamefont {Talkner}},\
  }\bibfield  {title} {\enquote {\bibinfo {title} {Colloquium: Quantum
  fluctuation relations: Foundations and applications},}\ }\href {\doibase
  10.1103/RevModPhys.83.771} {\bibfield  {journal} {\bibinfo  {journal} {Rev.
  Mod. Phys.}\ }\textbf {\bibinfo {volume} {83}},\ \bibinfo {pages} {771}
  (\bibinfo {year} {2011})}\BibitemShut {NoStop}%
\bibitem [{\citenamefont {Maslennikov}\ \emph {et~al.}(2019)\citenamefont
  {Maslennikov}, \citenamefont {Ding}, \citenamefont {Hablutzel}, \citenamefont
  {Gan}, \citenamefont {Roulet}, \citenamefont {Nimmrichter}, \citenamefont
  {Dai}, \citenamefont {Scarani},\ and\ \citenamefont
  {Matsukevich}}]{maslennikov:2019}%
  \BibitemOpen
  \bibfield  {author} {\bibinfo {author} {\bibfnamefont {G.}~\bibnamefont
  {Maslennikov}}, \bibinfo {author} {\bibfnamefont {S.}~\bibnamefont {Ding}},
  \bibinfo {author} {\bibfnamefont {R.}~\bibnamefont {Hablutzel}}, \bibinfo
  {author} {\bibfnamefont {J.}~\bibnamefont {Gan}}, \bibinfo {author}
  {\bibfnamefont {A.}~\bibnamefont {Roulet}}, \bibinfo {author} {\bibfnamefont
  {S.}~\bibnamefont {Nimmrichter}}, \bibinfo {author} {\bibfnamefont
  {J.}~\bibnamefont {Dai}}, \bibinfo {author} {\bibfnamefont {V.}~\bibnamefont
  {Scarani}}, \ and\ \bibinfo {author} {\bibfnamefont {D.}~\bibnamefont
  {Matsukevich}},\ }\bibfield  {title} {\enquote {\bibinfo {title} {Quantum
  absorption refrigerator with trapped ions},}\ }\href {\doibase
  10.1038/s41467-018-08090-0} {\bibfield  {journal} {\bibinfo  {journal} {Nat.
  Commun.}\ }\textbf {\bibinfo {volume} {10}},\ \bibinfo {pages} {202}
  (\bibinfo {year} {2019})}\BibitemShut {NoStop}%
\bibitem [{\citenamefont {Cottet}\ \emph {et~al.}(2017)\citenamefont {Cottet},
  \citenamefont {Jezouin}, \citenamefont {Bretheau}, \citenamefont
  {Campagne-Ibarcq}, \citenamefont {Ficheux}, \citenamefont {Anders},
  \citenamefont {Auff{\`e}ves}, \citenamefont {Azouit}, \citenamefont
  {Rouchon},\ and\ \citenamefont {Huard}}]{cottet:2017}%
  \BibitemOpen
  \bibfield  {author} {\bibinfo {author} {\bibfnamefont {N.}~\bibnamefont
  {Cottet}}, \bibinfo {author} {\bibfnamefont {S.}~\bibnamefont {Jezouin}},
  \bibinfo {author} {\bibfnamefont {L.}~\bibnamefont {Bretheau}}, \bibinfo
  {author} {\bibfnamefont {P.}~\bibnamefont {Campagne-Ibarcq}}, \bibinfo
  {author} {\bibfnamefont {Q.}~\bibnamefont {Ficheux}}, \bibinfo {author}
  {\bibfnamefont {J.}~\bibnamefont {Anders}}, \bibinfo {author} {\bibfnamefont
  {A.}~\bibnamefont {Auff{\`e}ves}}, \bibinfo {author} {\bibfnamefont
  {R.}~\bibnamefont {Azouit}}, \bibinfo {author} {\bibfnamefont
  {P.}~\bibnamefont {Rouchon}}, \ and\ \bibinfo {author} {\bibfnamefont
  {B.}~\bibnamefont {Huard}},\ }\bibfield  {title} {\enquote {\bibinfo {title}
  {Observing a quantum {M}axwell demon at work},}\ }\href {\doibase
  10.1073/pnas.1704827114} {\bibfield  {journal} {\bibinfo  {journal} {Proc.
  Natl. Acad. Sci. USA}\ }\textbf {\bibinfo {volume} {114}},\ \bibinfo {pages}
  {7561} (\bibinfo {year} {2017})}\BibitemShut {NoStop}%
\bibitem [{\citenamefont {Masuyama}\ \emph {et~al.}(2018)\citenamefont
  {Masuyama}, \citenamefont {Funo}, \citenamefont {Murashita}, \citenamefont
  {Noguchi}, \citenamefont {Kono}, \citenamefont {Tabuchi}, \citenamefont
  {Yamazaki}, \citenamefont {Ueda},\ and\ \citenamefont
  {Nakamura}}]{masuyama:2018}%
  \BibitemOpen
  \bibfield  {author} {\bibinfo {author} {\bibfnamefont {Y.}~\bibnamefont
  {Masuyama}}, \bibinfo {author} {\bibfnamefont {K.}~\bibnamefont {Funo}},
  \bibinfo {author} {\bibfnamefont {Y.}~\bibnamefont {Murashita}}, \bibinfo
  {author} {\bibfnamefont {A.}~\bibnamefont {Noguchi}}, \bibinfo {author}
  {\bibfnamefont {S.}~\bibnamefont {Kono}}, \bibinfo {author} {\bibfnamefont
  {Y.}~\bibnamefont {Tabuchi}}, \bibinfo {author} {\bibfnamefont
  {R.}~\bibnamefont {Yamazaki}}, \bibinfo {author} {\bibfnamefont
  {M.}~\bibnamefont {Ueda}}, \ and\ \bibinfo {author} {\bibfnamefont
  {Y.}~\bibnamefont {Nakamura}},\ }\bibfield  {title} {\enquote {\bibinfo
  {title} {Information-to-work conversion by {M}axwell’s demon in a
  superconducting circuit quantum electrodynamical system},}\ }\href {\doibase
  10.1038/s41467-018-03686-y} {\bibfield  {journal} {\bibinfo  {journal} {Nat.
  Commun.}\ }\textbf {\bibinfo {volume} {9}},\ \bibinfo {pages} {1291}
  (\bibinfo {year} {2018})}\BibitemShut {NoStop}%
\bibitem [{\citenamefont {Naghiloo}\ \emph {et~al.}(2018)\citenamefont
  {Naghiloo}, \citenamefont {Alonso}, \citenamefont {Romito}, \citenamefont
  {Lutz},\ and\ \citenamefont {Murch}}]{naghiloo:2018}%
  \BibitemOpen
  \bibfield  {author} {\bibinfo {author} {\bibfnamefont {M.}~\bibnamefont
  {Naghiloo}}, \bibinfo {author} {\bibfnamefont {J.~J.}\ \bibnamefont
  {Alonso}}, \bibinfo {author} {\bibfnamefont {A.}~\bibnamefont {Romito}},
  \bibinfo {author} {\bibfnamefont {E.}~\bibnamefont {Lutz}}, \ and\ \bibinfo
  {author} {\bibfnamefont {K.~W.}\ \bibnamefont {Murch}},\ }\bibfield  {title}
  {\enquote {\bibinfo {title} {Information gain and loss for a quantum
  {M}axwell's demon},}\ }\href {\doibase 10.1103/PhysRevLett.121.030604}
  {\bibfield  {journal} {\bibinfo  {journal} {Phys. Rev. Lett.}\ }\textbf
  {\bibinfo {volume} {121}},\ \bibinfo {pages} {030604} (\bibinfo {year}
  {2018})}\BibitemShut {NoStop}%
\bibitem [{\citenamefont {Szilard}(1929)}]{szilard:1929}%
  \BibitemOpen
  \bibfield  {author} {\bibinfo {author} {\bibfnamefont {L.}~\bibnamefont
  {Szilard}},\ }\bibfield  {title} {\enquote {\bibinfo {title} {{\"U}ber die
  {E}ntropieverminderung in einem thermodynamischen {S}ystem bei {E}ingriffen
  intelligenter {W}esen},}\ }\href {\doibase 10.1007/BF01341281} {\bibfield
  {journal} {\bibinfo  {journal} {Z. Phys.}\ }\textbf {\bibinfo {volume}
  {53}},\ \bibinfo {pages} {840} (\bibinfo {year} {1929})}\BibitemShut
  {NoStop}%
\bibitem [{\citenamefont {Schaller}\ \emph {et~al.}(2011)\citenamefont
  {Schaller}, \citenamefont {Emary}, \citenamefont {Kiesslich},\ and\
  \citenamefont {Brandes}}]{schaller:2011}%
  \BibitemOpen
  \bibfield  {author} {\bibinfo {author} {\bibfnamefont {G.}~\bibnamefont
  {Schaller}}, \bibinfo {author} {\bibfnamefont {C.}~\bibnamefont {Emary}},
  \bibinfo {author} {\bibfnamefont {G.}~\bibnamefont {Kiesslich}}, \ and\
  \bibinfo {author} {\bibfnamefont {T.}~\bibnamefont {Brandes}},\ }\bibfield
  {title} {\enquote {\bibinfo {title} {Probing the power of an electronic
  {M}axwell's demon: {S}ingle-electron transistor monitored by a quantum point
  contact},}\ }\href {\doibase 10.1103/PhysRevB.84.085418} {\bibfield
  {journal} {\bibinfo  {journal} {Phys. Rev. B}\ }\textbf {\bibinfo {volume}
  {84}},\ \bibinfo {pages} {085418} (\bibinfo {year} {2011})}\BibitemShut
  {NoStop}%
\bibitem [{\citenamefont {Esposito}\ and\ \citenamefont
  {Schaller}(2012)}]{esposito:2012}%
  \BibitemOpen
  \bibfield  {author} {\bibinfo {author} {\bibfnamefont {M.}~\bibnamefont
  {Esposito}}\ and\ \bibinfo {author} {\bibfnamefont {G.}~\bibnamefont
  {Schaller}},\ }\bibfield  {title} {\enquote {\bibinfo {title} {Stochastic
  thermodynamics for ``{M}axwell demon'' feedbacks},}\ }\href {\doibase
  10.1209/0295-5075/99/30003} {\bibfield  {journal} {\bibinfo  {journal} {EPL}\
  }\textbf {\bibinfo {volume} {99}},\ \bibinfo {pages} {30003} (\bibinfo {year}
  {2012})}\BibitemShut {NoStop}%
\bibitem [{\citenamefont {Bergli}\ \emph {et~al.}(2013)\citenamefont {Bergli},
  \citenamefont {Galperin},\ and\ \citenamefont {Kopnin}}]{bergli:2013}%
  \BibitemOpen
  \bibfield  {author} {\bibinfo {author} {\bibfnamefont {J.}~\bibnamefont
  {Bergli}}, \bibinfo {author} {\bibfnamefont {Y.~M.}\ \bibnamefont
  {Galperin}}, \ and\ \bibinfo {author} {\bibfnamefont {N.~B.}\ \bibnamefont
  {Kopnin}},\ }\bibfield  {title} {\enquote {\bibinfo {title} {Information flow
  and optimal protocol for a {M}axwell-demon single-electron pump},}\ }\href
  {\doibase 10.1103/PhysRevE.88.062139} {\bibfield  {journal} {\bibinfo
  {journal} {Phys. Rev. E}\ }\textbf {\bibinfo {volume} {88}},\ \bibinfo
  {pages} {062139} (\bibinfo {year} {2013})}\BibitemShut {NoStop}%
\bibitem [{\citenamefont {Sandberg}\ \emph {et~al.}(2014)\citenamefont
  {Sandberg}, \citenamefont {Delvenne}, \citenamefont {Newton},\ and\
  \citenamefont {Mitter}}]{sandberg:2014}%
  \BibitemOpen
  \bibfield  {author} {\bibinfo {author} {\bibfnamefont {H.}~\bibnamefont
  {Sandberg}}, \bibinfo {author} {\bibfnamefont {J.-C.}\ \bibnamefont
  {Delvenne}}, \bibinfo {author} {\bibfnamefont {N.~J.}\ \bibnamefont
  {Newton}}, \ and\ \bibinfo {author} {\bibfnamefont {S.~K.}\ \bibnamefont
  {Mitter}},\ }\bibfield  {title} {\enquote {\bibinfo {title} {Maximum work
  extraction and implementation costs for nonequilibrium {M}axwell's demons},}\
  }\href {\doibase 10.1103/PhysRevE.90.042119} {\bibfield  {journal} {\bibinfo
  {journal} {Phys. Rev. E}\ }\textbf {\bibinfo {volume} {90}},\ \bibinfo
  {pages} {042119} (\bibinfo {year} {2014})}\BibitemShut {NoStop}%
\bibitem [{\citenamefont {Kutvonen}\ \emph
  {et~al.}(2016{\natexlab{a}})\citenamefont {Kutvonen}, \citenamefont
  {Sagawa},\ and\ \citenamefont {Ala-Nissila}}]{kutvonen:2016pre}%
  \BibitemOpen
  \bibfield  {author} {\bibinfo {author} {\bibfnamefont {A.}~\bibnamefont
  {Kutvonen}}, \bibinfo {author} {\bibfnamefont {T.}~\bibnamefont {Sagawa}}, \
  and\ \bibinfo {author} {\bibfnamefont {T.}~\bibnamefont {Ala-Nissila}},\
  }\bibfield  {title} {\enquote {\bibinfo {title} {Thermodynamics of
  information exchange between two coupled quantum dots},}\ }\href {\doibase
  10.1103/PhysRevE.93.032147} {\bibfield  {journal} {\bibinfo  {journal} {Phys.
  Rev. E}\ }\textbf {\bibinfo {volume} {93}},\ \bibinfo {pages} {032147}
  (\bibinfo {year} {2016}{\natexlab{a}})}\BibitemShut {NoStop}%
\bibitem [{\citenamefont {Schaller}\ \emph {et~al.}(2018)\citenamefont
  {Schaller}, \citenamefont {Cerrillo}, \citenamefont {Engelhardt},\ and\
  \citenamefont {Strasberg}}]{schaller:2018}%
  \BibitemOpen
  \bibfield  {author} {\bibinfo {author} {\bibfnamefont {G.}~\bibnamefont
  {Schaller}}, \bibinfo {author} {\bibfnamefont {J.}~\bibnamefont {Cerrillo}},
  \bibinfo {author} {\bibfnamefont {G.}~\bibnamefont {Engelhardt}}, \ and\
  \bibinfo {author} {\bibfnamefont {P.}~\bibnamefont {Strasberg}},\ }\bibfield
  {title} {\enquote {\bibinfo {title} {Electronic {M}axwell demon in the
  coherent strong-coupling regime},}\ }\href {\doibase
  10.1103/PhysRevB.97.195104} {\bibfield  {journal} {\bibinfo  {journal} {Phys.
  Rev. B}\ }\textbf {\bibinfo {volume} {97}},\ \bibinfo {pages} {195104}
  (\bibinfo {year} {2018})}\BibitemShut {NoStop}%
\bibitem [{\citenamefont {Engelhardt}\ and\ \citenamefont
  {Schaller}(2018)}]{engelhardt:2018}%
  \BibitemOpen
  \bibfield  {author} {\bibinfo {author} {\bibfnamefont {G.}~\bibnamefont
  {Engelhardt}}\ and\ \bibinfo {author} {\bibfnamefont {G.}~\bibnamefont
  {Schaller}},\ }\bibfield  {title} {\enquote {\bibinfo {title} {Maxwell{'}s
  demon in the quantum-{Z}eno regime and beyond},}\ }\href {\doibase
  10.1088/1367-2630/aaa38d} {\bibfield  {journal} {\bibinfo  {journal} {New J.
  Phys.}\ }\textbf {\bibinfo {volume} {20}},\ \bibinfo {pages} {023011}
  (\bibinfo {year} {2018})}\BibitemShut {NoStop}%
\bibitem [{\citenamefont {Mandal}\ and\ \citenamefont
  {Jarzynski}(2012)}]{mandal:2012}%
  \BibitemOpen
  \bibfield  {author} {\bibinfo {author} {\bibfnamefont {D.}~\bibnamefont
  {Mandal}}\ and\ \bibinfo {author} {\bibfnamefont {C.}~\bibnamefont
  {Jarzynski}},\ }\bibfield  {title} {\enquote {\bibinfo {title} {Work and
  information processing in a solvable model of {M}axwell's demon},}\ }\href
  {\doibase 10.1073/pnas.1204263109} {\bibfield  {journal} {\bibinfo  {journal}
  {Proc. Natl. Acad. Sci. U.S.A.}\ }\textbf {\bibinfo {volume} {109}},\
  \bibinfo {pages} {11641} (\bibinfo {year} {2012})}\BibitemShut {NoStop}%
\bibitem [{\citenamefont {Mandal}\ \emph {et~al.}(2013)\citenamefont {Mandal},
  \citenamefont {Quan},\ and\ \citenamefont {Jarzynski}}]{mandal:2013}%
  \BibitemOpen
  \bibfield  {author} {\bibinfo {author} {\bibfnamefont {D.}~\bibnamefont
  {Mandal}}, \bibinfo {author} {\bibfnamefont {H.~T.}\ \bibnamefont {Quan}}, \
  and\ \bibinfo {author} {\bibfnamefont {C.}~\bibnamefont {Jarzynski}},\
  }\bibfield  {title} {\enquote {\bibinfo {title} {{M}axwell's refrigerator: An
  exactly solvable model},}\ }\href {\doibase 10.1103/PhysRevLett.111.030602}
  {\bibfield  {journal} {\bibinfo  {journal} {Phys. Rev. Lett.}\ }\textbf
  {\bibinfo {volume} {111}},\ \bibinfo {pages} {030602} (\bibinfo {year}
  {2013})}\BibitemShut {NoStop}%
\bibitem [{\citenamefont {Strasberg}\ \emph {et~al.}(2013)\citenamefont
  {Strasberg}, \citenamefont {Schaller}, \citenamefont {Brandes},\ and\
  \citenamefont {Esposito}}]{strasberg:2013}%
  \BibitemOpen
  \bibfield  {author} {\bibinfo {author} {\bibfnamefont {P.}~\bibnamefont
  {Strasberg}}, \bibinfo {author} {\bibfnamefont {G.}~\bibnamefont {Schaller}},
  \bibinfo {author} {\bibfnamefont {T.}~\bibnamefont {Brandes}}, \ and\
  \bibinfo {author} {\bibfnamefont {M.}~\bibnamefont {Esposito}},\ }\bibfield
  {title} {\enquote {\bibinfo {title} {Thermodynamics of a physical model
  implementing a {M}axwell demon},}\ }\href {\doibase
  10.1103/PhysRevLett.110.040601} {\bibfield  {journal} {\bibinfo  {journal}
  {Phys. Rev. Lett.}\ }\textbf {\bibinfo {volume} {110}},\ \bibinfo {pages}
  {040601} (\bibinfo {year} {2013})}\BibitemShut {NoStop}%
\bibitem [{\citenamefont {Barato}\ and\ \citenamefont
  {Seifert}(2013)}]{barato:2013}%
  \BibitemOpen
  \bibfield  {author} {\bibinfo {author} {\bibfnamefont {A.~C.}\ \bibnamefont
  {Barato}}\ and\ \bibinfo {author} {\bibfnamefont {U.}~\bibnamefont
  {Seifert}},\ }\bibfield  {title} {\enquote {\bibinfo {title} {An autonomous
  and reversible {M}axwell's demon},}\ }\href {\doibase
  10.1209/0295-5075/101/60001} {\bibfield  {journal} {\bibinfo  {journal}
  {EPL}\ }\textbf {\bibinfo {volume} {101}},\ \bibinfo {pages} {60001}
  (\bibinfo {year} {2013})}\BibitemShut {NoStop}%
\bibitem [{\citenamefont {Deffner}(2013)}]{deffner:2013pre}%
  \BibitemOpen
  \bibfield  {author} {\bibinfo {author} {\bibfnamefont {S.}~\bibnamefont
  {Deffner}},\ }\bibfield  {title} {\enquote {\bibinfo {title}
  {Information-driven current in a quantum {M}axwell demon},}\ }\href {\doibase
  10.1103/PhysRevE.88.062128} {\bibfield  {journal} {\bibinfo  {journal} {Phys.
  Rev. E}\ }\textbf {\bibinfo {volume} {88}},\ \bibinfo {pages} {062128}
  (\bibinfo {year} {2013})}\BibitemShut {NoStop}%
\bibitem [{\citenamefont {Hartich}\ \emph {et~al.}(2014)\citenamefont
  {Hartich}, \citenamefont {Barato},\ and\ \citenamefont
  {Seifert}}]{hartich:2014}%
  \BibitemOpen
  \bibfield  {author} {\bibinfo {author} {\bibfnamefont {D.}~\bibnamefont
  {Hartich}}, \bibinfo {author} {\bibfnamefont {A.~C.}\ \bibnamefont {Barato}},
  \ and\ \bibinfo {author} {\bibfnamefont {U.}~\bibnamefont {Seifert}},\
  }\bibfield  {title} {\enquote {\bibinfo {title} {Stochastic thermodynamics of
  bipartite systems: transfer entropy inequalities and a {M}axwell{'}s demon
  interpretation},}\ }\href {http://stacks.iop.org/1742-5468/2014/i=2/a=P02016}
  {\bibfield  {journal} {\bibinfo  {journal} {J. Stat. Mech. Theor. Exp.}\
  }\textbf {\bibinfo {volume} {2014}},\ \bibinfo {pages} {P02016} (\bibinfo
  {year} {2014})}\BibitemShut {NoStop}%
\bibitem [{\citenamefont {Cao}\ \emph {et~al.}(2015)\citenamefont {Cao},
  \citenamefont {Gong},\ and\ \citenamefont {Quan}}]{cao:2015}%
  \BibitemOpen
  \bibfield  {author} {\bibinfo {author} {\bibfnamefont {Y.}~\bibnamefont
  {Cao}}, \bibinfo {author} {\bibfnamefont {Z.}~\bibnamefont {Gong}}, \ and\
  \bibinfo {author} {\bibfnamefont {H.~T.}\ \bibnamefont {Quan}},\ }\bibfield
  {title} {\enquote {\bibinfo {title} {Thermodynamics of information processing
  based on enzyme kinetics: {A}n exactly solvable model of an information
  pump},}\ }\href {\doibase 10.1103/PhysRevE.91.062117} {\bibfield  {journal}
  {\bibinfo  {journal} {Phys. Rev. E}\ }\textbf {\bibinfo {volume} {91}},\
  \bibinfo {pages} {062117} (\bibinfo {year} {2015})}\BibitemShut {NoStop}%
\bibitem [{\citenamefont {Shiraishi}\ \emph {et~al.}(2015)\citenamefont
  {Shiraishi}, \citenamefont {Ito}, \citenamefont {Kawaguchi},\ and\
  \citenamefont {Sagawa}}]{shiraishi:2015njp}%
  \BibitemOpen
  \bibfield  {author} {\bibinfo {author} {\bibfnamefont {N.}~\bibnamefont
  {Shiraishi}}, \bibinfo {author} {\bibfnamefont {S.}~\bibnamefont {Ito}},
  \bibinfo {author} {\bibfnamefont {K.}~\bibnamefont {Kawaguchi}}, \ and\
  \bibinfo {author} {\bibfnamefont {T.}~\bibnamefont {Sagawa}},\ }\bibfield
  {title} {\enquote {\bibinfo {title} {Role of measurement-feedback separation
  in autonomous {M}axwell's demons},}\ }\href {\doibase
  10.1088/1367-2630/17/4/045012} {\bibfield  {journal} {\bibinfo  {journal}
  {New J. Phys.}\ }\textbf {\bibinfo {volume} {17}},\ \bibinfo {pages} {045012}
  (\bibinfo {year} {2015})}\BibitemShut {NoStop}%
\bibitem [{\citenamefont {Rana}\ and\ \citenamefont
  {Jayannavar}(2016)}]{rana:2016}%
  \BibitemOpen
  \bibfield  {author} {\bibinfo {author} {\bibfnamefont {S.}~\bibnamefont
  {Rana}}\ and\ \bibinfo {author} {\bibfnamefont {A.~M.}\ \bibnamefont
  {Jayannavar}},\ }\bibfield  {title} {\enquote {\bibinfo {title} {A
  multipurpose information engine that can go beyond the {C}arnot limit},}\
  }\href {\doibase 10.1088/1742-5468/2016/10/103207} {\bibfield  {journal}
  {\bibinfo  {journal} {J. Stat. Mech. Theor. Exp.}\ }\textbf {\bibinfo
  {volume} {2016}},\ \bibinfo {pages} {103207} (\bibinfo {year}
  {2016})}\BibitemShut {NoStop}%
\bibitem [{\citenamefont {Kutvonen}\ \emph
  {et~al.}(2016{\natexlab{b}})\citenamefont {Kutvonen}, \citenamefont {Koski},\
  and\ \citenamefont {Ala-Nissila}}]{kutvonen:2016}%
  \BibitemOpen
  \bibfield  {author} {\bibinfo {author} {\bibfnamefont {A.}~\bibnamefont
  {Kutvonen}}, \bibinfo {author} {\bibfnamefont {J.}~\bibnamefont {Koski}}, \
  and\ \bibinfo {author} {\bibfnamefont {T.}~\bibnamefont {Ala-Nissila}},\
  }\bibfield  {title} {\enquote {\bibinfo {title} {Thermodynamics and
  efficiency of an autonomous on-chip {M}axwell’s demon},}\ }\href {\doibase
  10.1038/srep21126} {\bibfield  {journal} {\bibinfo  {journal} {Sci. Rep.}\
  }\textbf {\bibinfo {volume} {6}},\ \bibinfo {pages} {21126} (\bibinfo {year}
  {2016}{\natexlab{b}})}\BibitemShut {NoStop}%
\bibitem [{\citenamefont {Boyd}\ \emph {et~al.}(2016)\citenamefont {Boyd},
  \citenamefont {Mandal},\ and\ \citenamefont {Crutchfield}}]{boyd:2016}%
  \BibitemOpen
  \bibfield  {author} {\bibinfo {author} {\bibfnamefont {A.~B.}\ \bibnamefont
  {Boyd}}, \bibinfo {author} {\bibfnamefont {D.}~\bibnamefont {Mandal}}, \ and\
  \bibinfo {author} {\bibfnamefont {J.~P.}\ \bibnamefont {Crutchfield}},\
  }\bibfield  {title} {\enquote {\bibinfo {title} {Identifying functional
  thermodynamics in autonomous {M}axwellian ratchets},}\ }\href {\doibase
  10.1088/1367-2630/18/2/023049} {\bibfield  {journal} {\bibinfo  {journal}
  {New J. Phys.}\ }\textbf {\bibinfo {volume} {18}},\ \bibinfo {pages} {023049}
  (\bibinfo {year} {2016})}\BibitemShut {NoStop}%
\bibitem [{\citenamefont {Whitney}\ \emph {et~al.}(2016)\citenamefont
  {Whitney}, \citenamefont {S\'anchez}, \citenamefont {Haupt},\ and\
  \citenamefont {Splettstoesser}}]{whitney:2016}%
  \BibitemOpen
  \bibfield  {author} {\bibinfo {author} {\bibfnamefont {R.~S.}\ \bibnamefont
  {Whitney}}, \bibinfo {author} {\bibfnamefont {R.}~\bibnamefont {S\'anchez}},
  \bibinfo {author} {\bibfnamefont {F.}~\bibnamefont {Haupt}}, \ and\ \bibinfo
  {author} {\bibfnamefont {J.}~\bibnamefont {Splettstoesser}},\ }\bibfield
  {title} {\enquote {\bibinfo {title} {Thermoelectricity without absorbing
  energy from the heat sources},}\ }\href {\doibase
  10.1016/j.physe.2015.09.025} {\bibfield  {journal} {\bibinfo  {journal}
  {Physica E}\ }\textbf {\bibinfo {volume} {75}},\ \bibinfo {pages} {257}
  (\bibinfo {year} {2016})}\BibitemShut {NoStop}%
\bibitem [{\citenamefont {Rossell\'o}\ \emph {et~al.}(2017)\citenamefont
  {Rossell\'o}, \citenamefont {L\'opez},\ and\ \citenamefont
  {Platero}}]{rossello:2017}%
  \BibitemOpen
  \bibfield  {author} {\bibinfo {author} {\bibfnamefont {G.}~\bibnamefont
  {Rossell\'o}}, \bibinfo {author} {\bibfnamefont {R.}~\bibnamefont {L\'opez}},
  \ and\ \bibinfo {author} {\bibfnamefont {G.}~\bibnamefont {Platero}},\
  }\bibfield  {title} {\enquote {\bibinfo {title} {Chiral {M}axwell demon in a
  quantum {H}all system with a localized impurity},}\ }\href {\doibase
  10.1103/PhysRevB.96.075305} {\bibfield  {journal} {\bibinfo  {journal} {Phys.
  Rev. B}\ }\textbf {\bibinfo {volume} {96}},\ \bibinfo {pages} {075305}
  (\bibinfo {year} {2017})}\BibitemShut {NoStop}%
\bibitem [{\citenamefont {Spinney}\ \emph {et~al.}(2018)\citenamefont
  {Spinney}, \citenamefont {Prokopenko},\ and\ \citenamefont
  {Chu}}]{spinney:2018}%
  \BibitemOpen
  \bibfield  {author} {\bibinfo {author} {\bibfnamefont {R.~E.}\ \bibnamefont
  {Spinney}}, \bibinfo {author} {\bibfnamefont {M.}~\bibnamefont {Prokopenko}},
  \ and\ \bibinfo {author} {\bibfnamefont {D.}~\bibnamefont {Chu}},\ }\bibfield
   {title} {\enquote {\bibinfo {title} {Information ratchets exploiting
  spatially structured information reservoirs},}\ }\href {\doibase
  10.1103/PhysRevE.98.022124} {\bibfield  {journal} {\bibinfo  {journal} {Phys.
  Rev. E}\ }\textbf {\bibinfo {volume} {98}},\ \bibinfo {pages} {022124}
  (\bibinfo {year} {2018})}\BibitemShut {NoStop}%
\bibitem [{\citenamefont {Ptaszy\'{n}ski}(2018)}]{ptaszynski:2018}%
  \BibitemOpen
  \bibfield  {author} {\bibinfo {author} {\bibfnamefont {K.}~\bibnamefont
  {Ptaszy\'{n}ski}},\ }\bibfield  {title} {\enquote {\bibinfo {title}
  {Autonomous quantum {M}axwell's demon based on two exchange-coupled quantum
  dots},}\ }\href {\doibase 10.1103/PhysRevE.97.012116} {\bibfield  {journal}
  {\bibinfo  {journal} {Phys. Rev. E}\ }\textbf {\bibinfo {volume} {97}},\
  \bibinfo {pages} {012116} (\bibinfo {year} {2018})}\BibitemShut {NoStop}%
\bibitem [{\citenamefont {Strasberg}\ \emph {et~al.}(2018)\citenamefont
  {Strasberg}, \citenamefont {Schaller}, \citenamefont {Schmidt},\ and\
  \citenamefont {Esposito}}]{strasberg:2018}%
  \BibitemOpen
  \bibfield  {author} {\bibinfo {author} {\bibfnamefont {P.}~\bibnamefont
  {Strasberg}}, \bibinfo {author} {\bibfnamefont {G.}~\bibnamefont {Schaller}},
  \bibinfo {author} {\bibfnamefont {T.~L.}\ \bibnamefont {Schmidt}}, \ and\
  \bibinfo {author} {\bibfnamefont {M.}~\bibnamefont {Esposito}},\ }\bibfield
  {title} {\enquote {\bibinfo {title} {Fermionic reaction coordinates and their
  application to an autonomous {M}axwell demon in the strong-coupling
  regime},}\ }\href {\doibase 10.1103/PhysRevB.97.205405} {\bibfield  {journal}
  {\bibinfo  {journal} {Phys. Rev. B}\ }\textbf {\bibinfo {volume} {97}},\
  \bibinfo {pages} {205405} (\bibinfo {year} {2018})}\BibitemShut {NoStop}%
\bibitem [{\citenamefont {Erdman}\ \emph {et~al.}(2018)\citenamefont {Erdman},
  \citenamefont {Bhandari}, \citenamefont {Fazio}, \citenamefont {Pekola},\
  and\ \citenamefont {Taddei}}]{erdman_absorption_2018}%
  \BibitemOpen
  \bibfield  {author} {\bibinfo {author} {\bibfnamefont {P.~A.}\ \bibnamefont
  {Erdman}}, \bibinfo {author} {\bibfnamefont {B.}~\bibnamefont {Bhandari}},
  \bibinfo {author} {\bibfnamefont {R.}~\bibnamefont {Fazio}}, \bibinfo
  {author} {\bibfnamefont {J.~P.}\ \bibnamefont {Pekola}}, \ and\ \bibinfo
  {author} {\bibfnamefont {F.}~\bibnamefont {Taddei}},\ }\bibfield  {title}
  {\enquote {\bibinfo {title} {{Absorption refrigerators based on
  Coulomb-coupled single-electron systems}},}\ }\href {\doibase
  10.1103/PhysRevB.98.045433} {\bibfield  {journal} {\bibinfo  {journal} {Phys.
  Rev. B}\ }\textbf {\bibinfo {volume} {98}},\ \bibinfo {pages} {045433}
  (\bibinfo {year} {2018})}\BibitemShut {NoStop}%
\bibitem [{\citenamefont {S\'anchez}\ \emph {et~al.}()\citenamefont
  {S\'anchez}, \citenamefont {Splettstoesser},\ and\ \citenamefont
  {Whitney}}]{sanchez:2018}%
  \BibitemOpen
  \bibfield  {author} {\bibinfo {author} {\bibfnamefont {R.}~\bibnamefont
  {S\'anchez}}, \bibinfo {author} {\bibfnamefont {J.}~\bibnamefont
  {Splettstoesser}}, \ and\ \bibinfo {author} {\bibfnamefont {R.~S.}\
  \bibnamefont {Whitney}},\ }\href@noop {} {\enquote {\bibinfo {title}
  {Nonequilibrium system as a demon},}\ }\Eprint
  {http://arxiv.org/abs/1811.02453} {arXiv:1811.02453 [cond-mat]} \BibitemShut
  {NoStop}%
\bibitem [{\citenamefont {Horowitz}\ \emph {et~al.}(2013)\citenamefont
  {Horowitz}, \citenamefont {Sagawa},\ and\ \citenamefont
  {Parrondo}}]{horowitz:2013}%
  \BibitemOpen
  \bibfield  {author} {\bibinfo {author} {\bibfnamefont {J.~M.}\ \bibnamefont
  {Horowitz}}, \bibinfo {author} {\bibfnamefont {T.}~\bibnamefont {Sagawa}}, \
  and\ \bibinfo {author} {\bibfnamefont {J.~M.~R.}\ \bibnamefont {Parrondo}},\
  }\bibfield  {title} {\enquote {\bibinfo {title} {Imitating chemical motors
  with optimal information motors},}\ }\href {\doibase
  10.1103/PhysRevLett.111.010602} {\bibfield  {journal} {\bibinfo  {journal}
  {Phys. Rev. Lett.}\ }\textbf {\bibinfo {volume} {111}},\ \bibinfo {pages}
  {010602} (\bibinfo {year} {2013})}\BibitemShut {NoStop}%
\bibitem [{\citenamefont {Strasberg}\ \emph {et~al.}(2014)\citenamefont
  {Strasberg}, \citenamefont {Schaller}, \citenamefont {Brandes},\ and\
  \citenamefont {Jarzynski}}]{strasberg:2014}%
  \BibitemOpen
  \bibfield  {author} {\bibinfo {author} {\bibfnamefont {P.}~\bibnamefont
  {Strasberg}}, \bibinfo {author} {\bibfnamefont {G.}~\bibnamefont {Schaller}},
  \bibinfo {author} {\bibfnamefont {T.}~\bibnamefont {Brandes}}, \ and\
  \bibinfo {author} {\bibfnamefont {C.}~\bibnamefont {Jarzynski}},\ }\bibfield
  {title} {\enquote {\bibinfo {title} {Second laws for an information driven
  current through a spin valve},}\ }\href {\doibase 10.1103/PhysRevE.90.062107}
  {\bibfield  {journal} {\bibinfo  {journal} {Phys. Rev. E}\ }\textbf {\bibinfo
  {volume} {90}},\ \bibinfo {pages} {062107} (\bibinfo {year}
  {2014})}\BibitemShut {NoStop}%
\bibitem [{\citenamefont {Shiraishi}\ \emph {et~al.}(2016)\citenamefont
  {Shiraishi}, \citenamefont {Matsumoto},\ and\ \citenamefont
  {Sagawa}}]{shiraishi:2016}%
  \BibitemOpen
  \bibfield  {author} {\bibinfo {author} {\bibfnamefont {N.}~\bibnamefont
  {Shiraishi}}, \bibinfo {author} {\bibfnamefont {T.}~\bibnamefont
  {Matsumoto}}, \ and\ \bibinfo {author} {\bibfnamefont {T.}~\bibnamefont
  {Sagawa}},\ }\bibfield  {title} {\enquote {\bibinfo {title}
  {Measurement-feedback formalism meets information reservoirs},}\ }\href
  {\doibase 10.1088/1367-2630/18/1/013044} {\bibfield  {journal} {\bibinfo
  {journal} {New J. Phys.}\ }\textbf {\bibinfo {volume} {18}},\ \bibinfo
  {pages} {013044} (\bibinfo {year} {2016})}\BibitemShut {NoStop}%
\bibitem [{\citenamefont {Strasberg}\ \emph {et~al.}(2017)\citenamefont
  {Strasberg}, \citenamefont {Schaller}, \citenamefont {Brandes},\ and\
  \citenamefont {Esposito}}]{strasberg:2017}%
  \BibitemOpen
  \bibfield  {author} {\bibinfo {author} {\bibfnamefont {P.}~\bibnamefont
  {Strasberg}}, \bibinfo {author} {\bibfnamefont {G.}~\bibnamefont {Schaller}},
  \bibinfo {author} {\bibfnamefont {T.}~\bibnamefont {Brandes}}, \ and\
  \bibinfo {author} {\bibfnamefont {M.}~\bibnamefont {Esposito}},\ }\bibfield
  {title} {\enquote {\bibinfo {title} {Quantum and information thermodynamics:
  A unifying framework based on repeated interactions},}\ }\href {\doibase
  10.1103/PhysRevX.7.021003} {\bibfield  {journal} {\bibinfo  {journal} {Phys.
  Rev. X}\ }\textbf {\bibinfo {volume} {7}},\ \bibinfo {pages} {021003}
  (\bibinfo {year} {2017})}\BibitemShut {NoStop}%
\bibitem [{\citenamefont {Sothmann}\ \emph {et~al.}(2015)\citenamefont
  {Sothmann}, \citenamefont {S\'anchez},\ and\ \citenamefont
  {Jordan}}]{sothmann:2015}%
  \BibitemOpen
  \bibfield  {author} {\bibinfo {author} {\bibfnamefont {B.}~\bibnamefont
  {Sothmann}}, \bibinfo {author} {\bibfnamefont {R.}~\bibnamefont {S\'anchez}},
  \ and\ \bibinfo {author} {\bibfnamefont {A.~N.}\ \bibnamefont {Jordan}},\
  }\bibfield  {title} {\enquote {\bibinfo {title} {Thermoelectric energy
  harvesting with quantum dots},}\ }\href
  {http://stacks.iop.org/0957-4484/26/i=3/a=032001} {\bibfield  {journal}
  {\bibinfo  {journal} {Nanotechnology}\ }\textbf {\bibinfo {volume} {26}},\
  \bibinfo {pages} {032001} (\bibinfo {year} {2015})}\BibitemShut {NoStop}%
\bibitem [{\citenamefont {S\'anchez}\ and\ \citenamefont
  {B\"uttiker}(2011)}]{sanchez:2011}%
  \BibitemOpen
  \bibfield  {author} {\bibinfo {author} {\bibfnamefont {R.}~\bibnamefont
  {S\'anchez}}\ and\ \bibinfo {author} {\bibfnamefont {M.}~\bibnamefont
  {B\"uttiker}},\ }\bibfield  {title} {\enquote {\bibinfo {title} {Optimal
  energy quanta to current conversion},}\ }\href {\doibase
  10.1103/PhysRevB.83.085428} {\bibfield  {journal} {\bibinfo  {journal} {Phys.
  Rev. B}\ }\textbf {\bibinfo {volume} {83}},\ \bibinfo {pages} {085428}
  (\bibinfo {year} {2011})}\BibitemShut {NoStop}%
\bibitem [{\citenamefont {Staring}\ \emph {et~al.}(1993)\citenamefont
  {Staring}, \citenamefont {Molenkamp}, \citenamefont {Alphenaar},
  \citenamefont {van Houten}, \citenamefont {Buyk}, \citenamefont {Mabesoone},
  \citenamefont {Beenakker},\ and\ \citenamefont
  {Foxon}}]{staring_coulomb-blockade_1993}%
  \BibitemOpen
  \bibfield  {author} {\bibinfo {author} {\bibfnamefont {A.~A.~M.}\
  \bibnamefont {Staring}}, \bibinfo {author} {\bibfnamefont {L.~W.}\
  \bibnamefont {Molenkamp}}, \bibinfo {author} {\bibfnamefont {B.~W.}\
  \bibnamefont {Alphenaar}}, \bibinfo {author} {\bibfnamefont {H.}~\bibnamefont
  {van Houten}}, \bibinfo {author} {\bibfnamefont {O.~J.~A.}\ \bibnamefont
  {Buyk}}, \bibinfo {author} {\bibfnamefont {M.~A.~A.}\ \bibnamefont
  {Mabesoone}}, \bibinfo {author} {\bibfnamefont {C.~W.~J.}\ \bibnamefont
  {Beenakker}}, \ and\ \bibinfo {author} {\bibfnamefont {C.~T.}\ \bibnamefont
  {Foxon}},\ }\bibfield  {title} {\enquote {\bibinfo {title} {Coulomb-blockade
  oscillations in the thermopower of a quantum dotz},}\ }\href {\doibase
  10.1209/0295-5075/22/1/011} {\bibfield  {journal} {\bibinfo  {journal} {EPL}\
  }\textbf {\bibinfo {volume} {22}},\ \bibinfo {pages} {57} (\bibinfo {year}
  {1993})}\BibitemShut {NoStop}%
\bibitem [{\citenamefont {Dzurak}\ \emph {et~al.}(1993)\citenamefont {Dzurak},
  \citenamefont {Smith}, \citenamefont {Pepper}, \citenamefont {Ritchie},
  \citenamefont {Frost}, \citenamefont {Jones},\ and\ \citenamefont
  {Hasko}}]{dzurak_observation_1993}%
  \BibitemOpen
  \bibfield  {author} {\bibinfo {author} {\bibfnamefont {A.~S.}\ \bibnamefont
  {Dzurak}}, \bibinfo {author} {\bibfnamefont {C.~G.}\ \bibnamefont {Smith}},
  \bibinfo {author} {\bibfnamefont {M.}~\bibnamefont {Pepper}}, \bibinfo
  {author} {\bibfnamefont {D.~A.}\ \bibnamefont {Ritchie}}, \bibinfo {author}
  {\bibfnamefont {J.~E.~F.}\ \bibnamefont {Frost}}, \bibinfo {author}
  {\bibfnamefont {G.~A.~C.}\ \bibnamefont {Jones}}, \ and\ \bibinfo {author}
  {\bibfnamefont {D.~G.}\ \bibnamefont {Hasko}},\ }\bibfield  {title} {\enquote
  {\bibinfo {title} {{Observation of Coulomb blockade oscillations in the
  thermopower of a quantum dot}},}\ }\href {\doibase
  10.1016/0038-1098(93)90819-9} {\bibfield  {journal} {\bibinfo  {journal}
  {Solid State Commun.}\ }\textbf {\bibinfo {volume} {87}},\ \bibinfo {pages}
  {1145} (\bibinfo {year} {1993})}\BibitemShut {NoStop}%
\bibitem [{\citenamefont {Dzurak}\ \emph {et~al.}(1997)\citenamefont {Dzurak},
  \citenamefont {Smith}, \citenamefont {Barnes}, \citenamefont {Pepper},
  \citenamefont {Martín-Moreno}, \citenamefont {Liang}, \citenamefont
  {Ritchie},\ and\ \citenamefont {Jones}}]{dzurak_thermoelectric_1997}%
  \BibitemOpen
  \bibfield  {author} {\bibinfo {author} {\bibfnamefont {A.~S.}\ \bibnamefont
  {Dzurak}}, \bibinfo {author} {\bibfnamefont {C.~G.}\ \bibnamefont {Smith}},
  \bibinfo {author} {\bibfnamefont {C.~H.~W.}\ \bibnamefont {Barnes}}, \bibinfo
  {author} {\bibfnamefont {M.}~\bibnamefont {Pepper}}, \bibinfo {author}
  {\bibfnamefont {L.}~\bibnamefont {Martín-Moreno}}, \bibinfo {author}
  {\bibfnamefont {C.~T.}\ \bibnamefont {Liang}}, \bibinfo {author}
  {\bibfnamefont {D.~A.}\ \bibnamefont {Ritchie}}, \ and\ \bibinfo {author}
  {\bibfnamefont {G.~A.~C.}\ \bibnamefont {Jones}},\ }\bibfield  {title}
  {\enquote {\bibinfo {title} {Thermoelectric signature of the excitation
  spectrum of a quantum dot},}\ }\href {\doibase 10.1103/PhysRevB.55.R10197}
  {\bibfield  {journal} {\bibinfo  {journal} {Phys. Rev. B}\ }\textbf {\bibinfo
  {volume} {55}},\ \bibinfo {pages} {R10197} (\bibinfo {year}
  {1997})}\BibitemShut {NoStop}%
\bibitem [{\citenamefont {Scheibner}\ \emph {et~al.}(2007)\citenamefont
  {Scheibner}, \citenamefont {Novik}, \citenamefont {Borzenko}, \citenamefont
  {K\"onig}, \citenamefont {Reuter}, \citenamefont {Wieck}, \citenamefont
  {Buhmann},\ and\ \citenamefont {Molenkamp}}]{scheibner_sequential_2007}%
  \BibitemOpen
  \bibfield  {author} {\bibinfo {author} {\bibfnamefont {R.}~\bibnamefont
  {Scheibner}}, \bibinfo {author} {\bibfnamefont {E.~G.}\ \bibnamefont
  {Novik}}, \bibinfo {author} {\bibfnamefont {T.}~\bibnamefont {Borzenko}},
  \bibinfo {author} {\bibfnamefont {M.}~\bibnamefont {K\"onig}}, \bibinfo
  {author} {\bibfnamefont {D.}~\bibnamefont {Reuter}}, \bibinfo {author}
  {\bibfnamefont {A.~D.}\ \bibnamefont {Wieck}}, \bibinfo {author}
  {\bibfnamefont {H.}~\bibnamefont {Buhmann}}, \ and\ \bibinfo {author}
  {\bibfnamefont {L.~W.}\ \bibnamefont {Molenkamp}},\ }\bibfield  {title}
  {\enquote {\bibinfo {title} {Sequential and cotunneling behavior in the
  temperature-dependent thermopower of few-electron quantum dots},}\ }\href
  {\doibase 10.1103/PhysRevB.75.041301} {\bibfield  {journal} {\bibinfo
  {journal} {Phys. Rev. B}\ }\textbf {\bibinfo {volume} {75}},\ \bibinfo
  {pages} {041301} (\bibinfo {year} {2007})}\BibitemShut {NoStop}%
\bibitem [{\citenamefont {Thierschmann}\ \emph {et~al.}(2013)\citenamefont
  {Thierschmann}, \citenamefont {Henke}, \citenamefont {Knorr}, \citenamefont
  {Maier}, \citenamefont {Heyn}, \citenamefont {Hansen}, \citenamefont
  {Buhmann},\ and\ \citenamefont {Molenkamp}}]{thierschmann_diffusion_2013}%
  \BibitemOpen
  \bibfield  {author} {\bibinfo {author} {\bibfnamefont {H.}~\bibnamefont
  {Thierschmann}}, \bibinfo {author} {\bibfnamefont {M.}~\bibnamefont {Henke}},
  \bibinfo {author} {\bibfnamefont {J.}~\bibnamefont {Knorr}}, \bibinfo
  {author} {\bibfnamefont {L.}~\bibnamefont {Maier}}, \bibinfo {author}
  {\bibfnamefont {C.}~\bibnamefont {Heyn}}, \bibinfo {author} {\bibfnamefont
  {W.}~\bibnamefont {Hansen}}, \bibinfo {author} {\bibfnamefont
  {H.}~\bibnamefont {Buhmann}}, \ and\ \bibinfo {author} {\bibfnamefont
  {L.~W.}\ \bibnamefont {Molenkamp}},\ }\bibfield  {title} {\enquote {\bibinfo
  {title} {Diffusion thermopower of a serial double quantum dot},}\ }\href
  {\doibase 10.1088/1367-2630/15/12/123010} {\bibfield  {journal} {\bibinfo
  {journal} {New J. Phys.}\ }\textbf {\bibinfo {volume} {15}},\ \bibinfo
  {pages} {123010} (\bibinfo {year} {2013})}\BibitemShut {NoStop}%
\bibitem [{\citenamefont {Svensson}\ \emph {et~al.}(2013)\citenamefont
  {Svensson}, \citenamefont {Hoffmann}, \citenamefont {Nakpathomkun},
  \citenamefont {Wu}, \citenamefont {Xu}, \citenamefont {Nilsson},
  \citenamefont {S{\ifmmode\acute{a}\else\'{a}\fi}nchez}, \citenamefont
  {Kashcheyevs},\ and\ \citenamefont {Linke}}]{svensson_nonlinear_2013}%
  \BibitemOpen
  \bibfield  {author} {\bibinfo {author} {\bibfnamefont {S.~F.}\ \bibnamefont
  {Svensson}}, \bibinfo {author} {\bibfnamefont {E.~A.}\ \bibnamefont
  {Hoffmann}}, \bibinfo {author} {\bibfnamefont {N.}~\bibnamefont
  {Nakpathomkun}}, \bibinfo {author} {\bibfnamefont {P.~M.}\ \bibnamefont
  {Wu}}, \bibinfo {author} {\bibfnamefont {H.~Q.}\ \bibnamefont {Xu}}, \bibinfo
  {author} {\bibfnamefont {H.~A.}\ \bibnamefont {Nilsson}}, \bibinfo {author}
  {\bibfnamefont {D.}~\bibnamefont {S{\ifmmode\acute{a}\else\'{a}\fi}nchez}},
  \bibinfo {author} {\bibfnamefont {V.}~\bibnamefont {Kashcheyevs}}, \ and\
  \bibinfo {author} {\bibfnamefont {H.}~\bibnamefont {Linke}},\ }\bibfield
  {title} {\enquote {\bibinfo {title} {{Nonlinear thermovoltage and
  thermocurrent in quantum dots}},}\ }\href {\doibase
  10.1088/1367-2630/15/10/105011} {\bibfield  {journal} {\bibinfo  {journal}
  {New J. Phys.}\ }\textbf {\bibinfo {volume} {15}},\ \bibinfo {pages} {105011}
  (\bibinfo {year} {2013})}\BibitemShut {NoStop}%
\bibitem [{\citenamefont {Thierschmann}\ \emph
  {et~al.}(2015{\natexlab{b}})\citenamefont {Thierschmann}, \citenamefont
  {Arnold}, \citenamefont {Mitterm\"uller}, \citenamefont {Maier},
  \citenamefont {Heyn}, \citenamefont {Hansen}, \citenamefont {Buhmann},\ and\
  \citenamefont {Molenkamp}}]{thierschmann_thermal_2015}%
  \BibitemOpen
  \bibfield  {author} {\bibinfo {author} {\bibfnamefont {H.}~\bibnamefont
  {Thierschmann}}, \bibinfo {author} {\bibfnamefont {F.}~\bibnamefont
  {Arnold}}, \bibinfo {author} {\bibfnamefont {M.}~\bibnamefont
  {Mitterm\"uller}}, \bibinfo {author} {\bibfnamefont {L.}~\bibnamefont
  {Maier}}, \bibinfo {author} {\bibfnamefont {C.}~\bibnamefont {Heyn}},
  \bibinfo {author} {\bibfnamefont {W.}~\bibnamefont {Hansen}}, \bibinfo
  {author} {\bibfnamefont {H.}~\bibnamefont {Buhmann}}, \ and\ \bibinfo
  {author} {\bibfnamefont {L.~W.}\ \bibnamefont {Molenkamp}},\ }\bibfield
  {title} {\enquote {\bibinfo {title} {Thermal gating of charge currents with
  {Coulomb} coupled quantum dots},}\ }\href {\doibase
  10.1088/1367-2630/17/11/113003} {\bibfield  {journal} {\bibinfo  {journal}
  {New J. Phys.}\ }\textbf {\bibinfo {volume} {17}},\ \bibinfo {pages} {113003}
  (\bibinfo {year} {2015}{\natexlab{b}})}\BibitemShut {NoStop}%
\bibitem [{\citenamefont {Josefsson}\ \emph {et~al.}(2018)\citenamefont
  {Josefsson}, \citenamefont {Svilans}, \citenamefont {Burke}, \citenamefont
  {Hoffmann}, \citenamefont {Fahlvik}, \citenamefont {Thelander}, \citenamefont
  {Leijnse},\ and\ \citenamefont {Linke}}]{josefsson_quantum-dot_2018}%
  \BibitemOpen
  \bibfield  {author} {\bibinfo {author} {\bibfnamefont {M.}~\bibnamefont
  {Josefsson}}, \bibinfo {author} {\bibfnamefont {A.}~\bibnamefont {Svilans}},
  \bibinfo {author} {\bibfnamefont {A.~M.}\ \bibnamefont {Burke}}, \bibinfo
  {author} {\bibfnamefont {E.~A.}\ \bibnamefont {Hoffmann}}, \bibinfo {author}
  {\bibfnamefont {S.}~\bibnamefont {Fahlvik}}, \bibinfo {author} {\bibfnamefont
  {C.}~\bibnamefont {Thelander}}, \bibinfo {author} {\bibfnamefont
  {M.}~\bibnamefont {Leijnse}}, \ and\ \bibinfo {author} {\bibfnamefont
  {H.}~\bibnamefont {Linke}},\ }\bibfield  {title} {\enquote {\bibinfo {title}
  {{A quantum-dot heat engine operating close to the thermodynamic efficiency
  limits}},}\ }\href {\doibase 10.1038/s41565-018-0200-5} {\bibfield  {journal}
  {\bibinfo  {journal} {Nat. Nanotechnol.}\ }\textbf {\bibinfo {volume} {13}},\
  \bibinfo {pages} {920--924} (\bibinfo {year} {2018})}\BibitemShut {NoStop}%
\bibitem [{\citenamefont {S\'anchez}\ \emph {et~al.}(2010)\citenamefont
  {S\'anchez}, \citenamefont {L\'opez}, \citenamefont {S\'anchez},\ and\
  \citenamefont {B\"uttiker}}]{sanchez:2010}%
  \BibitemOpen
  \bibfield  {author} {\bibinfo {author} {\bibfnamefont {R.}~\bibnamefont
  {S\'anchez}}, \bibinfo {author} {\bibfnamefont {R.}~\bibnamefont {L\'opez}},
  \bibinfo {author} {\bibfnamefont {D.}~\bibnamefont {S\'anchez}}, \ and\
  \bibinfo {author} {\bibfnamefont {M.}~\bibnamefont {B\"uttiker}},\ }\bibfield
   {title} {\enquote {\bibinfo {title} {Mesoscopic {C}oulomb drag, broken
  detailed balance, and fluctuation relations},}\ }\href {\doibase
  10.1103/PhysRevLett.104.076801} {\bibfield  {journal} {\bibinfo  {journal}
  {Phys. Rev. Lett.}\ }\textbf {\bibinfo {volume} {104}},\ \bibinfo {pages}
  {076801} (\bibinfo {year} {2010})}\BibitemShut {NoStop}%
\bibitem [{\citenamefont {Bischoff}\ \emph {et~al.}(2015)\citenamefont
  {Bischoff}, \citenamefont {Eich}, \citenamefont {Zilberberg}, \citenamefont
  {Rössler}, \citenamefont {Ihn},\ and\ \citenamefont
  {Ensslin}}]{bischoff:2015}%
  \BibitemOpen
  \bibfield  {author} {\bibinfo {author} {\bibfnamefont {D.}~\bibnamefont
  {Bischoff}}, \bibinfo {author} {\bibfnamefont {M.}~\bibnamefont {Eich}},
  \bibinfo {author} {\bibfnamefont {O.}~\bibnamefont {Zilberberg}}, \bibinfo
  {author} {\bibfnamefont {C.}~\bibnamefont {Rössler}}, \bibinfo {author}
  {\bibfnamefont {T.}~\bibnamefont {Ihn}}, \ and\ \bibinfo {author}
  {\bibfnamefont {K.}~\bibnamefont {Ensslin}},\ }\bibfield  {title} {\enquote
  {\bibinfo {title} {Measurement back-action in stacked graphene quantum
  dots},}\ }\href {\doibase 10.1021/acs.nanolett.5b02167} {\bibfield  {journal}
  {\bibinfo  {journal} {Nano Lett.}\ }\textbf {\bibinfo {volume} {15}},\
  \bibinfo {pages} {6003} (\bibinfo {year} {2015})}\BibitemShut {NoStop}%
\bibitem [{\citenamefont {Kaasbjerg}\ and\ \citenamefont
  {Jauho}(2016)}]{kaasbjerg:2016}%
  \BibitemOpen
  \bibfield  {author} {\bibinfo {author} {\bibfnamefont {K.}~\bibnamefont
  {Kaasbjerg}}\ and\ \bibinfo {author} {\bibfnamefont {A.-P.}\ \bibnamefont
  {Jauho}},\ }\bibfield  {title} {\enquote {\bibinfo {title} {Correlated
  coulomb drag in capacitively coupled quantum-dot structures},}\ }\href
  {\doibase 10.1103/PhysRevLett.116.196801} {\bibfield  {journal} {\bibinfo
  {journal} {Phys. Rev. Lett.}\ }\textbf {\bibinfo {volume} {116}},\ \bibinfo
  {pages} {196801} (\bibinfo {year} {2016})}\BibitemShut {NoStop}%
\bibitem [{\citenamefont {Keller}\ \emph {et~al.}(2016)\citenamefont {Keller},
  \citenamefont {Lim}, \citenamefont {S\'anchez}, \citenamefont {L\'opez},
  \citenamefont {Amasha}, \citenamefont {Katine}, \citenamefont {Shtrikman},\
  and\ \citenamefont {Goldhaber-Gordon}}]{keller:2016}%
  \BibitemOpen
  \bibfield  {author} {\bibinfo {author} {\bibfnamefont {A.~J.}\ \bibnamefont
  {Keller}}, \bibinfo {author} {\bibfnamefont {J.~S.}\ \bibnamefont {Lim}},
  \bibinfo {author} {\bibfnamefont {D.}~\bibnamefont {S\'anchez}}, \bibinfo
  {author} {\bibfnamefont {R.}~\bibnamefont {L\'opez}}, \bibinfo {author}
  {\bibfnamefont {S.}~\bibnamefont {Amasha}}, \bibinfo {author} {\bibfnamefont
  {J.~A.}\ \bibnamefont {Katine}}, \bibinfo {author} {\bibfnamefont
  {H.}~\bibnamefont {Shtrikman}}, \ and\ \bibinfo {author} {\bibfnamefont
  {D.}~\bibnamefont {Goldhaber-Gordon}},\ }\bibfield  {title} {\enquote
  {\bibinfo {title} {Cotunneling drag effect in {C}oulomb-coupled quantum
  dots},}\ }\href {\doibase 10.1103/PhysRevLett.117.066602} {\bibfield
  {journal} {\bibinfo  {journal} {Phys. Rev. Lett.}\ }\textbf {\bibinfo
  {volume} {117}},\ \bibinfo {pages} {066602} (\bibinfo {year}
  {2016})}\BibitemShut {NoStop}%
\bibitem [{\citenamefont {Dar\'e}\ and\ \citenamefont
  {Lombardo}(2017)}]{dare:2017}%
  \BibitemOpen
  \bibfield  {author} {\bibinfo {author} {\bibfnamefont {A.-M.}\ \bibnamefont
  {Dar\'e}}\ and\ \bibinfo {author} {\bibfnamefont {P.}~\bibnamefont
  {Lombardo}},\ }\bibfield  {title} {\enquote {\bibinfo {title} {Powerful
  {C}oulomb-drag thermoelectric engine},}\ }\href {\doibase
  10.1103/PhysRevB.96.115414} {\bibfield  {journal} {\bibinfo  {journal} {Phys.
  Rev. B}\ }\textbf {\bibinfo {volume} {96}},\ \bibinfo {pages} {115414}
  (\bibinfo {year} {2017})}\BibitemShut {NoStop}%
\bibitem [{\citenamefont {Walldorf}\ \emph {et~al.}(2017)\citenamefont
  {Walldorf}, \citenamefont {Jauho},\ and\ \citenamefont
  {Kaasbjerg}}]{walldorf:2017}%
  \BibitemOpen
  \bibfield  {author} {\bibinfo {author} {\bibfnamefont {N.}~\bibnamefont
  {Walldorf}}, \bibinfo {author} {\bibfnamefont {A.-P.}\ \bibnamefont {Jauho}},
  \ and\ \bibinfo {author} {\bibfnamefont {K.}~\bibnamefont {Kaasbjerg}},\
  }\bibfield  {title} {\enquote {\bibinfo {title} {Thermoelectrics in
  {C}oulomb-coupled quantum dots: Cotunneling and energy-dependent lead
  couplings},}\ }\href {\doibase 10.1103/PhysRevB.96.115415} {\bibfield
  {journal} {\bibinfo  {journal} {Phys. Rev. B}\ }\textbf {\bibinfo {volume}
  {96}},\ \bibinfo {pages} {115415} (\bibinfo {year} {2017})}\BibitemShut
  {NoStop}%
\bibitem [{\citenamefont {Schaller}\ \emph {et~al.}(2010)\citenamefont
  {Schaller}, \citenamefont {Kie\ss{}lich},\ and\ \citenamefont
  {Brandes}}]{schaller:2010}%
  \BibitemOpen
  \bibfield  {author} {\bibinfo {author} {\bibfnamefont {G.}~\bibnamefont
  {Schaller}}, \bibinfo {author} {\bibfnamefont {G.}~\bibnamefont
  {Kie\ss{}lich}}, \ and\ \bibinfo {author} {\bibfnamefont {T.}~\bibnamefont
  {Brandes}},\ }\bibfield  {title} {\enquote {\bibinfo {title} {Low-dimensional
  detector model for full counting statistics: Trajectories, back action, and
  fidelity},}\ }\href {\doibase 10.1103/PhysRevB.82.041303} {\bibfield
  {journal} {\bibinfo  {journal} {Phys. Rev. B}\ }\textbf {\bibinfo {volume}
  {82}},\ \bibinfo {pages} {041303} (\bibinfo {year} {2010})}\BibitemShut
  {NoStop}%
\bibitem [{\citenamefont {Bulnes~Cuetara}\ \emph {et~al.}(2011)\citenamefont
  {Bulnes~Cuetara}, \citenamefont {Esposito},\ and\ \citenamefont
  {Gaspard}}]{bulnes:2011}%
  \BibitemOpen
  \bibfield  {author} {\bibinfo {author} {\bibfnamefont {G.}~\bibnamefont
  {Bulnes~Cuetara}}, \bibinfo {author} {\bibfnamefont {M.}~\bibnamefont
  {Esposito}}, \ and\ \bibinfo {author} {\bibfnamefont {P.}~\bibnamefont
  {Gaspard}},\ }\bibfield  {title} {\enquote {\bibinfo {title} {Fluctuation
  theorems for capacitively coupled electronic currents},}\ }\href {\doibase
  10.1103/PhysRevB.84.165114} {\bibfield  {journal} {\bibinfo  {journal} {Phys.
  Rev. B}\ }\textbf {\bibinfo {volume} {84}},\ \bibinfo {pages} {165114}
  (\bibinfo {year} {2011})}\BibitemShut {NoStop}%
\bibitem [{\citenamefont {Bhandari}\ \emph {et~al.}(2018)\citenamefont
  {Bhandari}, \citenamefont {Chiriac\`o}, \citenamefont {Erdman}, \citenamefont
  {Fazio},\ and\ \citenamefont {Taddei}}]{bhandari:2018}%
  \BibitemOpen
  \bibfield  {author} {\bibinfo {author} {\bibfnamefont {B.}~\bibnamefont
  {Bhandari}}, \bibinfo {author} {\bibfnamefont {G.}~\bibnamefont
  {Chiriac\`o}}, \bibinfo {author} {\bibfnamefont {P.~A.}\ \bibnamefont
  {Erdman}}, \bibinfo {author} {\bibfnamefont {R.}~\bibnamefont {Fazio}}, \
  and\ \bibinfo {author} {\bibfnamefont {F.}~\bibnamefont {Taddei}},\
  }\bibfield  {title} {\enquote {\bibinfo {title} {Thermal drag in electronic
  conductors},}\ }\href {\doibase 10.1103/PhysRevB.98.035415} {\bibfield
  {journal} {\bibinfo  {journal} {Phys. Rev. B}\ }\textbf {\bibinfo {volume}
  {98}},\ \bibinfo {pages} {035415} (\bibinfo {year} {2018})}\BibitemShut
  {NoStop}%
\bibitem [{\citenamefont {Zhang}\ \emph {et~al.}(2015)\citenamefont {Zhang},
  \citenamefont {Lin},\ and\ \citenamefont {Chen}}]{zhang:2015}%
  \BibitemOpen
  \bibfield  {author} {\bibinfo {author} {\bibfnamefont {Y.}~\bibnamefont
  {Zhang}}, \bibinfo {author} {\bibfnamefont {G.}~\bibnamefont {Lin}}, \ and\
  \bibinfo {author} {\bibfnamefont {J.}~\bibnamefont {Chen}},\ }\bibfield
  {title} {\enquote {\bibinfo {title} {Three-terminal quantum-dot
  refrigerators},}\ }\href {\doibase 10.1103/PhysRevE.91.052118} {\bibfield
  {journal} {\bibinfo  {journal} {Phys. Rev. E}\ }\textbf {\bibinfo {volume}
  {91}},\ \bibinfo {pages} {052118} (\bibinfo {year} {2015})}\BibitemShut
  {NoStop}%
\bibitem [{\citenamefont
  {S{\ifmmode\acute{a}\else\'{a}\fi}nchez}(2017)}]{sanchez:2017}%
  \BibitemOpen
  \bibfield  {author} {\bibinfo {author} {\bibfnamefont {R.}~\bibnamefont
  {S{\ifmmode\acute{a}\else\'{a}\fi}nchez}},\ }\bibfield  {title} {\enquote
  {\bibinfo {title} {{Correlation-induced refrigeration with superconducting
  single-electron transistors}},}\ }\href {\doibase 10.1063/1.5008481}
  {\bibfield  {journal} {\bibinfo  {journal} {Appl. Phys. Lett.}\ }\textbf
  {\bibinfo {volume} {111}},\ \bibinfo {pages} {223103} (\bibinfo {year}
  {2017})}\BibitemShut {NoStop}%
\bibitem [{\citenamefont {Kouwenhoven}\ \emph {et~al.}(1997)\citenamefont
  {Kouwenhoven}, \citenamefont {Marcus}, \citenamefont {McEuen}, \citenamefont
  {Tarucha}, \citenamefont {Westervelt},\ and\ \citenamefont
  {Wingreen}}]{kouwenhoven:1997}%
  \BibitemOpen
  \bibfield  {author} {\bibinfo {author} {\bibfnamefont {L.~P.}\ \bibnamefont
  {Kouwenhoven}}, \bibinfo {author} {\bibfnamefont {C.~M.}\ \bibnamefont
  {Marcus}}, \bibinfo {author} {\bibfnamefont {P.~L.}\ \bibnamefont {McEuen}},
  \bibinfo {author} {\bibfnamefont {S.}~\bibnamefont {Tarucha}}, \bibinfo
  {author} {\bibfnamefont {R.~M.}\ \bibnamefont {Westervelt}}, \ and\ \bibinfo
  {author} {\bibfnamefont {N.~S.}\ \bibnamefont {Wingreen}},\ }\enquote
  {\bibinfo {title} {Electron transport in quantum dots},}\ in\ \href {\doibase
  10.1007/978-94-015-8839-3_4} {\emph {\bibinfo {booktitle} {Mesoscopic
  Electron Transport}}},\ \bibinfo {editor} {edited by\ \bibinfo {editor}
  {\bibfnamefont {L.~L.}\ \bibnamefont {Sohn}}, \bibinfo {editor}
  {\bibfnamefont {L.~P.}\ \bibnamefont {Kouwenhoven}}, \ and\ \bibinfo {editor}
  {\bibfnamefont {G.}~\bibnamefont {Sch{\"o}n}}}\ (\bibinfo  {publisher}
  {Springer Netherlands},\ \bibinfo {address} {Dordrecht},\ \bibinfo {year}
  {1997})\ pp.\ \bibinfo {pages} {105--214}\BibitemShut {NoStop}%
\bibitem [{\citenamefont {S\'anchez}\ \emph {et~al.}(2017)\citenamefont
  {S\'anchez}, \citenamefont {Thierschmann},\ and\ \citenamefont
  {Molenkamp}}]{sanchez_all-thermal_2017}%
  \BibitemOpen
  \bibfield  {author} {\bibinfo {author} {\bibfnamefont {R.}~\bibnamefont
  {S\'anchez}}, \bibinfo {author} {\bibfnamefont {H.}~\bibnamefont
  {Thierschmann}}, \ and\ \bibinfo {author} {\bibfnamefont {L.~W.}\
  \bibnamefont {Molenkamp}},\ }\bibfield  {title} {\enquote {\bibinfo {title}
  {All-thermal transistor based on stochastic switching},}\ }\href {\doibase
  10.1103/PhysRevB.95.241401} {\bibfield  {journal} {\bibinfo  {journal} {Phys.
  Rev. B}\ }\textbf {\bibinfo {volume} {95}},\ \bibinfo {pages} {241401}
  (\bibinfo {year} {2017})}\BibitemShut {NoStop}%
\bibitem [{\citenamefont {Lu}\ \emph {et~al.}(2003)\citenamefont {Lu},
  \citenamefont {Ji}, \citenamefont {Pfeiffer}, \citenamefont {West},\ and\
  \citenamefont {Rimberg}}]{lu:2003}%
  \BibitemOpen
  \bibfield  {author} {\bibinfo {author} {\bibfnamefont {W.}~\bibnamefont
  {Lu}}, \bibinfo {author} {\bibfnamefont {Z.}~\bibnamefont {Ji}}, \bibinfo
  {author} {\bibfnamefont {L.}~\bibnamefont {Pfeiffer}}, \bibinfo {author}
  {\bibfnamefont {K.~W.}\ \bibnamefont {West}}, \ and\ \bibinfo {author}
  {\bibfnamefont {A.~J.}\ \bibnamefont {Rimberg}},\ }\bibfield  {title}
  {\enquote {\bibinfo {title} {Real-time detection of electron tunnelling in a
  quantum dot},}\ }\href {https://doi.org/10.1038/nature01642} {\bibfield
  {journal} {\bibinfo  {journal} {Nature}\ }\textbf {\bibinfo {volume} {423}},\
  \bibinfo {pages} {422} (\bibinfo {year} {2003})}\BibitemShut {NoStop}%
\bibitem [{\citenamefont {Vandersypen}\ \emph {et~al.}(2004)\citenamefont
  {Vandersypen}, \citenamefont {Elzerman}, \citenamefont {Schouten},
  \citenamefont {Willems~van Beveren}, \citenamefont {Hanson},\ and\
  \citenamefont {Kouwenhoven}}]{vandersypen:2004}%
  \BibitemOpen
  \bibfield  {author} {\bibinfo {author} {\bibfnamefont {L.~M.~K.}\
  \bibnamefont {Vandersypen}}, \bibinfo {author} {\bibfnamefont {J.~M.}\
  \bibnamefont {Elzerman}}, \bibinfo {author} {\bibfnamefont {R.~N.}\
  \bibnamefont {Schouten}}, \bibinfo {author} {\bibfnamefont {L.~H.}\
  \bibnamefont {Willems~van Beveren}}, \bibinfo {author} {\bibfnamefont
  {R.}~\bibnamefont {Hanson}}, \ and\ \bibinfo {author} {\bibfnamefont {L.~P.}\
  \bibnamefont {Kouwenhoven}},\ }\bibfield  {title} {\enquote {\bibinfo {title}
  {Real-time detection of single-electron tunneling using a quantum point
  contact},}\ }\href {\doibase 10.1063/1.1815041} {\bibfield  {journal}
  {\bibinfo  {journal} {Appl. Phys. Lett.}\ }\textbf {\bibinfo {volume} {85}},\
  \bibinfo {pages} {4394} (\bibinfo {year} {2004})}\BibitemShut {NoStop}%
\bibitem [{\citenamefont {Fujisawa}\ \emph {et~al.}(2006)\citenamefont
  {Fujisawa}, \citenamefont {Hayashi}, \citenamefont {Tomita},\ and\
  \citenamefont {Hirayama}}]{fujisawa:2006}%
  \BibitemOpen
  \bibfield  {author} {\bibinfo {author} {\bibfnamefont {T.}~\bibnamefont
  {Fujisawa}}, \bibinfo {author} {\bibfnamefont {T.}~\bibnamefont {Hayashi}},
  \bibinfo {author} {\bibfnamefont {R.}~\bibnamefont {Tomita}}, \ and\ \bibinfo
  {author} {\bibfnamefont {Y.}~\bibnamefont {Hirayama}},\ }\bibfield  {title}
  {\enquote {\bibinfo {title} {{Bidirectional Counting of Single Electrons}},}\
  }\href {\doibase 10.1126/science.1126788} {\bibfield  {journal} {\bibinfo
  {journal} {Science}\ }\textbf {\bibinfo {volume} {312}},\ \bibinfo {pages}
  {1634} (\bibinfo {year} {2006})}\BibitemShut {NoStop}%
\bibitem [{\citenamefont {Gustavsson}\ \emph {et~al.}(2006)\citenamefont
  {Gustavsson}, \citenamefont {Leturcq}, \citenamefont
  {Simovi\ifmmode~\check{c}\else \v{c}\fi{}}, \citenamefont {Schleser},
  \citenamefont {Ihn}, \citenamefont {Studerus}, \citenamefont {Ensslin},
  \citenamefont {Driscoll},\ and\ \citenamefont {Gossard}}]{gustavsson:2006}%
  \BibitemOpen
  \bibfield  {author} {\bibinfo {author} {\bibfnamefont {S.}~\bibnamefont
  {Gustavsson}}, \bibinfo {author} {\bibfnamefont {R.}~\bibnamefont {Leturcq}},
  \bibinfo {author} {\bibfnamefont {B.}~\bibnamefont
  {Simovi\ifmmode~\check{c}\else \v{c}\fi{}}}, \bibinfo {author} {\bibfnamefont
  {R.}~\bibnamefont {Schleser}}, \bibinfo {author} {\bibfnamefont
  {T.}~\bibnamefont {Ihn}}, \bibinfo {author} {\bibfnamefont {P.}~\bibnamefont
  {Studerus}}, \bibinfo {author} {\bibfnamefont {K.}~\bibnamefont {Ensslin}},
  \bibinfo {author} {\bibfnamefont {D.~C.}\ \bibnamefont {Driscoll}}, \ and\
  \bibinfo {author} {\bibfnamefont {A.~C.}\ \bibnamefont {Gossard}},\
  }\bibfield  {title} {\enquote {\bibinfo {title} {Counting statistics of
  single electron transport in a quantum dot},}\ }\href {\doibase
  10.1103/PhysRevLett.96.076605} {\bibfield  {journal} {\bibinfo  {journal}
  {Phys. Rev. Lett.}\ }\textbf {\bibinfo {volume} {96}},\ \bibinfo {pages}
  {076605} (\bibinfo {year} {2006})}\BibitemShut {NoStop}%
\bibitem [{\citenamefont {Ubbelohde}\ \emph {et~al.}(2012)\citenamefont
  {Ubbelohde}, \citenamefont {Fricke}, \citenamefont {Flindt}, \citenamefont
  {Hohls},\ and\ \citenamefont {Haug}}]{ubbelohde:2012}%
  \BibitemOpen
  \bibfield  {author} {\bibinfo {author} {\bibfnamefont {N.}~\bibnamefont
  {Ubbelohde}}, \bibinfo {author} {\bibfnamefont {C.}~\bibnamefont {Fricke}},
  \bibinfo {author} {\bibfnamefont {C.}~\bibnamefont {Flindt}}, \bibinfo
  {author} {\bibfnamefont {F.}~\bibnamefont {Hohls}}, \ and\ \bibinfo {author}
  {\bibfnamefont {R.~J.}\ \bibnamefont {Haug}},\ }\bibfield  {title} {\enquote
  {\bibinfo {title} {Measurement of finite-frequency current statistics in a
  single-electron transistor},}\ }\href {https://doi.org/10.1038/ncomms1620}
  {\bibfield  {journal} {\bibinfo  {journal} {Nat. Commun.}\ }\textbf {\bibinfo
  {volume} {3}},\ \bibinfo {pages} {612} (\bibinfo {year} {2012})}\BibitemShut
  {NoStop}%
\bibitem [{\citenamefont {Breuer}\ and\ \citenamefont
  {Petruccione}(2002)}]{breuer:book}%
  \BibitemOpen
  \bibfield  {author} {\bibinfo {author} {\bibfnamefont {H.-P.}\ \bibnamefont
  {Breuer}}\ and\ \bibinfo {author} {\bibfnamefont {F.}~\bibnamefont
  {Petruccione}},\ }\href {\doibase 10.1093/acprof:oso/9780199213900.001.0001}
  {\emph {\bibinfo {title} {The theory of open quantum systems}}}\ (\bibinfo
  {publisher} {Oxford University Press},\ \bibinfo {year} {2002})\BibitemShut
  {NoStop}%
\bibitem [{\citenamefont {Schaller}(2014)}]{schaller:book}%
  \BibitemOpen
  \bibfield  {author} {\bibinfo {author} {\bibfnamefont {G.}~\bibnamefont
  {Schaller}},\ }\href {\doibase 10.1007/978-3-319-03877-3} {\emph {\bibinfo
  {title} {Open Quantum Systems Far from Equilibrium}}}\ (\bibinfo  {publisher}
  {Springer},\ \bibinfo {year} {2014})\BibitemShut {NoStop}%
\bibitem [{\citenamefont {Dynes}\ \emph {et~al.}(1978)\citenamefont {Dynes},
  \citenamefont {Narayanamurti},\ and\ \citenamefont {Garno}}]{dynes:1978}%
  \BibitemOpen
  \bibfield  {author} {\bibinfo {author} {\bibfnamefont {R.~C.}\ \bibnamefont
  {Dynes}}, \bibinfo {author} {\bibfnamefont {V.}~\bibnamefont
  {Narayanamurti}}, \ and\ \bibinfo {author} {\bibfnamefont {J.~P.}\
  \bibnamefont {Garno}},\ }\bibfield  {title} {\enquote {\bibinfo {title}
  {Direct measurement of quasiparticle-lifetime broadening in a strong-coupled
  superconductor},}\ }\href {\doibase 10.1103/PhysRevLett.41.1509} {\bibfield
  {journal} {\bibinfo  {journal} {Phys. Rev. Lett.}\ }\textbf {\bibinfo
  {volume} {41}},\ \bibinfo {pages} {1509--1512} (\bibinfo {year}
  {1978})}\BibitemShut {NoStop}%
\bibitem [{\citenamefont {Schnakenberg}(1976)}]{schnakenberg:1976}%
  \BibitemOpen
  \bibfield  {author} {\bibinfo {author} {\bibfnamefont {J.}~\bibnamefont
  {Schnakenberg}},\ }\bibfield  {title} {\enquote {\bibinfo {title} {Network
  theory of microscopic and macroscopic behavior of master equation systems},}\
  }\href {\doibase 10.1103/RevModPhys.48.571} {\bibfield  {journal} {\bibinfo
  {journal} {Rev. Mod. Phys.}\ }\textbf {\bibinfo {volume} {48}},\ \bibinfo
  {pages} {571} (\bibinfo {year} {1976})}\BibitemShut {NoStop}%
\bibitem [{\citenamefont {S{\'{a}}nchez}\ and\ \citenamefont
  {B{\"u}ttiker}(2012)}]{sanchez_detection_2012}%
  \BibitemOpen
  \bibfield  {author} {\bibinfo {author} {\bibfnamefont {R.}~\bibnamefont
  {S{\'{a}}nchez}}\ and\ \bibinfo {author} {\bibfnamefont {M.}~\bibnamefont
  {B{\"u}ttiker}},\ }\bibfield  {title} {\enquote {\bibinfo {title} {Detection
  of single-electron heat transfer statistics},}\ }\href {\doibase
  10.1209/0295-5075/100/47008} {\bibfield  {journal} {\bibinfo  {journal}
  {{EPL}}\ }\textbf {\bibinfo {volume} {100}},\ \bibinfo {pages} {47008}
  (\bibinfo {year} {2012})}\BibitemShut {NoStop}%
\bibitem [{\citenamefont {S{\'{a}}nchez}\ and\ \citenamefont
  {B{\"u}ttiker}(2013)}]{sanchez_erratum_2013}%
  \BibitemOpen
  \bibfield  {author} {\bibinfo {author} {\bibfnamefont {R.}~\bibnamefont
  {S{\'{a}}nchez}}\ and\ \bibinfo {author} {\bibfnamefont {M.}~\bibnamefont
  {B{\"u}ttiker}},\ }\bibfield  {title} {\enquote {\bibinfo {title} {Erratum:
  Detection of single-electron heat transfer statistics},}\ }\href {\doibase
  10.1209/0295-5075/104/49901} {\bibfield  {journal} {\bibinfo  {journal}
  {{EPL}}\ }\textbf {\bibinfo {volume} {104}},\ \bibinfo {pages} {49901}
  (\bibinfo {year} {2013})}\BibitemShut {NoStop}%
\bibitem [{\citenamefont {S{\ifmmode\acute{a}\else\'{a}\fi}nchez}\ \emph
  {et~al.}(2017)\citenamefont {S{\ifmmode\acute{a}\else\'{a}\fi}nchez},
  \citenamefont {Thierschmann},\ and\ \citenamefont
  {Molenkamp}}]{sanchez_single-electron_2017}%
  \BibitemOpen
  \bibfield  {author} {\bibinfo {author} {\bibfnamefont {R.}~\bibnamefont
  {S{\ifmmode\acute{a}\else\'{a}\fi}nchez}}, \bibinfo {author} {\bibfnamefont
  {H.}~\bibnamefont {Thierschmann}}, \ and\ \bibinfo {author} {\bibfnamefont
  {L.~W.}\ \bibnamefont {Molenkamp}},\ }\bibfield  {title} {\enquote {\bibinfo
  {title} {{Single-electron thermal devices coupled to a mesoscopic gate}},}\
  }\href {\doibase 10.1088/1367-2630/aa8b94} {\bibfield  {journal} {\bibinfo
  {journal} {New J. Phys.}\ }\textbf {\bibinfo {volume} {19}},\ \bibinfo
  {pages} {113040} (\bibinfo {year} {2017})}\BibitemShut {NoStop}%
\bibitem [{\citenamefont {Benenti}\ \emph {et~al.}(2017)\citenamefont
  {Benenti}, \citenamefont {Casati}, \citenamefont {Saito},\ and\ \citenamefont
  {Whitney}}]{benenti:2017}%
  \BibitemOpen
  \bibfield  {author} {\bibinfo {author} {\bibfnamefont {G.}~\bibnamefont
  {Benenti}}, \bibinfo {author} {\bibfnamefont {G.}~\bibnamefont {Casati}},
  \bibinfo {author} {\bibfnamefont {K.}~\bibnamefont {Saito}}, \ and\ \bibinfo
  {author} {\bibfnamefont {R.~S.}\ \bibnamefont {Whitney}},\ }\bibfield
  {title} {\enquote {\bibinfo {title} {Fundamental aspects of steady-state
  conversion of heat to work at the nanoscale},}\ }\href {\doibase
  10.1016/j.physrep.2017.05.008} {\bibfield  {journal} {\bibinfo  {journal}
  {Phys. Rep.}\ }\textbf {\bibinfo {volume} {694}},\ \bibinfo {pages} {1}
  (\bibinfo {year} {2017})}\BibitemShut {NoStop}%
\bibitem [{\citenamefont {S\'anchez}\ \emph {et~al.}(2013)\citenamefont
  {S\'anchez}, \citenamefont {Sothmann}, \citenamefont {Jordan},\ and\
  \citenamefont {B\"uttiker}}]{sanchez:2013}%
  \BibitemOpen
  \bibfield  {author} {\bibinfo {author} {\bibfnamefont {R.}~\bibnamefont
  {S\'anchez}}, \bibinfo {author} {\bibfnamefont {B.}~\bibnamefont {Sothmann}},
  \bibinfo {author} {\bibfnamefont {A.~N.}\ \bibnamefont {Jordan}}, \ and\
  \bibinfo {author} {\bibfnamefont {M.}~\bibnamefont {B\"uttiker}},\ }\bibfield
   {title} {\enquote {\bibinfo {title} {Correlations of heat and charge
  currents in quantum-dot thermoelectric engines},}\ }\href
  {http://stacks.iop.org/1367-2630/15/i=12/a=125001} {\bibfield  {journal}
  {\bibinfo  {journal} {New J. Phys.}\ }\textbf {\bibinfo {volume} {15}},\
  \bibinfo {pages} {125001} (\bibinfo {year} {2013})}\BibitemShut {NoStop}%
\bibitem [{\citenamefont {Hofer}\ \emph {et~al.}(2016)\citenamefont {Hofer},
  \citenamefont {Souquet},\ and\ \citenamefont {Clerk}}]{hofer:2016prb}%
  \BibitemOpen
  \bibfield  {author} {\bibinfo {author} {\bibfnamefont {P.~P.}\ \bibnamefont
  {Hofer}}, \bibinfo {author} {\bibfnamefont {J.-R.}\ \bibnamefont {Souquet}},
  \ and\ \bibinfo {author} {\bibfnamefont {A.~A.}\ \bibnamefont {Clerk}},\
  }\bibfield  {title} {\enquote {\bibinfo {title} {Quantum heat engine based on
  photon-assisted {C}ooper pair tunneling},}\ }\href {\doibase
  10.1103/PhysRevB.93.041418} {\bibfield  {journal} {\bibinfo  {journal} {Phys.
  Rev. B}\ }\textbf {\bibinfo {volume} {93}},\ \bibinfo {pages} {041418(R)}
  (\bibinfo {year} {2016})}\BibitemShut {NoStop}%
\bibitem [{\citenamefont {Beenakker}\ and\ \citenamefont
  {Staring}(1992)}]{beenakker_theory_1992}%
  \BibitemOpen
  \bibfield  {author} {\bibinfo {author} {\bibfnamefont {C.~W.~J.}\
  \bibnamefont {Beenakker}}\ and\ \bibinfo {author} {\bibfnamefont {A.~A.~M.}\
  \bibnamefont {Staring}},\ }\bibfield  {title} {\enquote {\bibinfo {title}
  {{Theory of the thermopower of a quantum dot}},}\ }\href {\doibase
  10.1103/PhysRevB.46.9667} {\bibfield  {journal} {\bibinfo  {journal} {Phys.
  Rev. B}\ }\textbf {\bibinfo {volume} {46}},\ \bibinfo {pages} {9667}
  (\bibinfo {year} {1992})}\BibitemShut {NoStop}%
\bibitem [{\citenamefont {Andrieux}\ \emph {et~al.}(2009)\citenamefont
  {Andrieux}, \citenamefont {Gaspard}, \citenamefont {Monnai},\ and\
  \citenamefont {Tasaki}}]{andrieux:2009}%
  \BibitemOpen
  \bibfield  {author} {\bibinfo {author} {\bibfnamefont {D.}~\bibnamefont
  {Andrieux}}, \bibinfo {author} {\bibfnamefont {P.}~\bibnamefont {Gaspard}},
  \bibinfo {author} {\bibfnamefont {T.}~\bibnamefont {Monnai}}, \ and\ \bibinfo
  {author} {\bibfnamefont {S.}~\bibnamefont {Tasaki}},\ }\bibfield  {title}
  {\enquote {\bibinfo {title} {The fluctuation theorem for currents in open
  quantum systems},}\ }\href {\doibase 10.1088/1367-2630/11/4/043014}
  {\bibfield  {journal} {\bibinfo  {journal} {New Journal of Physics}\ }\textbf
  {\bibinfo {volume} {11}},\ \bibinfo {pages} {043014} (\bibinfo {year}
  {2009})}\BibitemShut {NoStop}%
\bibitem [{\citenamefont {Krogstrup}\ \emph {et~al.}(2015)\citenamefont
  {Krogstrup}, \citenamefont {Ziino}, \citenamefont {Chang}, \citenamefont
  {Albrecht}, \citenamefont {Madsen}, \citenamefont {Johnson}, \citenamefont
  {Nyg{\aa}rd}, \citenamefont {Marcus},\ and\ \citenamefont
  {Jespersen}}]{krogstrup:2015}%
  \BibitemOpen
  \bibfield  {author} {\bibinfo {author} {\bibfnamefont {P.}~\bibnamefont
  {Krogstrup}}, \bibinfo {author} {\bibfnamefont {N.~L.~B.}\ \bibnamefont
  {Ziino}}, \bibinfo {author} {\bibfnamefont {W.}~\bibnamefont {Chang}},
  \bibinfo {author} {\bibfnamefont {S.~M.}\ \bibnamefont {Albrecht}}, \bibinfo
  {author} {\bibfnamefont {M.~H.}\ \bibnamefont {Madsen}}, \bibinfo {author}
  {\bibfnamefont {E.}~\bibnamefont {Johnson}}, \bibinfo {author} {\bibfnamefont
  {J.}~\bibnamefont {Nyg{\aa}rd}}, \bibinfo {author} {\bibfnamefont {C.~M.}\
  \bibnamefont {Marcus}}, \ and\ \bibinfo {author} {\bibfnamefont {T.~S.}\
  \bibnamefont {Jespersen}},\ }\bibfield  {title} {\enquote {\bibinfo {title}
  {{Epitaxy of semiconductor{\textendash}superconductor nanowires}},}\ }\href
  {\doibase 10.1038/nmat4176} {\bibfield  {journal} {\bibinfo  {journal} {Nat.
  Mater.}\ }\textbf {\bibinfo {volume} {14}},\ \bibinfo {pages} {400--406}
  (\bibinfo {year} {2015})}\BibitemShut {NoStop}%
\bibitem [{\citenamefont {Shabani}\ \emph {et~al.}(2016)\citenamefont
  {Shabani}, \citenamefont {Kjaergaard}, \citenamefont {Suominen},
  \citenamefont {Kim}, \citenamefont {Nichele}, \citenamefont {Pakrouski},
  \citenamefont {Stankevic}, \citenamefont {Lutchyn}, \citenamefont
  {Krogstrup}, \citenamefont {Feidenhans'l}, \citenamefont {Kraemer},
  \citenamefont {Nayak}, \citenamefont {Troyer}, \citenamefont {Marcus},\ and\
  \citenamefont {Palmstr{\o}m}}]{shabani:2016}%
  \BibitemOpen
  \bibfield  {author} {\bibinfo {author} {\bibfnamefont {J.}~\bibnamefont
  {Shabani}}, \bibinfo {author} {\bibfnamefont {M.}~\bibnamefont {Kjaergaard}},
  \bibinfo {author} {\bibfnamefont {H.~J.}\ \bibnamefont {Suominen}}, \bibinfo
  {author} {\bibfnamefont {Y.}~\bibnamefont {Kim}}, \bibinfo {author}
  {\bibfnamefont {F.}~\bibnamefont {Nichele}}, \bibinfo {author} {\bibfnamefont
  {K.}~\bibnamefont {Pakrouski}}, \bibinfo {author} {\bibfnamefont
  {T.}~\bibnamefont {Stankevic}}, \bibinfo {author} {\bibfnamefont {R.~M.}\
  \bibnamefont {Lutchyn}}, \bibinfo {author} {\bibfnamefont {P.}~\bibnamefont
  {Krogstrup}}, \bibinfo {author} {\bibfnamefont {R.}~\bibnamefont
  {Feidenhans'l}}, \bibinfo {author} {\bibfnamefont {S.}~\bibnamefont
  {Kraemer}}, \bibinfo {author} {\bibfnamefont {C.}~\bibnamefont {Nayak}},
  \bibinfo {author} {\bibfnamefont {M.}~\bibnamefont {Troyer}}, \bibinfo
  {author} {\bibfnamefont {C.~M.}\ \bibnamefont {Marcus}}, \ and\ \bibinfo
  {author} {\bibfnamefont {C.~J.}\ \bibnamefont {Palmstr{\o}m}},\ }\bibfield
  {title} {\enquote {\bibinfo {title} {{Two-dimensional epitaxial
  superconductor-semiconductor heterostructures: A platform for topological
  superconducting networks}},}\ }\href {\doibase 10.1103/PhysRevB.93.155402}
  {\bibfield  {journal} {\bibinfo  {journal} {Phys. Rev. B}\ }\textbf {\bibinfo
  {volume} {93}},\ \bibinfo {pages} {155402} (\bibinfo {year}
  {2016})}\BibitemShut {NoStop}%
\bibitem [{\citenamefont {Kjaergaard}\ \emph {et~al.}(2016)\citenamefont
  {Kjaergaard}, \citenamefont {Nichele}, \citenamefont {Suominen},
  \citenamefont {Nowak}, \citenamefont {Wimmer}, \citenamefont {Akhmerov},
  \citenamefont {Folk}, \citenamefont {Flensberg}, \citenamefont {Shabani},
  \citenamefont {Palmstr{\o}m},\ and\ \citenamefont
  {Marcus}}]{kjaergaard:2016}%
  \BibitemOpen
  \bibfield  {author} {\bibinfo {author} {\bibfnamefont {M.}~\bibnamefont
  {Kjaergaard}}, \bibinfo {author} {\bibfnamefont {F.}~\bibnamefont {Nichele}},
  \bibinfo {author} {\bibfnamefont {H.~J.}\ \bibnamefont {Suominen}}, \bibinfo
  {author} {\bibfnamefont {M.~P.}\ \bibnamefont {Nowak}}, \bibinfo {author}
  {\bibfnamefont {M.}~\bibnamefont {Wimmer}}, \bibinfo {author} {\bibfnamefont
  {A.~R.}\ \bibnamefont {Akhmerov}}, \bibinfo {author} {\bibfnamefont {J.~A.}\
  \bibnamefont {Folk}}, \bibinfo {author} {\bibfnamefont {K.}~\bibnamefont
  {Flensberg}}, \bibinfo {author} {\bibfnamefont {J.}~\bibnamefont {Shabani}},
  \bibinfo {author} {\bibfnamefont {C.~J.}\ \bibnamefont {Palmstr{\o}m}}, \
  and\ \bibinfo {author} {\bibfnamefont {C.~M.}\ \bibnamefont {Marcus}},\
  }\bibfield  {title} {\enquote {\bibinfo {title} {{Quantized conductance
  doubling and hard gap in a two-dimensional
  semiconductor{\textendash}superconductor heterostructure}},}\ }\href
  {\doibase 10.1038/ncomms12841} {\bibfield  {journal} {\bibinfo  {journal}
  {Nat. Commun.}\ }\textbf {\bibinfo {volume} {7}},\ \bibinfo {pages} {1--6}
  (\bibinfo {year} {2016})}\BibitemShut {NoStop}%
\bibitem [{\citenamefont {Averin}\ and\ \citenamefont
  {Likharev}(1986)}]{averin:1986}%
  \BibitemOpen
  \bibfield  {author} {\bibinfo {author} {\bibfnamefont {D.~V.}\ \bibnamefont
  {Averin}}\ and\ \bibinfo {author} {\bibfnamefont {K.~K.}\ \bibnamefont
  {Likharev}},\ }\bibfield  {title} {\enquote {\bibinfo {title} {Coulomb
  blockade of single-electron tunneling, and coherent oscillations in small
  tunnel junctions},}\ }\href {\doibase 10.1007/BF00683469} {\bibfield
  {journal} {\bibinfo  {journal} {J. Low Temp. Phys.}\ }\textbf {\bibinfo
  {volume} {62}},\ \bibinfo {pages} {345} (\bibinfo {year} {1986})}\BibitemShut
  {NoStop}%
\end{thebibliography}%

\end{document}